\documentclass[journal, twoside]{IEEEtran}
\usepackage{indentfirst}
\usepackage{graphicx}
\usepackage{amsmath}
\usepackage{amssymb}
\usepackage{amsfonts}
\usepackage{mathrsfs}
\usepackage{leftidx}
\usepackage{color}
\usepackage{amsmath}
\usepackage{arydshln}
\usepackage{amsthm}
\usepackage{ragged2e}
\usepackage{cite}
\usepackage{enumerate}
\usepackage{longtable}
\usepackage{float}
\usepackage{stfloats}
\usepackage{hyperref}
\usepackage{algpseudocode}
\usepackage{algorithm}
\usepackage[caption=false,font=footnotesize]{subfig}
\usepackage{tabularx}
\usepackage{makecell}
\usepackage{url}
\usepackage{adjustbox} 

\usepackage{multirow} 
\usepackage{booktabs}
\theoremstyle{plain}

\usepackage{caption}

\usepackage{multirow}   
\usepackage{array}      

\usepackage{xltabular}

\newcolumntype{P}[1]{>{\raggedright\arraybackslash\footnotesize}m{#1}}
\newcolumntype{A}[1]{>{\centering\arraybackslash\footnotesize}m{#1}}

\usepackage[table,dvipsnames]{xcolor}
\definecolor{aa}{RGB}{175,238,238}
\definecolor{bb}{RGB}{255,255,255}

\usepackage{bm}
\usepackage{makecell}

\begin{document}

\title{Generative Diffusion Models for Wireless Networks: Fundamental, Architecture, and State-of-the-Art}

\author{Dayu Fan, Rui Meng,~\IEEEmembership{Member,~IEEE,} Xiaodong Xu,~\IEEEmembership{Senior Member,~IEEE,} Yiming Liu,~\IEEEmembership{Member,~IEEE,}

Guoshun Nan,~\IEEEmembership{Member,~IEEE,} Chenyuan Feng,~\IEEEmembership{Member,~IEEE,} Shujun Han,~\IEEEmembership{Member,~IEEE,}

Song Gao, Bingxuan Xu,~\IEEEmembership{Graduate Student Member,~IEEE,} 
Dusit Niyato,~\IEEEmembership{Fellow,~IEEE,} 

Tony Q. S. Quek,~\IEEEmembership{Fellow,~IEEE,} and Ping Zhang,~\IEEEmembership{Fellow,~IEEE}

\thanks{
This work was supported in part by the National Key Research and Development Program of China under Grant 2020YFB1806905; in part by the National Natural Science Foundation of China under Grant 62501066 and under Grant U24B20131; in part by the Beijing Municipal Natural Science Foundation under Grant L242012; and in part by the Long Term Science and Technology Plan for Broadcasting, Television, and Online Audiovisual Programs under Grant 2025AD0300. \textit{(Corresponding author: Rui Meng.)}

Dayu Fan, Rui Meng, Xiaodong Xu, Yiming Liu, Song Gao, Bingxuan Xu, and Ping Zhang are with State Key Laboratory of Networking and Switching Technology, Beijing University of Posts and Telecommunications, Beijing, China (e-mail: fandayu@bupt.edu.cn; buptmengrui@bupt.edu.cn; xuxiaodong@bupt.edu.cn; liuyiming@bupt.edu.cn;
wkd251292@bupt.edu.cn; xubingxuan@bupt.edu.cn; pzhang@bupt.edu.cn).

Guoshun Nan and Shujun Han are with National Engineering Research Center for Mobile Network Technologies, Beijing University of Posts and Telecommunications, Beijing, China (e-mail: nanguo2021@bupt.edu.cn; hanshujun@bupt.edu.cn).

Chenyuan Feng is with Department of Computer Science, University of Exeter, EX4 4QF Exeter, U.K. (e-mail: c.feng@exeter.ac.uk).

Dusit Niyato is with College of Computing and Data Science, Nanyang Technological University, Singapore (e-mail: dniyato@ntu.edu.sg).

Tony Q. S. Quek is with the Singapore University of Technology and Design, Singapore (e-mail: tonyquek@sutd.edu.sg).

}}

\maketitle

\begin{abstract}

With the rapid development of Generative Artificial Intelligence (GAI) technology, Generative Diffusion Models (GDMs) have shown significant empowerment potential in the field of wireless networks due to advantages, such as noise resistance, training stability, controllability, and multimodal generation. Although there have been multiple studies focusing on GDMs for wireless networks, there is still a lack of comprehensive reviews on their technological evolution. Motivated by this, we systematically explore the application of GDMs in wireless networks. Firstly, we identify the core challenges of wireless networks and argue why GDMs are uniquely suited to address them. We then introduce the mathematical principles of GDMs and representative models. Furthermore, we organize our comprehensive review through a structured taxonomy that categorizes GDM-based schemes into the sensing, transmission, and Applications, complemented by a security plane. For each representative scheme, we analyze its innovative points, the role of GDMs, strengths, and weaknesses. Ultimately, we extract key challenges and provide potential solutions, with the aim of providing directional guidance for future research in this field.

\end{abstract}


\section{Introduction}

\subsection{Motivation}

The global industrial and academic communities are engaged in comprehensive and profound explorations of 6G. As a revolutionary mobile communication paradigm, 6G will not only achieve performance leaps and scenario expansions beyond 5G capabilities, but also establish an innovation ecosystem integrating multi-domain technological convergence \cite{dang2020should}. Within the evolutionary trajectory of 6G, the deep fusion of communications and artificial intelligence (AI) has emerged as the core driver \cite{cui2025overview}. 
It will help construct intellicise (intelligent and concise) networks with self-perception, self-learning, and self-optimization capabilities, thus enhancing spectrum efficiency, assuring reliability assurance, and laying foundational support for cutting-edge applications \cite{zhang2024intellicise,meng2026intellicise}.
This profound integration is catalyzing a paradigm revolution across the entire communications domain, spanning breakthroughs in 
standardization framework development\footnote{\url{https://www.itu.int/en/ITU-R/study-groups/rsg5/rwp5d/imt-2030/pages/default.aspx}}, fundamental theories \cite{niu2025mathematical,shao2024theory}, architectural transformations \cite{duan20236g}, and application scenario expansions
\cite{xiang2024realizing,zhang2025industrial,li2022federated,prathiba2025digital}.

However, the realization of this vision is confronted by several fundamental and long-standing challenges that traditional methods struggle to overcome:
\begin{itemize}
    \item \textbf{The Challenge of Environmental Complexity:} The physics of wireless propagation creates an environment of extreme randomness, dynamism, and high dimensionality. Channel distributions are profoundly complex, often exhibiting non-Gaussian and non-linear characteristics that defy accurate mathematical modeling \cite{wei2022multi}.
    \item \textbf{The Dilemma of Data Acquisition:} While AI models thrive on data, acquiring large-scale, high-quality, and diverse datasets for wireless networks is exceptionally difficult and costly, leading to a persistent scarcity of training data \cite{lee2024generating,sengupta2023generative}.
    \item \textbf{The Demand for High-Fidelity and Controllable Generation:} Emerging applications like network digital twins\cite{li2025generative}, and semantic communications (SemComs) require the ability to generate synthetic data that is indistinguishable from reality and can be controlled by specific conditions.
\end{itemize}
These challenges call for a new generation of AI models capable of learning from complex distributions, even with limited data, while offering high fidelity and control. While early generative models like Generative Adversarial Networks (GANs)\cite{goodfellow2014generative7.21} have shown promise, they often suffer from training instability and mode collapse, failing to capture the full diversity of rare but critical network states.
Recently, Generative Diffusion Models (GDMs) \cite{ho2020denoising}, as a new paradigm in Generative AI (GAI), have attracted much attention due to their powerful ability to directly address these core challenges. More specifically, GDMs demonstrate the following advantages as a direct response to the needs of wireless networks:
\begin{itemize}
    \item \textbf{High-Fidelity Distribution Learning to Tame Complexity:} GDMs employ an iterative denoising process that excels at capturing the complete underlying distribution of data \cite{ho2020denoising}. This allows them to generate high-quality, diverse samples, effectively overcoming the mode collapse issue of GANs \cite{du2024enhancing}.
    \item \textbf{High Training Stability with Strong Interpretability:} The training objective of GDMs, rooted in non-equilibrium thermodynamics theories, is more stable and interpretable than the adversarial game of GANs, allowing them to learn robustly from smaller datasets.
    \item \textbf{Controllable and Flexible Generation:} By incorporating conditioning mechanisms, GDMs can achieve highly controllable sample generation. This includes generating Channel State Information (CSI) for a specific user location \cite{lee2024generating}, handling multimodal data \cite{liu2024cross}, and enabling personalized network services \cite{liang2024uav}.
    \item \textbf{Inherent Noise Resistance for Semantic Resilience:} The very nature of GDM training, injecting and then learning to remove noise, makes them exceptionally robust. This is particularly valuable for applications like SemCom, where recovering intent from noisy channel-corrupted features is paramount.
\end{itemize}

To this end, the main objective of this paper is to offer an in-depth exploration of why and how GDMs can be employed in the realm of wireless networks. Instead of proposing a rigid architecture, this survey systematically categorizes GDM-based schemes into a structured taxonomy covering the sensing layer, transmission layer, and vertical applications, complemented by a security plane. This survey will further comprehensively review related driving elements and key technologies in detail, with the hope that it will ignite future research endeavors within this burgeoning area.

\subsection{Preliminaries and State-of-the-Art Works}

\subsubsection{Generative AI for Wireless Networks}

\begin{table*}[tbp] 
    \centering
    \caption{Existing Survey/Review/Tutorial/Magazine papers on GAI for wireless networks}
    \label{gaisurvey}
    \renewcommand{\arraystretch}{1.2}  
    \normalsize  
    \begin{tabular}{|>{\arraybackslash}m{0.23\textwidth}|>{\arraybackslash}m{0.72\textwidth}|}
        \hline
        \textbf{\small{Topic}} & \textbf{\small{Survey/Review/Tutorial/Magazine Papers}} \\
        \hline
        \footnotesize GAI for physical layer design  & \footnotesize MIMO \cite{wang2024generative2} and physical layer communications \cite{van2024generative}.
        \\
        \hline

        \footnotesize GAI for wireless networks & \footnotesize mobile networks \cite{khoramnejad2025generative,vu2024applications,karapantelakis2024survey}, telecommunications \cite{bariah2024large}, wireless network management \cite{liu2024deep}, game-theory-based mobile networking \cite{he2025generative}, data augmentation in wireless networks \cite{wen2024generative}, Wi-Fi networks \cite{wang2024next}, end-to-end programmable networks \cite{he2025advancing}, and multimedia networks \cite{xu2024generative2}.
        \\
        \hline

        \footnotesize GAI for emerging techniques & \footnotesize SemCom \cite{nguyen2025contemporary,xia2025generative,liang2024generative,zhou2025generative2}, wireless intelligence \cite{celik2024dawn}, mobile edge networks \cite{lai2024resource}, and holographic communications \cite{zhang2024spatial}.
        \\
        \hline

        \footnotesize GAI for IoT  & \footnotesize Generative IoT \cite{wen2024generative2}, IoT computing \cite{mangione2025generative}, consumer IoT \cite{jiang2025generative}, energy harvesting IoT
        \cite{xie2024generative}, and IoT-healthcare \cite{chen2024generative}.
        \\
        \hline
        \footnotesize GAI for immersive communications  & \footnotesize Immersive communications \cite{sehad2024generative}, and wireless network digital twins \cite{li2025generative,tao2024wireless}.
        \\
        \hline
        
        \footnotesize GAI for UAV applications  & \footnotesize UAV networks \cite{sun2024generative,liu2024generative}, low-altitude economy networking \cite{zhao2025generative}, and UAV-assisted IoT networks \cite{sharif2024resource}.
        \\
        \hline
        
        \footnotesize GAI for other applications  & \footnotesize
        Vehicular networks \cite{zhang2024generative}, space-air-ground integrated networks \cite{zhang2024generative2}, and AIGC services \cite{xu2024unleashing,du2024age}.
        \\
        \hline
        
        \footnotesize GAI for securing wireless networks  & \footnotesize 
        Secure physical-layer communications \cite{zhao2024generative,zhao2025enhancing}, secure ISAC networks \cite{wang2024generative}, cross-layer covert communications \cite{liu2025generative}, and physical-layer authentication \cite{meng2025generative}.
        \\
        \hline

        \footnotesize LLM for wireless networks & \footnotesize Wireless networking \cite{boateng2025survey,qiao2025deepseek}, telecommunications \cite{zhou2024large}, intelligent network operations and performance optimization \cite{long2025survey,guo2025survey}, future communications \cite{jiang2025comprehensive,chen2024big}, and edge networks \cite{shen2024large,sharshar2025vision}.
        \\
        \hline
        
    \end{tabular}

\end{table*}

Compared to traditional Discriminative AI (DAI), which is confined to the framework of pattern recognition and logical judgment, GAI establishes a complete creative pipeline from abstract features to concrete content through profound analysis of inherent data patterns \cite{feuerriegel2024generative}. It transcends simple classification or prediction constraints, enabling the reconstruction of data elements based on probabilistic distributions to generate entirely novel content with original value \cite{du2024age}. Table \ref{gaisurvey} illustrates existing survey/review/tutorial/magazine papers on GAI for wireless networks, which can be divided into physical layer design 
including physical layer design \cite{wang2024generative2,van2024generative}, mobile and wireless networks \cite{khoramnejad2025generative,vu2024applications,karapantelakis2024survey,bariah2024large,liu2024deep,he2025generative,wen2024generative,wang2024next,he2025advancing,xu2024generative2}, emerging techniques \cite{nguyen2025contemporary,xia2025generative,liang2024generative,zhou2025generative2,celik2024dawn,lai2024resource,zhang2024spatial}, IoT \cite{wen2024generative2,mangione2025generative,jiang2025generative,xie2024generative,chen2024generative}, immersive communications \cite{sehad2024generative,li2025generative,tao2024wireless}, unmanned aerial vehicle (UAV) applications \cite{sun2024generative,liu2024generative,zhao2025generative,sharif2024resource}, vehicular networks \cite{zhang2024generative}, space-air-ground integrated networks \cite{zhang2024generative2}, AI-Generated Content (AIGC) services \cite{xu2024unleashing,du2024age}, and securing wireless networks \cite{zhao2024generative,zhao2025enhancing}, secure Integrated Sensing and Communication (ISAC) networks \cite{wang2024generative}, cross-layer covert communications \cite{liu2025generative}, and physical layer authentication \cite{meng2025generative}.
Notably, as a representative GAI technology, the large language model (LLM) has attracted much attention due to its powerful context learning and extensive task generalization ability. For example, LLMs can enable wireless networking \cite{boateng2025survey,qiao2025deepseek}, telecommunications \cite{zhou2024large}, intelligent network operations and performance optimization \cite{long2025survey,guo2025survey}, future communications \cite{jiang2025comprehensive,chen2024big}, and edge networks \cite{shen2024large,sharshar2025vision}.

\subsubsection{Generative Diffusion Model (GDM) for Wireless Networks}

\newcommand{\cfull}{\textcolor{black}{$\bullet$}}    
\newcommand{\cpart}{\textcolor{black}{$\circ$}}      
\newcommand{\cnone}{--}                             

\begin{table*}[tbp]
\centering
\caption{Comparison of existing Survey/Review/Tutorial/Magazine papers on GDMs for wireless networks.}
\label{gdmsurvey}
\begin{tabular}{p{0.8cm} c p{4.8cm} c c c c c}
\toprule
\textbf{Survey} & \textbf{Year} & \textbf{Main Topic} & \textbf{Theory} & \textbf{Sensing} & \textbf{Transmission} & \textbf{Security} & \textbf{Application} \\ \midrule

\cite{du2023exploring} & 2023 & GDM for distributed AIGC service & \cpart & \cnone & \cfull & \cfull & \cpart \\[2pt]

\cite{letafati2023diffusion} & 2023 & DDPM's applications in networks & \cpart & \cnone & \cpart & \cnone & \cnone \\[2pt]

\cite{jin2024gdm4mmimo} & 2024 &GDM for MIMO channel estimation & \cpart & \cfull & \cpart & \cnone & \cpart \\[2pt]

\cite{xu2024generative3} & 2024 & GDM-driven communication framework & \cpart & \cnone & \cpart & \cnone & \cfull \\[2pt]

\cite{du2024enhancing} & 2024 & GDM with DRL for network optimisation & \cpart & \cpart & \cfull & \cpart & \cfull \\[2pt]

\textbf{Ours} & 2025 & GDM for Wireless Networks & \cfull & \cfull & \cfull & \cfull & \cfull  
\\ 

\bottomrule
\end{tabular}

\begin{flushleft}\footnotesize
\textbf{Symbol Legend} — \cfull: comprehensive coverage; \cpart: moderate coverage; \cnone: not covered.\\
\textbf{Column Legend} — \emph{Theory}: A systematic review of the mathematical foundations of core GDMs and their major evolutionary variants; 
\emph{Sensing}: GDM-enabled sensing tasks, including channel estimation, channel generation, and radio map construction; 
\emph{Transmission}: GDM-enabled transmission tasks, such as SemCom;
\emph{Security}: GDM-enabled secure wireless networks;
\emph{Applications}: Applications of GDM-enabled wireless networks, such as intelligent healthcare, intelligent factory, intelligent transportation, immersive communication, and satellite communication.
\end{flushleft}
\end{table*}

We further list existing survey papers on GDM for wireless networks in Table \ref{gdmsurvey} in detail. Du et al. \cite{du2023exploring} provide an AIGC framework based on cooperative distributed GDM, which aims to solve the energy consumption and privacy problems of AIGC services on resource-constrained devices and optimize the utilization of computing resources in wireless networks. Letafati et al. \cite{letafati2023diffusion} validate the efficacy of Denoising Diffusion Probabilistic Models (DDPMs) in real-world wireless challenges, offering actionable insights for resilient and adaptive communication design. Jin et al. \cite{jin2024gdm4mmimo} discuss the potential application of GDM in massive multi-input multi-output (MIMO) communications, reveal its core technical characteristics, and deeply analyze the future research direction. Du et al. \cite{du2024enhancing} offer a detailed guide on applying GDMs to Deep Reinforcement Learning (DRL)-based network optimization tasks, bridging theory and practical implementation. Xu et al. \cite{xu2024generative3} introduce a GDM-driven communication framework for wireless data generation and GDM-enhanced DRL for communication management.

\subsection{Key Contributions and Outline}

Despite the fact that many researchers focus on GDMs for wireless networks \cite{du2024enhancing,xu2024generative,letafati2023diffusion,du2023exploring,jin2024gdm4mmimo}, notably, a comprehensive understanding of the state-of-the-art in GDMs for wireless networks remains preliminary. 
For instance, \cite{letafati2023diffusion} primarily discusses the applications of DDPMs, and \cite{jin2024gdm4mmimo} concentrates on the MIMO scenario. A systematic review that clarifies the evolutionary path of GDM technology itself, particularly the connections and distinctions in the mathematical principles of its core variants, is currently absent in the literature. This gap makes it challenging for wireless network researchers to fully grasp the trade-offs between different GDM variants and to select the most suitable model for a given communication task.
To address this gap, we present a comprehensive survey that analyzes representative GDMs and explores state-of-the-art GDM-driven approaches for enhancing wireless network performance. The main contributions are as follows.

\subsubsection{Systematically Clarify the Evolution of GDM's Core Models}

We first provide a systematic review of the mathematical principles behind eight representative GDM variants, including Denoising Diffusion Probabilistic Models (DDPMs), Score-based Generative Models (SGMs), Stochastic Differential Equations (SDEs), Ordinary Differential Equations (ODEs), Denoising Diffusion Implicit Models (DDIMs), Conditional Diffusion Models (CDMs), and Latent Diffusion Models (LDMs), as well as the cutting-edge Flow Matching (FM) and Consistency Models (CMs). Our analysis clarifies their evolutionary path and comparative differences. Then, we present a structured taxonomy that categorizes GDM-based schemes into the sensing layer, transmission layer, vertical applications, and security plane, systematically demonstrating the benefits of GDMs for wireless networks. For channel modeling and radio map construction in the sensing layer, we demonstrate that GDMs can effectively simulate multi-path superimposed channel environments and achieve high-precision predictions under sparse measurement scenarios. Regarding SemCom in the transmission layer, we establish that GDMs exhibit enhanced semantic learning capabilities and robust resilience against semantic noise and interference. In the context of vertical applications, we illustrate how GDMs evolve from passive data generators for network digital twins to active policy optimizers for complex network decisions and enablers for task-oriented generative communications. Finally, concerning the security plane, GDMs strengthen security across all layers, while privacy-preserving techniques further secure the overall network.

\subsubsection{Explore the State-of-the-Art in GDM-enabled Wireless Networks}
Based on the GDM-empowered multi-layer network architecture, we investigate recent advancements in GDM's applications. For the sensing layer, we provide a comprehensive review of existing GDM-based schemes for channel estimation, channel generation, and radio map construction. For the transmission layer, we survey existing GDM-based schemes for semantic denoising, auxiliary recovery, semantic-based generation, multimodal transmission, and resource allocation. For the vertical applications, distinct from previous works, we categorize applications based on the advanced role of GDM as a world model for network digital twins, as a solver for network policy optimization, and as a generator for vertical services. Concerning the security plane, we review existing GDM-enhanced schemes for securing sensing, transmission, and vertical applications.

\subsubsection{Discuss Challenges and Potential Solutions}
Although researchers have extensively investigated GDM-based approaches for wireless networks, several fundamental challenges persist. To this end, we systematically identify these core conflicts and propose potential solution strategies to guide future research. Specifically, we analyze the fundamental conflict between the iterative inference latency of GDMs and the coherence time of wireless channels, the model mismatch between Gaussian noise assumptions and physical channel impairments, and the representation gap between standard computer vision architectures such as U-Net and the physical structure of wireless data. Furthermore, we discuss unique security vulnerabilities including process amplification attacks and highlight the evaluation dilemma between generative fidelity metrics and wireless task utility.

\begin{figure}
\centering
\includegraphics[width=0.45\textwidth]{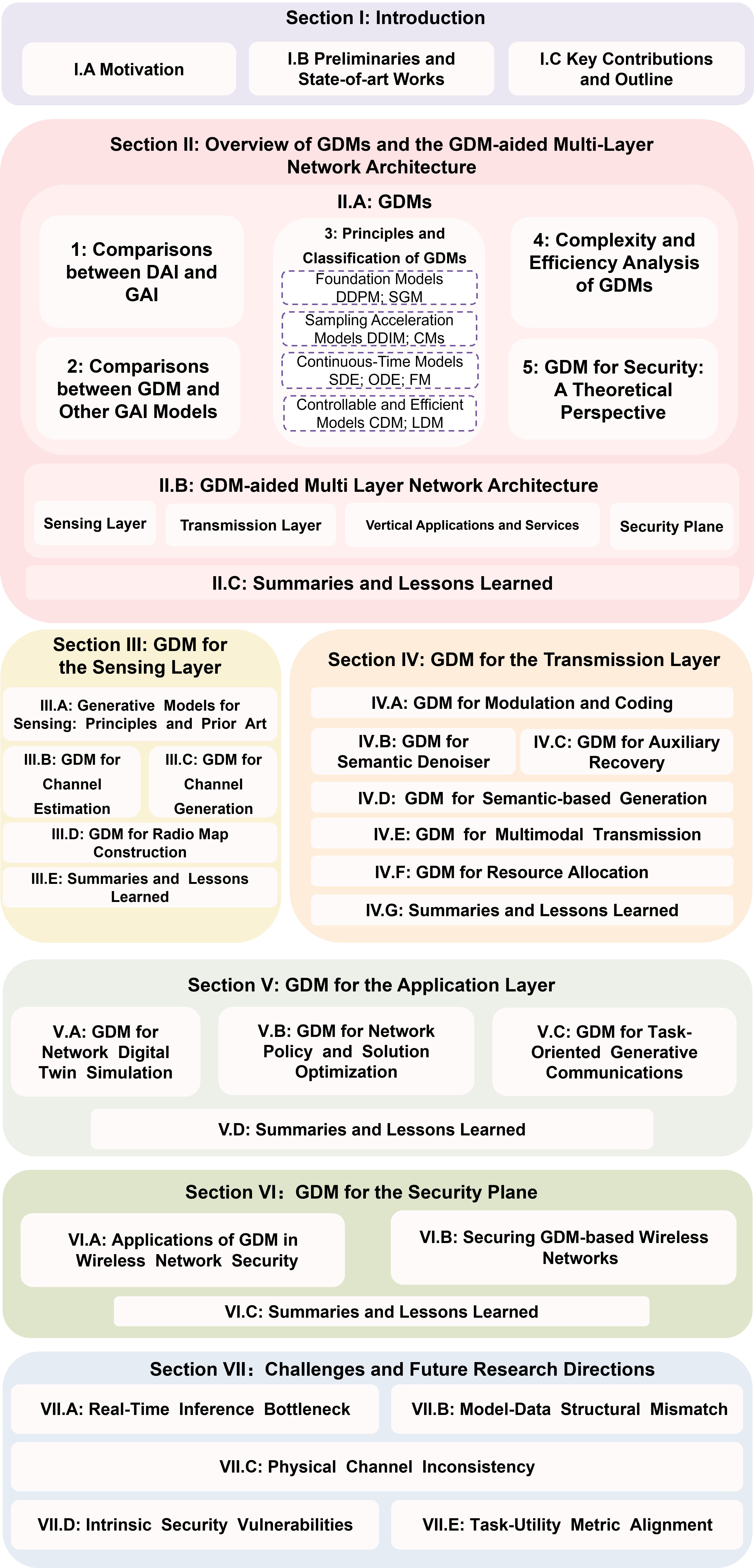}
\caption{The structure of this paper.}
\label{picture_paperarchitecture}
\end{figure}

\textit{Roadmap:} The outline of this survey is illustrated in Figure \ref{picture_paperarchitecture}. Specifically, Section \ref{section2} overviews GDMs and presents the categorization framework used to organize their applications in wireless networks. Sections \ref{section3}, \ref{section4}, \ref{section5}, and \ref{section6} provide insights into existing GDM-based schemes from the perspectives of the sensing layer, transmission layer, application layer, and security plane, respectively. Section \ref{section7} looks forward to future research directions. Finally, Section \ref{section8} concludes this paper.

\section{Overview of GDMs and the Structured Taxonomy for GDM-enabled Wireless Networks}
\label{section2}

\subsection{Overview of GDMs}
This subsection compares DAI and GAI, compares GDM and other GAI models, and introduces six typical GDMs.
\subsubsection{Comparisons Between DAI and GAI}

As foundational approaches in machine learning, DAI and GAI are defined by fundamentally different objectives and modeling paradigms. To clarify their distinct roles, it is essential to define each concept.

DAI learns the conditional probability $P(Y| X)$\footnote{$P(Y | X)$ represents the conditional probability of observing each $Y$ given $X$, establishing input-to-output mappings by supervised learning.}, focusing on modeling the decision boundary between different classes of data. Its primary goal is to map an observed input, $X$, to a desired output label, $Y$, making it highly effective for tasks such as classification and regression prediction\cite{bishop2006pattern7.20}. In essence, DAI is designed to distinguish between existing data patterns. In contrast, GAI aims to learn the underlying probability distribution of the data itself, modeling either the joint distribution $P(X,Y)$ \footnote{$P(X,Y)$ denotes the joint probability of input data $X$ and its corresponding labels or features $Y$.} or the marginal distribution $P(X)$ \footnote{$P(X)$ represents the probability distribution of observed data $X$ in the feature space.}. Rather than simply classifying inputs, GAI learns to represent inherent data patterns so profoundly that it can create entirely new, synthetic samples that are statistically consistent with the original data.

The core distinction, therefore, is that DAI learns to predict labels from given inputs, whereas GAI learns to generate novel data from underlying distributions. These foundational differences lead to significant distinctions in their capabilities and limitations, as summarized in Table \ref{Comparison Between DAI and GAI} and detailed below.

\begin{itemize}
\item \textbf{Relying on Labeled Data:}
DAI focuses on supervised learning-based feedforward conditional probability modeling $P(Y| X)$. While excelling in static, closed system scenarios such as image classification and regression prediction, DAI demonstrates critical limitations in generalization capability due to the heavy reliance on labeled data. 

\item \textbf{Difficulties in Internal Distribution Modeling:}
DAI's inherent modeling omits explicit data distribution modeling $P(X)$, thus restricting its capability to interpolate existing patterns rather than extrapolate unknown distributions or generate novel samples \cite{goodfellow2014generative7.21}. This is a critical constraint in scenarios demanding creative generation, such as data augmentation and simulation inference\cite{saharia2022photorealistic}.
\end{itemize}
In contrast, GAI show the following advantages.
\begin{itemize}
\item \textbf{Joint Distribution Modeling:}
GAI models the joint distribution $P(X,Y)$ or, in unsupervised settings, even the marginal $P(X)$. For example, in channel modeling, it can represent the joint probability of environmental characteristics and corresponding channel states, thus enabling the generation of new data samples \cite{chen2025generative}. 

\item \textbf{Controllable Generation:}
GAI offers precise control over the generative process. Through mechanisms like conditional injection, GAI can produce outputs that align with user defined constraints\cite{saharia2022palette}. For instance, conditional GANs can generate faces with specific attributes, such as smiling, male, wearing glasses, while CDMs\cite{ho2022classifier} can generate medical images with specific tumor shapes or create semantic consistent maps from wireless signal data. This capability contrasts sharply with DAI, which is limited to predicting predefined labels and cannot generate or manipulate new structured content.
\item \textbf{Multimodal Learning:}
GAI also possesses powerful multi modal reasoning capabilities\cite{ramesh2021zero}, enabling flexible transformation and mapping between different types of data. For example, GAI can perform tasks such as text-to-image synthesis, audio-to-text transcription, and image captioning. These tasks are highly challenging, as they require GAI not only to understand data in one modality but also to generate data in other modalities. 
\end{itemize}
These advantages make GAI play an irreplaceable role in wireless communications. Moreover, in the continuously evolving communication environments of the future, its importance will only increase. Although GAI models have issues such as high computational costs and potential instability during training, the continuous technological advancements in recent years are gradually improving these situations. GAI's outstanding capabilities in creative generation and cross modal adaptability have already established its position as a fundamental technology for future wireless networks.

\begin{table*}[tbp]
\centering
\caption{Comparisons Between DAI and GAI}
\label{Comparison Between DAI and GAI}
\renewcommand{\arraystretch}{1.2}
\begin{tabularx}{\textwidth}{@{}l|X|X@{}}
\toprule
\textbf{Dimension} & \textbf{DAI} & \textbf{GAI} \\
\midrule
Core Objective & Learns conditional probability $P(Y| X)$ \cite{bishop2006pattern7.20} & Models joint distribution $P(X, Y)$ to generate new samples\cite{goodfellow2014generative7.21} \\
Learning Paradigm & Supervised learning\cite{he2018deep,venugopal2017channel} & Unsupervised or self-supervised learning \\
Technical Methods & Logistic regression, SVM, CNN, and RNN & GDMs, GANs, VAEs, and Transformers\cite{rombach2022high} \\
Data Requirements & Relies on high quality labeled data\cite{goodfellow2016deep} & Works with unlabeled data, requires large scale training\cite{ho2020denoising} \\
Generation Capability & Cannot generate new data & Synthesizes high fidelity multimodal data (images, text, etc.) \\
Strengths & Efficient classification and precise prediction\cite{nayebi2017semi} & Captures data distributions and creative generation\cite{goodfellow2014generative7.21} \\
Limitations & Label dependency, weak generalization, and pattern rigidity & High training cost\cite{wang2024radiodiff} \\
Typical Applications & Image classification, regression, and object detection & Image synthesis, data augmentation, and cross modal reasoning \\
\bottomrule
\end{tabularx}
\end{table*}

\subsubsection{Comparisons Between GDM and Other GAI Models}
As illustrated in Table~\ref{tab:model_comparison_en}, modern GAI models exhibit distinct differences in their theoretical foundations, latent space design, and generation mechanisms. Below, we compare four representative GAI models: GANs, Variational Autoencoders (VAEs), Transformers, and GDMs.
\begin{itemize}
\item 
\textbf{GANs:}
GANs\cite{goodfellow2014generative7.21} use adversarial training between a generator and a discriminator. The generator learns to produce data that fools the discriminator, while the discriminator tries to distinguish between real and generated data. While GANs excel at producing high fidelity outputs, they often suffer from mode collapse and unstable training due to the adversarial objective.
In terms of extensibility, the inherent instability of adversarial training makes scaling GANs challenging, though their moderate training costs and high-fidelity output made them a popular choice in the communications field for several years\cite{navidan2021generative10.15}, particularly for applications like channel modeling and signal generation.

\item 
\textbf{VAEs:}
VAEs\cite{kingma2013auto,meng2023physical} adopt a probabilistic encoder-decoder framework, where the encoder maps input data to an approximated latent distribution, and the decoder reconstructs data from this latent representation. The training objective balances data reconstruction and regularization by introducing a Kullback–Leibler (KL) divergence term to encourage smooth latent structures. While VAEs offer stable training, the approximation of the latent distribution often leads to blurry generated outputs.
With their stable training and low training cost, VAEs exhibit good scalability. This has made them suitable for communication tasks involving efficient data representation, such as end-to-end communication system design and feature compression, where a robust latent space is more critical than perfect signal reconstruction.

\item 
\textbf{Transformers:}
Transformers\cite{vaswani2017attention} use self-attention to capture long range dependencies for sequence modeling. In generative tasks, they predict the next element step by step, enabling outputs to unfold progressively based on contextual information. Building upon the Transformer architecture, LLMs further extend the capacity through pre-training on trillion-token corpora, followed by instruction tuning or reinforcement learning with human feedback\cite{brown2020language7.20,fan2025kgrag}. As model size scales to hundreds of billions of parameters, these LLMs demonstrate emergent capabilities such as in-context learning, multilingual reasoning, and cross-modal generation, enabling generalization across a wide range of tasks with minimal supervision.
The exceptional scalability of Transformers is their defining feature, although it comes at a very high training cost. This has positioned them as the foundation\cite{long2025survey10.15} for large-scale models in communications, driving advancements in SemCom and network intelligence where contextual understanding is paramount.

\item 
\textbf{GDMs:}
GDMs redefine generation as a denoising process over a Markov chain \cite{ho2020denoising}. They gradually add noises to a data sample $x_0$ to obtain a fully noisy sample $x_T$, and then train a U-Net to reverse this process. The U-Net minimizes the mean squared error between predicted noises and true noises at each time step, aligning generation with the task of progressive denoising \cite{sohl2015deep}. Unlike GANs and VAEs, GDMs avoid mode collapse and posterior approximation, offering better coverage of the data distribution and stable training \cite{cao2024survey}.
While the architecture of GDMs scales well, the high training cost associated with their iterative process is a key consideration. Nonetheless, their ability to generate high-quality samples with full mode coverage has led to their rapid adoption in wireless communications for precision-critical tasks, such as channel estimation and radio map construction, where they have begun to outperform earlier generative models.

\end{itemize}

\begin{table*}[tbp]
\centering
\caption{Comparisons Between Representative Generative Models: GANs, VAEs, Transformers, and GDMs}
\label{tab:model_comparison_en}
\renewcommand{\arraystretch}{1.2}
\begin{tabularx}{\textwidth}{@{}l|p{2.2cm}|X|X|X@{}}
\toprule
\textbf{Dimension} & \textbf{GANs\cite{goodfellow2014generative7.21}} & \textbf{VAEs\cite{kingma2013auto}} & \textbf{Transformers\cite{vaswani2017attention}} & \textbf{GDMs\cite{ho2020denoising}} \\
\midrule

Modeling Paradigm & Adversarial game & Variational inference & Autoregressive likelihood & Stochastic differential equations \\

Mathematical Principles & Noise vector $z$ & Approximate posterior $q_\phi(z | x)$ & Discrete tokens & Noisy latents $\{x_t\}_{t=1}^T$ \\

Optimization Objective & Adversarial loss & ELBO maximization & Cross-entropy loss & Noise-prediction MSE\cite{song2020denoising} \\

\addlinespace
Training Stability and Cost & unstable $|$ Medium  & stable $|$ Low & Medium $|$ Very High & very stable $|$ High\\

Generation Process & Single-step (Fast) & Decoder sampling (Fast) & Iterative token prediction (Slow) & Multi-step denoising (Slow)\\

\addlinespace

Strengths & High fidelity & Stable training & Long-range coherence & Full mode coverage \cite{sengupta2023generative, kim2023diffusion} \\

Weaknesses & Mode collapse & Blurry outputs & Quadratic complexity & Slow sampling\cite{song2024lightweight} \\

Scalability            & Medium                  & High                  & Very High\\      
\bottomrule
\end{tabularx}
\end{table*}

\subsubsection{Principles and Classification of GDMs}

\begin{figure*}
\centering
\includegraphics[width=1\textwidth]{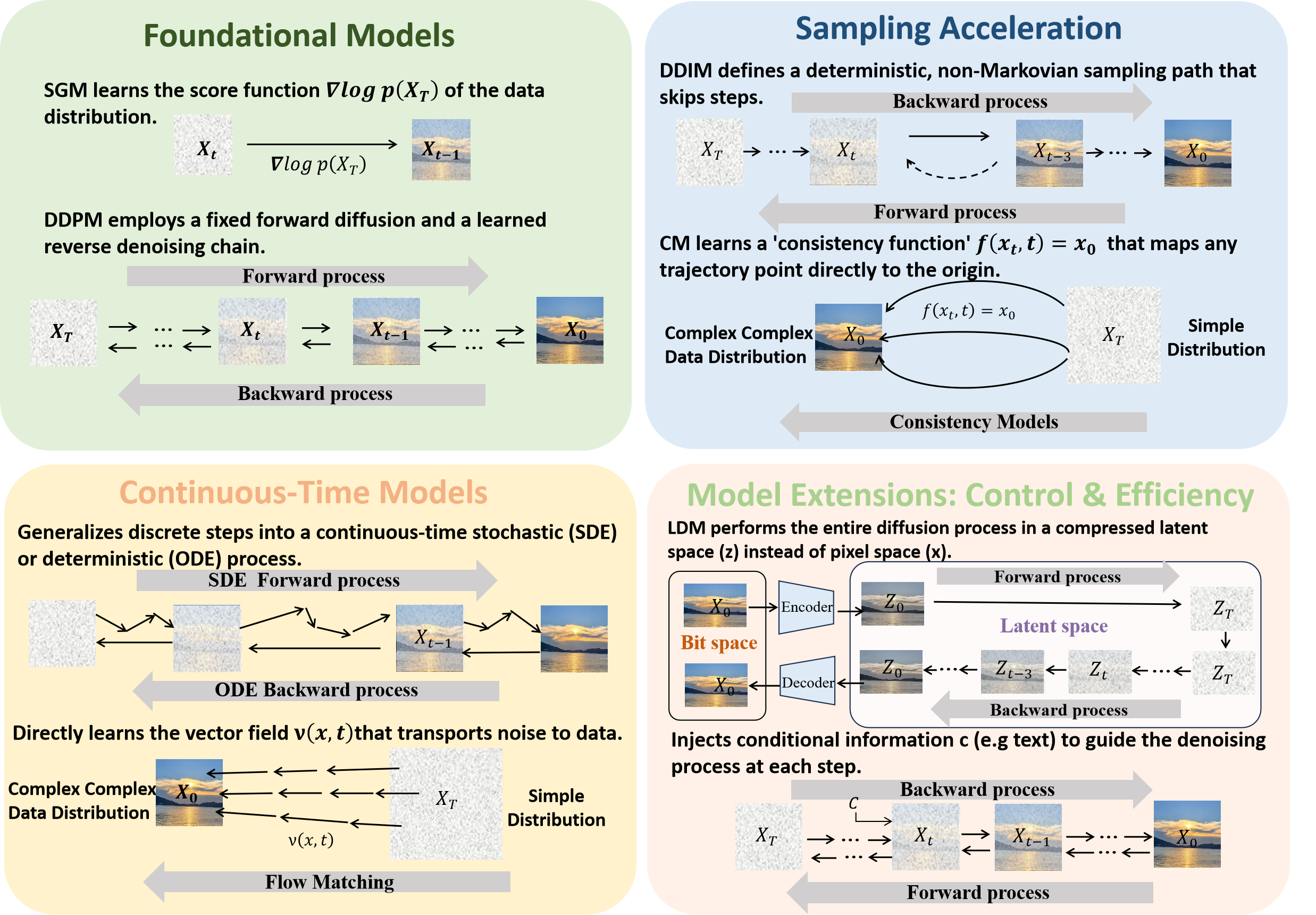}
\caption{Evolution of GDMs, where DDPM employs a fixed forward diffusion and a reverse denoising chain \cite{ho2020denoising}; SGM reframes denoising as direct gradient estimation over noise scales \cite{song2019generative, song2020score}; SDE and ODE supply a unifying limit that enables principled solver design and deterministic sampling \cite{song2020score}; FM offers a more stable training paradigm for these continuous flows; DDIM introduces a non Markovian deterministic shortcut that skips many intermediate noise levels, yielding substantial speed gains \cite{song2020denoising}; CMs further evolve this by enabling single-step generation; CDM augments the noise regression with external guidance signals \cite{nichol2021glide, ho2022classifier}; and LDM relocates the objective into a compressed space for scalability \cite{rombach2022high}.}
\label{picture_GDM}
\end{figure*}

As shown conceptually in Figure \ref{picture_GDM}, GDMs arise from ideas in nonequilibrium thermodynamics. An empirical data distribution is gradually perturbed toward an almost isotropic Gaussian by a sequence of small noise injections, then a learned reverse procedure reconstructs clean samples. The early formulation appears in \cite{sohl2015deep} but practical performance is limited until Ho et al. \cite{ho2020denoising} introduce the DDPM with a simplified variational objective and an effective U-Net backbone, reaching image quality beyond GANs \cite{dhariwal2021diffusion}. This milestone marks the official entry of GDMs into the mainstream technologies of GAI \cite{kingma2021variational}. Later work interprets or extends GDMs through score function learning \cite{song2019generative, song2020score} and continuous time stochastic/deterministic views \cite{song2020score}. Subsequent work has focused on more stable training paradigms like FM and crucial sampling acceleration techniques like CMs. Other principal extensions include conditional guidance \cite{ho2022classifier, nichol2021glide}, and latent space acceleration \cite{rombach2022high}. We now detail these key models, grouped by their core contribution.

\paragraph{\textbf{Foundation Models}}
\begin{itemize}

\item \textbf{DDPM:}
DDPM specifies a fixed forward diffusion and a learned reverse denoising chain. Let $x_0 \sim q(x_0)$ denote a data sample, where $x_0$ means the initial sample is randomly drawn from the true data distribution. The forward Markov chain produces noisy states $x_1, \dots, x_T$ by
\begin{equation}
q(x_t | x_{t-1}) = \mathcal{N}\!\big(x_t; \sqrt{1-\beta_t}\, x_{t-1}, \beta_t \mathbf{I} \big),
\end{equation}
where $q(x_t | x_{t-1})$ denotes the transition probability of the forward diffusion process, modeling the step-wise corruption of data by adding Gaussian noise conditioned on the previous state, $\mathcal{N}$ denotes a multivariate normal distribution, $\beta_t \in (0,1)$ is a user chosen variance increment at step $t$ that controls how much fresh Gaussian noise is injected, and $\mathbf{I}$ denotes the identity matrix, implying isotropic noise. A smaller $\beta_t$ means finer degradation and typically better empirical stability. Composing them yields the closed form
\begin{equation}
x_t = \sqrt{\bar\alpha_t}\, x_0 + \sqrt{1-\bar\alpha_t}\, \epsilon,\qquad \epsilon \sim \mathcal{N}(0,\mathbf{I}),
\end{equation}
where $ \bar\alpha_t = \prod_{s=1}^t (1-\beta_s)$, so $x_t$ is a linear interpolation between the clean sample and independent noise with weights determined by the cumulative product $\bar\alpha_t$. $\epsilon$ denotes a standard Gaussian noise variable independent of the data, sampled from $\mathcal{N}(0,\mathbf{I})$. This analytic expression allows direct sampling of any intermediate noisy state without iterating through all earlier steps.
The reverse chain is defined by learned Gaussian transitions as
\begin{equation}
p_\theta(x_{t-1} | x_t) = \mathcal{N}\!\big(x_{t-1}; \mu_\theta(x_t,t), \Sigma_\theta(x_t,t)\big),
\end{equation}
where a time conditioned U-Net predicts parameters that determine the mean $\mu_\theta(x_t,t)$ and either fixes or predicts the variance $\Sigma_\theta(x_t,t)$. Intuitively $\mu_\theta$ points back toward a region more consistent with earlier less noisy states while $\Sigma_\theta$ controls stochastic diversity.
Variational analysis constructs an evidence lower bound (ELBO) whose summands are Kullback Leibler divergences between true r everse conditionals and model conditionals across steps \cite{sohl2015deep, karras2022elucidating}.
The forward noise component is predicted through the simple loss as
\begin{equation}
\mathcal{L}_{\text{DDPM}} = \mathbb{E}_{t, x_0, \epsilon} \big[ \| \epsilon - \epsilon_\theta(x_t, t) \|_2^2 \big],
\end{equation}
where $\epsilon$ is the actual Gaussian noise that produced $x_t$ and $\epsilon_\theta$ is the noise predicted by the network. Minimizing this mean squared error drives $\epsilon_\theta$ toward the true conditional expectation of the noise which in turn specifies an optimal reverse mean through closed form algebra derived from the forward distribution. Sampling then alternates between predicting the noise and constructing $x_{t-1}$ from $x_t$. This training target concentrates learning signal, reduces gradient variance, and preserves the essential variational objective.

\item 
\textbf{SGM:} SGM reinterprets DDPM by learning the score $\nabla_x \log p_t(x)$ of progressively noisier data distributions rather than explicit reverse Gaussians \cite{song2019generative, song2020score}. For a noise level (or scale) $\sigma$, the model observes perturbed samples $x_\sigma = x_0 + \sigma \epsilon$ and trains the network $s_\theta(x_\sigma,\sigma)$ to approximate the gradient of the log density at that scale. The training objective uses denoising score matching, which states that the score can be recovered by regressing the added noise under suitable weighting. This avoids specifying reverse transition variances and unifies all timesteps in one continuous family of scales. Generation starts from a standard Gaussian $\mathcal{N}(0,\mathbf{I})$ and applies Langevin style\cite{welling2011bayesian7.22} updates as
\begin{equation}
x \leftarrow x + \frac{\eta}{2}\, s_\theta(x,\sigma) + \sqrt{\eta}\, z,\qquad z \sim \mathcal{N}(0,\mathbf{I}),
\end{equation}
where the step size $\eta$ controls refinement and the injected noise $z$ preserves exploration. As $\eta$ decreases the iteration approximates sampling from the learned density. This gradient field view lays groundwork for continuous time stochastic differential formulations.
\end{itemize}

\paragraph{\textbf{Continuous-Time Models}}
\begin{itemize}
\item 
\textbf{SDE and ODE:} Continuous time formulations generalize discrete chains into It\^o stochastic differential equations
\begin{equation}
d x = f(x,t)\, dt + g(t)\, d w,
\end{equation}
where $f(x,t)$ is the drift governing deterministic decay of structure, $g(t)$ is a scalar or schedule controlling instantaneous noise amplitude, and $w$ is standard Brownian motion \cite{song2020score}. Choosing $f(x,t)$ and $g(t)$ recovers families analogous to variance preserving or variance exploding diffusion. The forward SDE defines a family of perturbed densities $p_t$. Time reversal of stochastic processes gives the reverse SDE as
\begin{equation}
\label{SDE}
d x = \big[f(x,t) - g(t)^2 \nabla_x \log p_t(x)\big] dt + g(t)\, d \bar w,
\end{equation}
where $\bar w$ is Brownian motion in reverse time and the unknown score term $\nabla_x \log p_t(x)$ is estimated by a neural network $s_\theta(x,t)$. This unifies DDPM, which involve discrete steps, and score methods, which focus on direct gradient learning, under one equation. Numerical simulation uses discretization methods such as Euler, predictor corrector, and higher order solvers that trade accuracy and cost.
Unlike SDE which introduces randomness through noise at each step, the corresponding ODE eliminates such stochasticity and follows a fixed trajectory.
Removing stochasticity yields the probability flow ODE as
\begin{equation}
\frac{d x}{d t} = f(x,t) - \frac{1}{2} g(t)^2 \nabla_x \log p_t(x),
\end{equation}
which describes a smooth evolution without any noise term. This is because the random fluctuation term $ g(t)\, d \bar w$ in (\ref{SDE}) has been removed, leaving only the deterministic drift adjusted by the score. As a result, each starting point produces a fixed and repeatable path. SDE adds randomness via $ g(t)\, d \bar w$, producing diverse samples, while ODE eliminates this noise to trace a single most likely trajectory. This makes ODE useful for likelihood estimation and fast sampling.
This deterministic viewpoint underlies accelerated samplers conceptually related to DDIM, the following elaborates in detail, and enables exact change of variables formulas needed for probability evaluation.
\item 
\textbf{FM:} While SDE and ODE formulations provide a powerful continuous-time perspective, their training, particularly via score matching, can be complex and computationally intensive. FM\cite{lipman2022flow} emerges as a more recent and highly effective training paradigm designed to directly learn the continuous probability flow, or vector field, that transports a simple noise distribution $p_0$ to the complex data distribution $q(x_0)$.

Instead of learning the score function $\nabla_x \log p_t(x)$, FM aims to directly regress this vector field $v(x, t)$. A key innovation is the formulation of a simulation-free training objective. This is often achieved by constructing a simple, deterministic path between a noise sample $z \sim p_0(z)$ and a data sample $x_0 \sim q(x_0)$, for instance, a linear interpolation path $x_t = (1-t)z + t x_0$. The target vector field $u_t$ is then simply the time derivative of this path, $u_t = \frac{dx_t}{dt} = x_0 - z$. A neural network $v_{\theta}(x_t, t)$ is then trained to predict this constant vector field using a straightforward regression loss:
\begin{equation}
\mathcal{L}_{FM} = \mathbb{E}_{t, q(x_0), p_0(z)} [ || v_{\theta}(x_t, t) - (x_0 - z) ||_2^2 ].
\label{eq:fm_loss}
\end{equation}
This approach, particularly in its conditional forms, simplifies the training objective significantly compared to score matching. By directly learning the vector field of a deterministic ODE, FM provides a more stable training process and generates high-fidelity results, laying a robust foundation for subsequent acceleration techniques.
\end{itemize}

\paragraph{\textbf{Sampling Acceleration Models}}
\begin{itemize}
\item 
\textbf{DDIM:} DDIM \cite{song2020denoising} accelerates a pretrained DDPM by defining a non Markovian deterministic mapping over a sparse subset of timesteps that preserves the forward marginals.
From a noisy state $x_t$ the model first estimates the original clean sample by inverting the closed form forward relation as
\begin{equation}
\hat{x}_0(x_t,t) = \frac{x_t - \sqrt{1-\bar{\alpha}_t}\,\epsilon_\theta(x_t,t)}{\sqrt{\bar{\alpha}_t}},
\end{equation}
then jumps directly to an earlier time index $\tau < t$ chosen from a reduced schedule via
\begin{equation}
x_\tau = \sqrt{\bar{\alpha}_\tau}\,\hat{x}_0(x_t,t) + \sqrt{1-\bar{\alpha}_\tau}\,\epsilon_\theta(x_t,t),
\end{equation}
which restores the signal proportion appropriate to time $\tau$ and supplies the residual noise required by the forward marginal. This dependence on $\hat{x}_0$ across non adjacent times concentrates denoising information into fewer steps.

\item 
\textbf{CMs:}
While DDIM significantly accelerates sampling by defining a deterministic path, it still requires multiple inference steps to achieve high fidelity. CMs\cite{song2023consistencymodels} emerge as a new class of generative models designed to overcome this fundamental latency barrier by enabling high-fidelity, single-step generation.
The core principle is the enforcement of a consistency property. CMs learn a function $f_{\theta}(x_t, t)$ that maps any point $x_t$ along a continuous probability flow ODE trajectory back to its origin $x_0$. This function must satisfy $f_{\theta}(x_t, t) = x_0$ for all $t$ on the trajectory. This property is enforced during training. While early methods used Consistency Distillation from a pre-trained GDM, Consistency Training allows training from scratch. CT minimizes a loss that ensures the model's outputs are consistent between two adjacent points on the ODE path, $x_t$ and $x_{t'}$, using an exponential moving average target network $f_{\theta^{-}}$:
\begin{equation}
\mathcal{L}_{CT} = \mathbb{E}_{t, t', x_0} [ || f_{\theta}(x_t, t) - f_{\theta^{-}}(x_{t'}, t') ||_2^2 ].
\label{eq:cm_loss}
\end{equation}
Once trained, the model can generate a sample in a single step by computing $\hat{x}_0 = f_{\theta}(x_T, T)$ from pure noise $x_T$. This capability for one-step synthesis represents a significant breakthrough for practical wireless applications, such as real-time SemCom, that cannot tolerate the high latency of iterative sampling.
\end{itemize}

\paragraph{\textbf{Controllable and Efficient Models}}

To enable multimodal generation in wireless networks, GDMs must incorporate external condition information $y$ into the denoising process.
The core enabler for this is the cross-attention mechanism. Unlike standard convolution layers that process local features, cross-attention allows the GDM to dynamically attend to multimodal guidance signals at various resolution levels of the U-Net\cite{ronneberger2015u213}.

Mathematically, the noisy signal features act as Queries ($Q$), while the multimodal conditions are encoded as Keys ($K$) and Values ($V$).
The interaction is modeled as $\text{Attention}(Q, K, V) = \text{softmax}\left(\frac{QK^T}{\sqrt{d}}\right)V$, which\cite{vaswani2017attention} forces the generated wireless data to be structurally aligned with the semantic intent.

Furthermore, classifier-free guidance is widely employed to regulate the strength of this control, balancing the trade-off between sample diversity and fidelity to the condition $y$.

Based on these mechanisms, two representative models are derived:

\begin{itemize}
    \item \textbf{CDM:} CDM explicitly integrates the aforementioned conditioning mechanisms to direct the generative process \cite{nichol2021glide, ho2022classifier}. In wireless contexts, the condition $y$ extends beyond simple class labels to include complex physical constraints such as pilot signals, user locations, or CSI. By injecting these conditions via concatenation or cross-attention, CDMs can generate specialized wireless data that strictly adheres to environmental constraints, albeit sometimes at the cost of reduced diversity compared to unconditional models.
    
    \item \textbf{LDM:} LDM improves efficiency by shifting the diffusion process from the high-dimensional pixel/signal space to a compressed latent space\cite{rombach2022high}. A pretrained variational encoder first maps the input into a lower-dimensional latent representation. Crucially, this compressed space facilitates more efficient cross-modal alignment, as the heavy computation of cross-attention is performed on smaller feature maps. This makes LDMs particularly suitable for multimodal SemCom, enabling the fusion of rich semantic features, such as text or segmentation maps, with wireless signals at a significantly lower computational cost.
\end{itemize}

\subsubsection{Complexity and Efficiency Analysis of GDMs}
\begin{table*}[tbp]
\centering
\caption{Quantitative Complexity and Resource Comparison of GDM Variants.}
\label{table_complexity_resource_comparison}
\small
\renewcommand{\arraystretch}{1.3} 
\newcolumntype{L}{>{\RaggedRight\arraybackslash}X}

\begin{tabularx}{\textwidth}{l L L l L L}
\toprule
\textbf{GDM Variant} & \textbf{Core Strategy} & \textbf{Training Overhead} & \textbf{Inference Steps ($N$)} & \textbf{Computational Cost (FLOPs)} & \textbf{Memory Efficiency} \\ 
\midrule

DDPM/SGM & Baseline: Iterative refinement in data space. & High: Requires learning the full noise spectrum. & $\sim$1000 \newline ($O(N)$) & $1000 \times C_{Pixel}$ & Low: Operates on high-dimensional raw data. \\ 

DDIM & Non-Markovian: Deterministic step-skipping. & Low: Directly utilizes pre-trained weights. & 10--50 \newline ($O(N/S)$) & $(10 \sim 50) \times C_{Pixel}$ & Medium: No additional overhead over DDPM. \\ 

CM & Consistency Mapping: Single-step origin projection. & Very High: Complex distillation or specialized loss. & 1--2 \newline ($O(1)$) & $(1 \sim 2) \times C_{Pixel}$ & Medium: Fast state recovery. \\ 

FM & Vector Field: Learning straight ODE paths. & Medium: Fast convergence due to linear trajectories. & 10--50 \newline (Flexible) & $(10 \sim 50) \times C_{Pixel}$ & Medium: Lower training variance. \\ 

LDM & Dimension Reduction: Diffusion in compressed domain. & Low: Operates in low-dimensional space. & 10--50 \newline (with DDIM) & $(10 \sim 50) \times C_{Latent}$ ($C_{Latent} \ll C_{Pixel}$) & High: Compressed semantic features. \\ 

\bottomrule
\end{tabularx}

\vspace{0.5em}
\begin{minipage}{\textwidth}
\footnotesize
\textit{Notation:} $C_{Pixel}$ represents the computational cost per forward pass in the raw data domain; $C_{Latent}$ represents the cost in the compressed semantic domain.
\end{minipage}
\end{table*}

While GDMs offer superior generation fidelity, their iterative inference mechanism introduces non-negligible computational burdens. Different variants have evolved to optimize sampling paths or compress operational spaces, exhibiting distinct complexity profiles.

\paragraph{Training Efficiency: Stability at the Cost of Compute}
Unlike the min-max game of GANs, GDMs training is rooted in maximum likelihood estimation or score matching, effectively optimizing a mean squared errorloss function. This convex-like optimization avoids common GAN issues such as mode collapse and non-convergence, thereby reducing hyperparameter tuning difficulty and enhancing stability.

However, GDM training overhead is typically higher. The model must learn the de-noising score function across the entire diffusion time range $t \in [0, T]$ \cite{ho2020denoising,song2020denoising} rather than just the raw data distribution $P(x)$. This requires the network to process samples with diverse noise intensities in each iteration, resulting in total computational requirements that often exceed those of same-scale GANs or VAEs.

\begin{itemize}
\item \textbf{FM:} To optimize training, FM abandons complex SDE paths in favor of deterministic linear trajectories between noise and data. Due to reduced gradient variance, FM typically converges faster than standard diffusion models\cite{lipman2022flow}.

\item \textbf{LDM:} LDM\cite{rombach2022high} shifts complexity from the physical pixel space to the semantic latent space. By operating within a low-dimensional compressed domain, the per-step computational load decreases quadratically with the compression ratio, significantly alleviating memory bottlenecks.
\end{itemize}

\paragraph{Execution Complexity: Scaling the Latency-Fidelity Curve}

The execution efficiency during inference is governed by the product of the number of iterations \(N\) and the per-step network cost \(C_{Net}\).

\begin{itemize}
    \item \textbf{Iterative Solvers (DDPM/SDE)}: As baseline models, their inference complexity is \(O(N)\), with \(N \approx 1000\) steps required to ensure high-fidelity outputs. Such sequential overhead is prohibitive relative to the millisecond-level coherence time of wireless channels. 
    \item \textbf{Skip-step Acceleration (DDIM)}: DDIM\cite{song2020denoising} optimizes inference complexity to \(O(N/S)\) via non-Markovian sampling, allowing step reduction to 10–50. Its primary advantage is the ability to adjust the speed-quality trade-off during inference without re-training.
    \item \textbf{One-shot Generation (CMs)}: CMs\cite{song2023consistencymodels} aim for the theoretical limit of \(O(1)\) complexity by learning to map any point on the trajectory directly back to the origin. This compresses generation latency to a single pass, meeting the requirements of real-time control loops such as high-frequency beam tracking.
\end{itemize}

Finally, we summarize the quantitative complexity and resource comparison of these GDM variants in Table \ref{table_complexity_resource_comparison}.

\subsubsection{GDM for Security}
The mathematical architecture of GDMs offers unique theoretical advantages that align directly with the security requirements of wireless networks as follows.
\begin{itemize}
    \item \textbf{Forward Diffusion as Semantic Encryption:} The forward process, which incrementally maps structured data to isotropic Gaussian noise, is mathematically analogous to physical-layer encryption and artificial noise injection. By transforming meaningful semantic features into high-entropy noise distributions, GDMs inherently perform semantic obfuscation, rendering sensitive user information unintelligible to eavesdroppers who lack the specific reverse denoising ``key" or conditions \cite{qiu2025plugging}.
    \item \textbf{Reverse Denoising as Adversarial Purification:} The reverse denoising process functions as a robust purification mechanism against malicious interference\cite{nie2022diffusion}. Unlike discriminative models that are brittle to perturbations, GDMs effectively project data corrupted by adversarial attacks back onto the learned legitimate data manifold. This allows them to neutralize adversarial perturbations and recover clean semantic information, acting as a natural defense against jamming and spoofing.
    \item \textbf{Conditional Generation and Likelihood for Authentication:} GDM's sensitivity to conditional inputs enables coverless steganography, where specific semantic prompts act as private keys to control the generation of hidden information. Furthermore, the model's explicit likelihood estimation capabilities allow for rigorous out-of-distribution detection\cite{liu2025survey213}. By evaluating reconstruction errors, GDMs can identify unauthorized devices or anomalous signal patterns that deviate from the learned legitimate distribution, thereby enhancing identity management and authentication.
\end{itemize}

\subsection{Overview of the Structured Taxonomy for GDM-enabled Wireless Networks}
\label{section22}

\begin{figure*}
\centering
\includegraphics[width=1\textwidth]{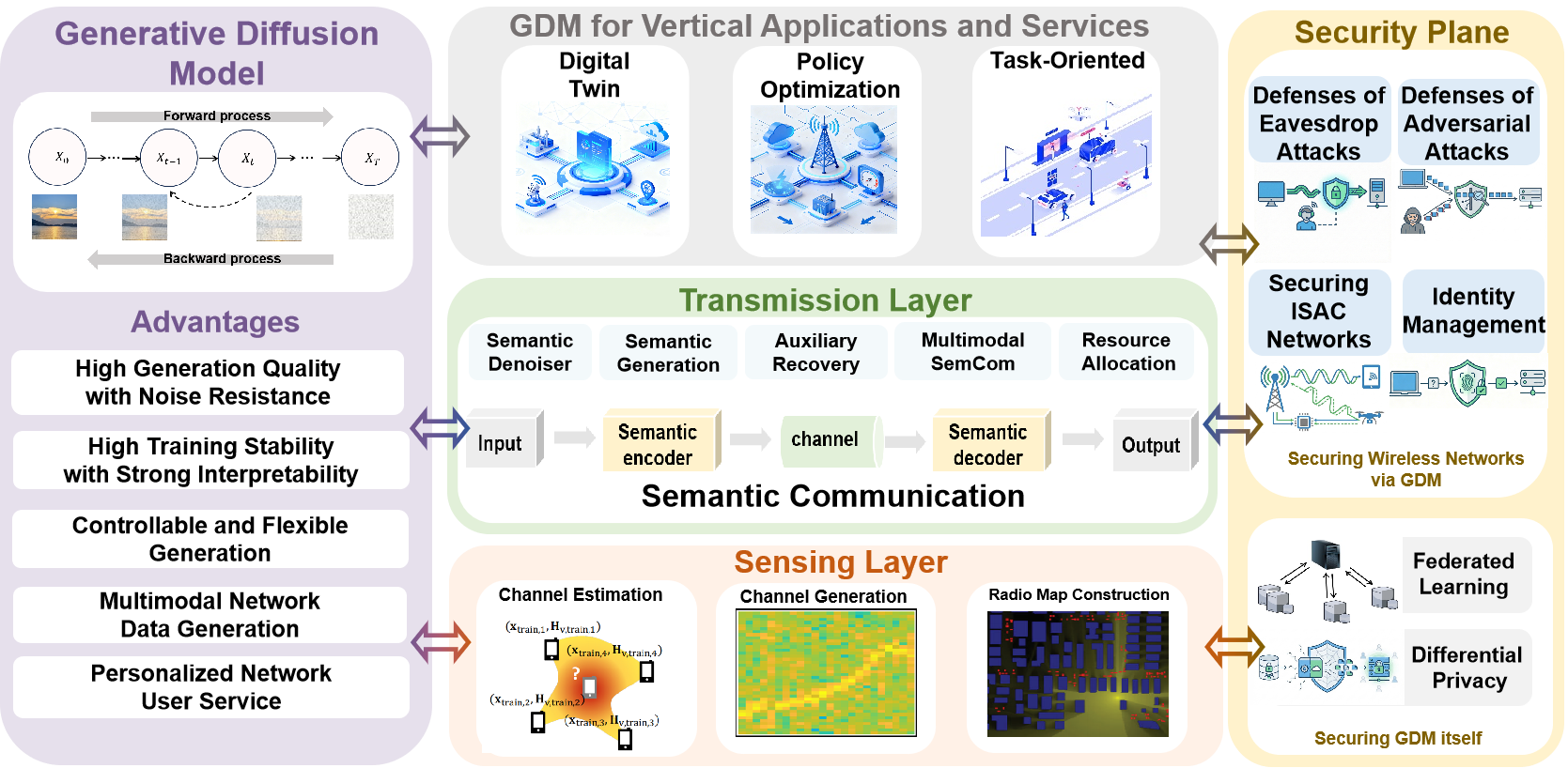}
\caption{Illustration of the GDM-aided multi-layer network taxonomy. This framework organizes the application of GDMs into three functional layers and a cross-cutting security plane. The GDM module on the left serves as the foundational engine that provides high-quality generation and noise resilience to the entire system. At the bottom level, the sensing layer utilizes these generative priors to accurately model the physical environment through tasks such as channel estimation and radio map construction. Above this lies the transmission Layer which leverages the iterative denoising capabilities of GDMs to optimize end-to-end data delivery and enhance SemCom. The top level depicts vertical applications and services where the GDM functions as a world model for network digital twins or a solver for policy optimization. Finally, the security plane on the right interacts with all layers to defend against eavesdropping and adversarial attacks while simultaneously employing privacy-preserving techniques to secure the deployment of GDMs themselves.}
\label{picture_multilayerarchitecture}
\end{figure*}

Before detailing the framework, it is crucial to clarify its rationale and scope. The presented multi-layer structure is not a new network protocol stack like the OSI or TCP/IP model. Instead, it serves as a conceptual framework, or taxonomy, designed to classify and organize the diverse applications of GDMs within the wireless network ecosystem. As illustrated in Figure \ref{picture_multilayerarchitecture}, the sensing layer encompasses GDM's role in perceiving and modeling the physical radio environment; the transmission layer focuses on GDM's role in optimizing the end-to-end data delivery process, particularly SemComs; and the vertical applications represents GDM's role in directly enabling or generating end-user services and vertical use cases.
Furthermore, in the security plane, GDM can enhance security across all three of the aforementioned layers. Such GDM-enabled multi-layer network framework allows us to systematically review the state-of-the-art and identify challenges based on the distinct problems GDMs can solve.

\subsubsection{Sensing Layer}
The wireless channel serves as the transmission medium for signals from the transmitter to the receiver, and its characteristics are often influenced by multiple factors such as environmental conditions, transmission distance, and terminal mobility \cite{wei2022multi}. The objective of the sensing layer is to abstract and refine these complex factors, thereby facilitating precise performance evaluation for the system design of wireless networks \cite{lopez2022survey}. In the sensing layer, channel estimation focuses on obtaining the characteristics and parameters of the actual channel \cite{chen2025generative}, channel generation focuses on simulating the channel environment \cite{kim2023diffusion}, and radio map construction focuses on presenting the distribution of radio signals in a specific area \cite{zhang2021intelligent,yue2024channel}.

\begin{itemize}

\item \textbf{High Quality Channel Acquisition:}
By learning the posterior channel data distribution and applying posterior sampling techniques in the reverse denoising processes of GDM, the true channel response can be recovered, thereby significantly improving channel estimation performance under low SNR environments.

\item \textbf{Generation Capability Covering Various Channel Conditions:}
Although GAN shows good performance for sample channel, it may face limitations for more complex channels, such as tapped delay line channel \cite{dorner2020wgan}. 
GAN can produce high-quality samples, but its mode coverage is very poor, which is called mode collapse \cite{kim2023diffusion}. 
In contrast, GDM can effectively solve this problem and generate channel samples covering a variety of environments \cite{sengupta2023generative,chi2024rf}.

\item \textbf{Controllable Generation of Channel Samples:}
GDM has the ability to integrate prior conditions such as transmitter locations and RSS fragments, enabling high-precision predictions under sparse measurement conditions and bringing new breakthroughs and possibilities to the field of radio map construction \cite{jia2025rmdm}.

\end{itemize}

\subsubsection{Transmission Layer}

In the context of this survey, the transmission layer adopts a generalized definition, encompassing the end-to-end data delivery process as well as associated optimization tasks such as resource allocation. While this layer theoretically spans various transmission paradigms, our literature review reveals that the application of GDMs is predominantly concentrated on SemCom \cite{lu2025important,cao2025importance}. This trend arises because the generative and denoising capabilities of GDMs align intrinsically with the core objectives of SemCom. Consequently, we primarily focus on GDM-based SemCom schemes, leveraging SemCom’s established advantages in efficient transmission without redundancy\cite{liu2024survey7.7}, high transmission accuracy\cite{dong2022semantic7.7}, and robust anti-interference capability\cite{grassucci2023generative}. Specifically, GDMs offer the following enhancements for SemCom:

\begin{itemize}
\item \textbf{Stronger Semantic Learning Capability:}
As previously discussed, the theoretical framework and architecture of GDMs enable multi-scale feature capture, resulting in superior distribution modeling and semantic learning capabilities \cite{zeng2024dmce}.
\item \textbf{Enhanced Anti-Interference Performance:}
The training process of GDMs inherently involves noise injection and denoising, granting them natural adaptability to noise interference. In SemCom, the receiver can iteratively denoise and recover semantic features corrupted by channel noises or other disturbances \cite{grassucci2023generative}.
\item \textbf{Improved Cross-Modal Communication Ability:}
GDMs support multimodal generation, such as text-to-image and speech-to-text conversions, enabling cross-modal SemCom. For example, the sender encodes textual semantics into a latent vector, while the receiver generates corresponding images or speech. This capability breaks through modal limitations, facilitating semantic interactions across heterogeneous devices \cite{wei2024language}.
\end{itemize}

\subsubsection{Vertical Applications and Services}

After reviewing the foundational roles of GDMs in the sensing and transmission layers, this section details how these capabilities are integrated to empower large-scale vertical applications and services. The focus shifts from single-layer functions to GDM's enhanced role as a key enabling technology, integrating its generative, modeling, and restoration capabilities to meet specific end-to-end goals.
\begin{itemize}

\item \textbf{GDM for Network Digital Twin Simulation:}
GDM's role evolves from passive channel modeling to the generative simulation of the entire network ecosystem. As a ``world model" or dynamic engine for the network digital twin, they learn complex network dynamics, topologies, and user behaviors to generate high-fidelity, interactive virtual network environments for testing and network introspection.

\item \textbf{GDM for Network Policy and Solution Optimization:}
GDM is used as an active decision-making component and policy optimizer, serving as a direct end-to-end optimization solver or as a highly efficient policy network within DRL frameworks. They excel at optimizing NP-hard problems, such as secure beamforming or UAV trajectory planning, by sampling from the complex distribution of high-quality solutions.

\item \textbf{GDM for Task-Oriented Generative Communications:}
In vertical domains such as vehicular networks, UAV swarms, and immersive communication, GDM enables task-oriented communication. The goal is no longer perfect reconstruction but to use minimal semantic inputs, such as a segmentation map, to generate entirely new, high-fidelity content that satisfies a specific task requirement, such as a photorealistic road scene.

\end{itemize}

\subsubsection{Security Plane}
The security plane serves as a cross-cutting layer that safeguards the entire network ecosystem. Building upon the intrinsic security properties discussed in Section II-A, this plane leverages the forward diffusion process for active semantic encryption and the reverse denoising trajectory for adversarial purification against eavesdropping and jamming. Beyond defense, it also encompasses internal security mechanisms, such as differential privacy and robust federated aggregation, to protect the training and deployment of the GDM infrastructure itself from privacy leakage and backdoor attacks.

\begin{itemize}
\item \textbf{Applications of GDM in Wireless Network Security:} 
GDMs function as robust defenders against diverse threats across network layers. 
For defenses against eavesdropping, GDMs enable pluggable semantic encryption and coverless steganography to conceal information from unauthorized interceptors \cite{he2025diffusion, gao2025semstediff}. 
Against adversarial attacks, they act as purification modules, utilizing asymmetric diffusion or pluggable protectors to eliminate malicious perturbations \cite{ren2023asymmetric, qiu2025plugging}. 
In ISAC networks, GDMs generate protective signals to mask user activities and prevent illegitimate sensing \cite{wang2024generative}, while in identity management, they enhance physical layer authentication through generative fingerprint modeling and open-set identification \cite{meng2025generative, wang2023diffusion}.

\item \textbf{Securing GDM-based Wireless Networks:} 
To mitigate security and privacy risks inherent to the deployment of GDMs, specific protective measures are adopted. 
For federated deployment, techniques such as dynamic quantization and trajectory analysis are employed to filter backdoor triggers and secure collaborative training \cite{he2024securing}. 
To preserve data privacy, hybrid training frameworks combine differential privacy with GDMs, effectively preventing membership inference attacks while maintaining the high fidelity of generated wireless data \cite{wangprivacy}.
\end{itemize}

\subsection{Summaries and Lessons Learned}

\begin{table*}[tbp]
\centering
\caption{Taxonomy of GDM Variants and Their Suitability for Wireless Network Tasks.}
\small
\renewcommand{\arraystretch}{1.2} 
\begin{tabular}{p{2.3cm} p{5.8cm} p{6.2cm} p{2.5cm}}
\toprule
\textbf{GDM Variant} & \textbf{Key Strengths in Wireless Context} & \textbf{Best-Suited Wireless Scenarios} & \textbf{Representative Works} \\ 
\midrule

DDPM & 
High-fidelity baseline providing the most stable training and accurate distribution coverage that effectively overcomes mode collapse. & 
High-quality data synthesis and offline training tasks including channel generation, radio map construction, and data augmentation where inference latency is not critical. & 
\cite{sengupta2023generative, qiu2023irdm, he2025diffusion, zeng2023radio} \\ 
\addlinespace[1.5ex] 

SGM, SDE, and ODE & 
Unified inverse solver that models generation as a continuous-time process to enable flexible estimation from noisy observations. & 
Sensing and estimation tasks such as channel estimation under complex noise, joint channel-data recovery, and signal denoising via score-based priors. & 
\cite{chen2025generative, tong2024diffusion, zilberstein2024joint, fesl2024diffusion} \\ 
\addlinespace[1.5ex]

FM & 
Stable and efficient training that directly learns the vector field with simulation-free objectives for faster convergence than standard diffusion. & 
Complex distribution modeling for efficient radio map construction and active learning tasks requiring rapid training on large-scale environmental data. & 
\cite{jia2025radioflow11.13, sun2025flow11.13} \\ 
\addlinespace[1.5ex]

DDIM and CM & 
Ultra-fast sampling that accelerates inference by skipping steps in DDIM or enabling single-step generation in CMs. & 
Real-time applications covering latency-sensitive SemCom, fast beam tracking, and real-time network optimization policies. & 
\cite{ma2024diffusion, pei2025latent, zhao2025secdiff, liu2025generative} \\ 
\addlinespace[1.5ex]

CDM & 
Controllable generation that directs the generative process using external guidance signals including pilots, text, and location. & 
Conditional and inverse tasks such as location-based channel generation, resource allocation based on channel state, and secure beamforming. & 
\cite{lee2024generating, luo2024rm, du2023ai, zhang2025enhanced11.8} \\ 
\addlinespace[1.5ex]

LDM & 
Resource efficiency optimization by performing diffusion in a compressed latent space to significantly reduce computational costs. & 
Bandwidth-constrained transmission including SemCom for images and video, edge-side generation, and multimodal transmission. & 
\cite{xu2023latent, li2025goal, fu2024multimodal, yan2025semantic}\\ 

\bottomrule
\end{tabular}
\label{table_model_scenario_mapping}
\end{table*}

In this section, we present an overview of GDMs and the GDM-aided multi-layer network framework. From this section, the lessons learned are as follows:
\begin{itemize}
    \item As a breakthrough in GAI, GDMs significantly outperform traditional models such as GANs in training stability and mode coverage. The ecosystem has evolved from foundational DDPMs to continuous-time SGMs and SDEs for flexible modeling. Recent advances focus on efficiency and control where LDMs reduce computational dimensionality and CDMs enable precise conditional guidance. Furthermore, FM and CMs are emerging to solve the critical inference latency bottleneck which paves the way for real-time wireless applications.
    
    \item The GDM-aided multi-layer wireless network framework presented in Section \ref{section22} systematically organizes the enhancements in functionalities and security across network domains. Instead of a rigid protocol stack, this framework emphasizes how GDMs address distinct problems including physically modeling the environment in the sensing domain, optimizing end-to-end data delivery in the transmission domain, and enabling generative services for vertical industries in the application domain.
    
    \item Table \ref{table_model_scenario_mapping} provides a strategic guide for mapping GDM variants to specific wireless tasks. As observed, there is no one size fits all model. Tasks requiring maximum fidelity such as channel modeling benefit from DDPMs while latency critical tasks including real-time semantic decoding necessitate fast samplers like DDIM and CM. Meanwhile, resource-constrained edge scenarios are best served by LDMs. This mapping aims to assist researchers in selecting the appropriate generative backbone that aligns with their specific constraints on latency, accuracy, and computational resources.
\end{itemize}

\section{GDM for the Sensing Layer}
\label{section3}

This section presents GDM-based channel estimation, channel generation, and radio map construction schemes for the sensing layer. Before detailing the specific GDM applications, we first establish the foundational principle of using generative models for sensing tasks and review the prior art, namely GANs, to contextualize the advantages of GDMs.

\subsection{Generative Models for Sensing: Principles and Prior Art}

\subsubsection{The Bayesian Principle: Generative Models as Learned Priors}

The underlying principle of using generative models for sensing tasks, such as channel estimation, is to reframe the task as a Bayesian inverse problem. The goal is to estimate the unknown channel $H$ given the received signal $Y$, which is governed by the posterior probability $P(H|Y)$. According to Bayes' theorem, this is proportional to the likelihood $P(Y|H)$ multiplied by the prior $P(H)$.

The likelihood $P(Y|H)$ is the physical forward model (e.g., $Y=HX+N$), which is often known. The critical challenge is the channel prior $P(H)$. Classical methods like LMMSE assume a simple, mathematically convenient Gaussian prior, which poorly represents the complex, non-Gaussian nature of real-world channels.

This is the exact gap that generative models fill. A generative model, pre-trained on channel data, serves as a powerful, non-Gaussian learned prior that implicitly models the high-dimensional distribution of all plausible channels. The estimation (or generation) task thus becomes an optimization or sampling process: finding a channel $H$ that is consistent with the physical observation $Y$ (maximizing the likelihood) and structurally realistic according to the learned generative prior $P(H)$.

\subsubsection{Prior Art: Limitations of Classical Methods and GANs}
The need for powerful generative priors is highlighted by the limitations of preceding methods. For instance, in channel estimation, classical linear channel estimation methods, such as Least Squares (LS) \cite{lin2008least}, Linear Minimum Mean Square Error (LMMSE) \cite{vithanage2010linear}, and compressed sensing \cite{berger2010application}, have the limitations in training overhead \cite{zhou2025generative}, reliance on acquired CSI samples \cite{chen2025generative}, and the assumption of channel sparsity \cite{tong2024diffusion}.

DAI methods were introduced to overcome these issues but brought their own limitations. They are highly dependent on specific measurement configurations, such as fixed numbers of pilots and antennas, which restricts their generalization ability. Additionally, these methods require a large amount of labeled datasets for support, resulting in high costs for data acquisition and annotation\cite{zhou2025generative}.

As an alternative to discriminative mapping, GANs were the first models to effectively act as a powerful learned prior. In both channel estimation and generation, a GAN generator could learn the complex channel distribution from data. However, GANs suffer from fundamental limitations that hinder their effectiveness in wireless sensing:
\begin{itemize}
    \item \textbf{Training Instability:} The core issue is the adversarial training process. The dynamic game between the generator and discriminator is prone to non-convergence and is difficult to stabilize.
    \item \textbf{Mode Collapse:} This instability often leads to mode collapse, where the generator fails to capture the full diversity of the channel distribution. This is a critical failure for wireless sensing, which must account for a wide variety of channel states, including rare events\cite{sengupta2023generative}.
    \item \textbf{Poor Generalization:} Consequently, GAN-based models show weaker generalization. For example, in end-to-end learning, WGAN-based models have been shown to diverge and perform poorly in high-SNR regions, whereas GDM-based approaches remain robust\cite{chen2025generative}.
\end{itemize}

GDMs emerged as a direct solution to these challenges. As established in Section II, GDM's training objective is based on a stable, mathematically interpretable maximum likelihood estimation, by reversing a diffusion process. This stable learning process allows them to overcome mode collapse and learn the entire distribution of complex data, leading to superior generalization and noise robustness. The following subsections detail how GDMs are applied to specific sensing tasks, leveraging this superior generative capability.

\subsection{GDM for Channel Estimation}

Building on the generative prior principle established in Section III-A, GDMs bring specific advantages to channel estimation. Researchers have adapted the GDM framework to serve as a powerful learned prior $P(H)$ for various specific and challenging estimation tasks. By learning the score function of the posterior channel data distribution, GDMs can recover the true channel response by applying posterior sampling techniques during the reverse denoising process.
  
Table \ref{channelmodeling1} illustrates existing GDM-based channel estimation schemes, and the detailed descriptions are as follows.

\subsubsection{Channel Estimation for MIMO Systems}
In massive MIMO systems, the primary challenges are the high dimensionality of the channel matrix $H$ and the desire to reduce pilot overhead. GDMs are applied here as powerful priors to regularize this ill-posed inverse problem.
Ma et al.\cite{ma2024diffusion}  directly apply DDPM and DDIM as posterior samplers to recover the channel. Their work compares the two, demonstrating that the deterministic DDIM process can achieve superior performance to the stochastic DDPM while reducing computational complexity by 80\%.
Addressing the critical challenge of complexity, Fesl et al.\cite{fesl2024diffusion}  focus on reducing complexity and memory overhead. They propose a lightweight CNN-based DM that learns the channel distribution in the sparse angular domain. Their key innovation is a deterministic estimation strategy that avoids stochastic resampling. It truncates the reverse diffusion process by matching the observation's SNR to a corresponding DM step, an approach shown to be asymptotically MSE-optimal.
Alternatively, Chen et al.\cite{chen2025generative}  frame the problem as variational inference using a pre-trained SGM as the prior. A central contribution is a novel weighting mechanism that reweights the prior term during posterior sampling. This regularizes the channel recovery process and, critically, avoids the need for backpropagation through the complex SGM, significantly accelerating the estimation.

\subsubsection{Channel Estimation for Low-resolution ADCs}
The challenge in systems with low-resolution ADCs is the severe non-linear quantization noise, which makes the likelihood term $P(Y|H)$ analytically intractable.
Zhou et al.\cite{zhou2025generative}  tackle this by adapting the GDM-based posterior inference. Their core contribution is to modify the likelihood information in the posterior sampling step to account for the non-linear quantization function. Furthermore, to enhance practical viability, they integrate Stein's Unbiased Risk Estimator, enabling the GDM prior to be learned directly from noisy channel observations without requiring a clean ground-truth dataset.

\subsubsection{Channel Estimation for RIS-aided Systems}

For RIS-aided systems, the estimation is complicated by the high-dimensional cascaded channel and the introduction of hardware-induced phase noise at the RIS elements.
Tong et al.\cite{tong2024diffusion} propose a GDM-based method specifically designed to be robust against both receiver noise and this RIS phase noise. The key innovation is to incorporate the gradient descent value of the RIS phase directly into the reverse sampling process, actively enhancing resilience against phase noise. To address the high inference overhead of GDMs, they employ a progressive distillation framework to reduce the sampling steps to 32, making the method computationally comparable to greedy algorithms.

\subsubsection{Joint Channel Estimation and Data Detection}
This task represents a blind inverse problem where both the channel $H$ and the data symbols $X_D$ are unknown, requiring the model to untangle both simultaneously.
Zilberstein et al.\cite{zilberstein2024joint} solve this by using an SGM to sample from the joint posterior distribution $P(H, X_D | Y)$. The core of their method is handling the hybrid (continuous and discrete) nature of the variables. It uses a learned SGM prior for the continuous channel $H$ and an analytically derived score based on Tweedie's identity for the discrete data symbols $X_D$. This allows the algorithm to efficiently explore the joint search space and reuse data symbols as pilots.

\begin{table}[tbp] 
    \centering
    \caption{Existing GDM-based channel estimation schemes, where \textcolor{blue}{$\diamondsuit$}, \textcolor{blue}{$\sphericalangle$}, \textcolor{green}{$\checkmark$}, and \textcolor{red}{$\times$} respectively are contributions, the role of GDM, pros, and cons.}
    \label{channelmodeling1}
    \renewcommand{\arraystretch}{1.2}  
    \normalsize  
    \begin{tabular}{|>{\arraybackslash}m{0.025\textwidth}|>{\arraybackslash}m{0.42\textwidth}|}
        \hline
        \textbf{\small{Ref.}} & \textbf{\small{Descriptions}} \\

        \hline
        \footnotesize \cite{ma2024diffusion}  & \footnotesize
         \textcolor{blue}{$\diamondsuit$}: Introduce DDPM and DDIM to enhance channel estimation performance.
  
         \textcolor{blue}{$\sphericalangle$}: Employs DDPM/DDIM as posterior samplers to learn the score function of the posterior channel data distribution and recover the channel.

         \textcolor{green}{$\checkmark$}: The deterministic DDIM process achieves superior performance to DDPM while reducing computational complexity by 80\%.

         \textcolor{red}{$\times$}: The performance under high-mobility conditions and complex environments is not verified.
        \\
        \hline
        \footnotesize \cite{fesl2024diffusion}  & \footnotesize
         \textcolor{blue}{$\diamondsuit$}: Propose a lightweight SDE-based method for low-complexity and low-memory overhead massive MIMO channel estimation.
  
         \textcolor{blue}{$\sphericalangle$}: A lightweight CNN learns the channel distribution in the sparse angular domain. Uses a deterministic strategy, truncating the reverse process by matching the observation's SNR to a corresponding DM step.

         \textcolor{green}{$\checkmark$}: Achieves low complexity and memory overhead; shows superior performance across different SNRs.

         \textcolor{red}{$\times$}: The performance is highly dependent on the sparse angular domain transformation.
        \\
        \hline
        \footnotesize \cite{chen2025generative}  & \footnotesize
         \textcolor{blue}{$\diamondsuit$}: Propose a variational inference method using a pre-trained SGM as a prior.
  
         \textcolor{blue}{$\sphericalangle$}: The SGM serves as the prior. A novel weighting mechanism reweights the prior term during posterior sampling.

         \textcolor{green}{$\checkmark$}: Significantly accelerates estimation by avoiding backpropagation through the SGM; outperforms baselines in convergence and complexity.

         \textcolor{red}{$\times$}: It relies on a pre-trained model and its performance slightly declines as the scale of antennas increases.
        \\
        \hline
        \footnotesize \cite{zhou2025generative}  & \footnotesize
         \textcolor{blue}{$\diamondsuit$}: Propose a GDM-based posterior inference method for systems with low-resolution ADCs.
  
         \textcolor{blue}{$\sphericalangle$}: Modifies the likelihood information in the posterior sampling step to account for non-linear quantization. Integrates SURE to enable learning the prior from noisy observations.

         \textcolor{green}{$\checkmark$}: Achieves high-precision channel recovery; can be learned directly from noisy observations without clean ground-truth data.

         \textcolor{red}{$\times$}: Its rapid adaptability under limited data samples needs analysis.
        \\
        \hline
        \footnotesize \cite{tong2024diffusion}  & \footnotesize
         \textcolor{blue}{$\diamondsuit$}: Propose an SGM-based method robust against receiver noise and RIS phase noise.
  
         \textcolor{blue}{$\sphericalangle$}: Incorporates the gradient descent value of the RIS phase into the reverse sampling process. Employs knowledge distillation to accelerate sampling.

         \textcolor{green}{$\checkmark$}: Robust against RIS phase noise; does not require retraining for different SNRs; computationally comparable to greedy algorithms.

         \textcolor{red}{$\times$}: The trade-off between estimation performance and complexity.
        \\
        \hline
        \footnotesize \cite{zilberstein2024joint}  & \footnotesize
         \textcolor{blue}{$\diamondsuit$}: Propose a joint channel estimation and data detection algorithm for massive MIMO systems.
  
         \textcolor{blue}{$\sphericalangle$}: Uses an SGM to sample from the joint posterior distribution $P(H, X_D | Y)$. It combines a learned SGM prior for the continuous channel (H) and an analytical score for the discrete symbols ($X_D$).

         \textcolor{green}{$\checkmark$}: Efficiently explores the joint search space; improves estimation performance by reusing data symbols as pilots.

         \textcolor{red}{$\times$}: The architecture of the model needs to be further simplified; performance under low SNR conditions needs improvement.
        \\
        \hline
    \end{tabular}
\end{table}

\subsection{GDM for Channel Generation}

While channel estimation focuses on inferring a specific channel from a noisy observation, channel generation addresses the parallel challenge of {synthesizing} new, realistic channel realizations from scratch. This capability is essential for creating large-scale, high-quality datasets for training and testing communication systems, a process that is otherwise prohibitively expensive and time-consuming \cite{sengupta2023generative, lee2024generating}. The following schemes demonstrate how GDMs are used for both high-fidelity synthesis and advanced, conditional channel generation.
Table \ref{channelmodeling2} presents existing GDM-based channel generation schemes, which are described in detail as follows.

\begin{table}[tbp] 
    \centering
    \caption{Existing GDM-based channel generation schemes, where \textcolor{blue}{$\diamondsuit$}, \textcolor{blue}{$\sphericalangle$}, \textcolor{green}{$\checkmark$}, and \textcolor{red}{$\times$} respectively are contributions, the role of GDM, pros, and cons.}
    \label{channelmodeling2}
    \renewcommand{\arraystretch}{1.2}  
    \normalsize  
    \begin{tabular}{|>{\arraybackslash}m{0.025\textwidth}|>{\arraybackslash}m{0.42\textwidth}|}
        \hline
        \textbf{\small{Ref.}} & \textbf{\small{Descriptions}} \\
        \hline
        \footnotesize \cite{sengupta2023generative}  & \footnotesize
         \textcolor{blue}{$\diamondsuit$}: Propose a DDPM for high-fidelity and diverse channel synthesis to overcome GAN limitations.
  
         \textcolor{blue}{$\sphericalangle$}: A DDPM with a U-Net architecture operates in the frequency-space domain to learn the channel distribution.

         \textcolor{green}{$\checkmark$}: Achieves stable training and high recall, successfully capturing full channel diversity where GANs suffer mode collapse. Validated for data augmentation in data-scarce scenarios.

         \textcolor{red}{$\times$}: The generation process is iterative and computationally complex.
        \\
        \hline
        \footnotesize \cite{kim2023diffusion}  & \footnotesize
         \textcolor{blue}{$\diamondsuit$}: Demonstrate an ODE-based model for accurate channel distribution generation, especially for complex channels where GANs fail.
  
         \textcolor{blue}{$\sphericalangle$}: An ODE-based model learns complex channel distributions, including correlated fading.

         \textcolor{green}{$\checkmark$}: Accurately captures complex channel properties like covariance structure, outperforming WGANs. Generated channels show negligible SER deviations in an end-to-end system.

         \textcolor{red}{$\times$}: A trade-off exists between the sampling speed reduction and the generation quality.
        \\
        \hline
        \footnotesize \cite{lee2024generating}  & \footnotesize
         \textcolor{blue}{$\diamondsuit$}: Propose a conditional DDIM framework to generate high-dimensional, user-specific mmWave channels based on user position.
  
         \textcolor{blue}{$\sphericalangle$}: A conditional DDIM learns the mapping from user position to the channel matrix, operating in the sparse beamspace domain. Explores consistency training for acceleration.

         \textcolor{green}{$\checkmark$}: Enables generation of user-specific channels based on location. Identifies that operating in the beamspace domain is crucial for success.

         \textcolor{red}{$\times$}: The iterative inference process has high computational complexity.
        \\
        \hline
        \footnotesize \cite{gong2025digital}  & \footnotesize
         \textcolor{blue}{$\diamondsuit$}: Introduce the Digital Twin of Channel DToC concept, mapping user location to statistical CSI.
  
         \textcolor{blue}{$\sphericalangle$}: A CDM is trained to learn the generative mapping from user position $p$ to statistical CSI $\Omega$.

         \textcolor{green}{$\checkmark$}: Bypasses pilot-based estimation by predicting statistical CSI from location alone. Supports parallel generation for multiple users.

         \textcolor{red}{$\times$}: The scalability of this framework to new environments requires further verification.
        \\
        \hline
        \footnotesize \cite{chi2024rf}  & \footnotesize
         \textcolor{blue}{$\diamondsuit$}: Propose a Time-Frequency Diffusion TFD theory and Hierarchical Diffusion Transformer HDT to generate raw, time-series RF signals.
  
         \textcolor{blue}{$\sphericalangle$}: A novel GDM combining time-series noise and frequency blur in the forward process, with an HDT backbone that decouples denoising and deblurring.

         \textcolor{green}{$\checkmark$}: Generates high-fidelity, complex-valued time-series signals, which standard GDMs designed for static images cannot do.

         \textcolor{red}{$\times$}: The new HDT architecture introduces significant computational complexity.
        \\
        \hline
    \end{tabular}
\end{table}

\subsubsection{High-Fidelity and Diverse Channel Synthesis} 
A primary motivation for using GDMs over GANs is to achieve stable training and generate a diverse set of high-fidelity samples, thereby avoiding the mode collapse that plagues adversarial models.
Sengupta et al. \cite{sengupta2023generative} directly tackle this issue by proposing a DDPM with a U-Net architecture operating in the frequency-space domain. They provide a rigorous comparison against a WGAN, ChannelGAN, using Wasserstein distance and generative model metrics of precision and recall. The results show that while the GAN's training is unstable, the DDPM converges stably. Critically, while the WGAN achieves decent precision, it suffers from extremely low recall, confirming severe mode collapse. The DDPM, in contrast, achieves high recall, proving it successfully captures the channel distribution's full diversity. This work also validates the use of GDM for data augmentation in data-scarce scenarios by pre-training on a large simulated dataset and successfully fine-tuning on a small, different dataset.
Similarly, Kim et al. \cite{kim2023diffusion} demonstrate that ODE-based models offer superior generative performance and can accurately learn complex channel distributions where GANs fail. Their evaluation in an end-to-end framework shows that DM-generated channels, such as AWGN, Rayleigh, SSPA, result in negligible SER deviations compared to the exact channel models. A key finding is that the GDM successfully learns a correlated fading channel, accurately capturing its covariance structure. In contrast, a strong WGAN variant completely fails to learn the correlation, highlighting GDM's superior ability to model complex, statistically-dependent channel properties.

\subsubsection{Conditional and Specialized Channel Generation}
Beyond sampling from the entire channel distribution $P(H)$, GDMs excel at conditional generation, $P(H|c)$, where new channels are synthesized based on specific physical conditions, such as user location or signal properties. Lee et al. \cite{lee2024generating} propose a conditional DDIM framework to generate high-dimensional, user-specific mmWave channels. The key condition is the user's position $x$, allowing the model to learn the specific mapping $p(H_v|x)$. A critical insight from this work is that the model's success hinges on operating in the sparse beamspace domain, by applying a DFT to the channel matrix, which is crucial for the GDM to effectively capture the underlying distribution. The work also explores consistency training to address the high computational complexity of GDM inference.

Gong et al. \cite{gong2025digital} expand this location-conditional concept into a full system architecture, which they term the ``Digital Twin of Channel" (DToC). In this framework, the UT's position is the "physical object," and the statistical CSI (e.g., the beam domain channel power matrix, $\Omega$) is the "virtual digital object". A conditional DM is trained on a dataset of pairs collected from reference points. This creates a learned generative mapping $p(\Omega|p)$. The DToC framework can then predict a new UT's statistical CSI based only on its location, entirely bypassing the need for pilot-based estimation.

Addressing a different aspect, Chi et al. \cite{chi2024rf} focus on generating raw, time-series RF signals, such as Wi-Fi or FMCW signals, rather than static channel matrices. Recognizing that standard GDMs are designed for static images and ignore the time and frequency domains, they propose a novel Time-Frequency Diffusion theory. The TFD forward process innovatively combines adding time-series noise with applying frequency blur. To implement this, they design a Hierarchical Diffusion Transformer. The reverse process in the HDT explicitly decouples the task into a spatial denoise stage and a time-frequency deblur stage, allowing it to generate high-fidelity, complex-valued time-series signals.

\subsection{GDM for Radio Map Construction}

Radio map obtains radio signal features such as path loss through location information. It presents radio signal features in a specific area in the form of a map, reflecting the distribution of radio signals in the area. Traditional radio map construction approaches are divided into sampling-based and non-sampling-based construction methods, which have the limitations in measurement costs, computational complexity, and dynamic environments. In comparison, GDM demonstrates unique advantages. GDM can more delicately capture the dynamic details of complex signal propagation. In addition, GDM adopts a step-by-step generation strategy, effectively simulating the signal strength distribution after multipath superposition. GDM-based approaches generally fall into two categories: (1) sampling-based interpolation, which uses sparse measurements to in-paint a complete map, and (2) sampling-free generation, which synthesizes a map based only on environmental features.
Table \ref{radiomap} illustrates existing GDM-based radio map construction schemes, with the detailed descriptions as follows.

\begin{table}[tbp] 
    \centering
    \caption{Existing GDM-based radio map construction schemes, where \textcolor{blue}{$\diamondsuit$}, \textcolor{blue}{$\sphericalangle$}, \textcolor{green}{$\checkmark$}, and \textcolor{red}{$\times$} respectively are contributions, the role of GDM, pros, and cons.}
    \label{radiomap}
    \renewcommand{\arraystretch}{1.2}  
    \normalsize  
    \begin{tabular}{|>{\arraybackslash}m{0.025\textwidth}|>{\arraybackslash}m{0.42\textwidth}|}
        \hline
        \textbf{\small{Ref.}} & \textbf{\small{Descriptions}} \\
        \hline
        \footnotesize \cite{qiu2023irdm}  & \footnotesize
         \textcolor{blue}{$\diamondsuit$}: Propose a DDPM-based radio map interpolation method for the first time to address the task of indoor path loss map interpolation. 
  
         \textcolor{blue}{$\sphericalangle$}: It generates the complete indoor path loss map layer based on the geometric layer, positional encoding layer, and sparse map layer to achieve end-to-end mapping.

         \textcolor{green}{$\checkmark$}: It can generate a complete indoor path loss map with only 10\% of reference points, and it enhances adaptability to unknown environments through online data augmentation.

         \textcolor{red}{$\times$}: Trade-off between generation performance and complexity.
        \\
        \hline
        \footnotesize \cite{luo2024rm}  & \footnotesize
         \textcolor{blue}{$\diamondsuit$}: Employ CDM to generate radio maps for mmWave WLANs and 5G cellular networks.
  
         \textcolor{blue}{$\sphericalangle$}: Based on two low-cost conditional inputs: sparse RSS segments and transmitter locations, this solution is capable of generating radio maps for complex scenarios.

         \textcolor{green}{$\checkmark$}: It reduces the amount of measurement data required.

         \textcolor{red}{$\times$}: The optimization effects of multi-condition joint inputs and complexity requirements are not analyzed.
        \\
        \hline
        \footnotesize \cite{liu2025wifi}  & \footnotesize
         \textcolor{blue}{$\diamondsuit$}: Combine DDIM and physical propagation models to generate and screen candidate radio maps that best conform to the laws of propagation.
  
         \textcolor{blue}{$\sphericalangle$}: It generates a rich variety of radio maps from a noise distribution using DDIM and combines physical priors provided by the Boost Block to enhance the rationality of the generated maps.

         \textcolor{green}{$\checkmark$}: It is capable of generating fine-grained radio maps, and each module is scalable.

         \textcolor{red}{$\times$}: Relies on prior propagation models to improve generation performance and does not analyze complexity.
        \\
        \hline
        \footnotesize \cite{wang2024radiodiff}  & \footnotesize
         \textcolor{blue}{$\diamondsuit$}: Present a sampling-free radio map construction scheme based on LDM and verify its performance under static and dynamic environments.
  
         \textcolor{blue}{$\sphericalangle$}: The decoupled GDM is combined with Fast Fourier Transform to extract features from dynamic environments better.

         \textcolor{green}{$\checkmark$}: Better generating flexibility in dynamic environment.

         \textcolor{red}{$\times$}: The generation performance between inference complexity.
        \\
        \hline
        \footnotesize \cite{jia2025rmdm}  & \footnotesize
         \textcolor{blue}{$\diamondsuit$}: Combine PIIN and CDM to enhance radio map construction performance.
  
         \textcolor{blue}{$\sphericalangle$}: It enforces physical consistency by adhering to constraints like the Helmholtz equation and refines predictions through CDM.

         \textcolor{green}{$\checkmark$}: Better construction accuracy and multi-scene generalization.

         \textcolor{red}{$\times$}: The setting of hyperparameter needs further analysis.
        \\
        \hline

        \footnotesize \cite{jia2025radioflow11.13}  & \footnotesize
         \textcolor{blue}{$\diamondsuit$}: Propose RadioFlow, a framework based on FM to solve the high latency of GDMs.

        \textcolor{blue}{$\sphericalangle$}: Directly learns the continuous transport trajectories  from noise to data, enabling single-step generation. The ODE formulation is analogous to energy conservation in EM propagation.

        \textcolor{green}{$\checkmark$}: Achieves SOTA performance with 8x fewer parameters and 4x faster inference than the RadioDiff baseline.

        \textcolor{red}{$\times$}: While much faster than diffusion, it is still slower than non-generative models like RadioUNet.
        \\
         \hline
        \footnotesize \cite{sun2025flow11.13}  & \footnotesize
        \textcolor{blue}{$\diamondsuit$}: Propose a novel active learning framework for UAV-based RM construction using limited measurements.

         \textcolor{blue}{$\sphericalangle$}: Uses PnP-refined Flow Matching as a generative prior. Leverages the model to generate an ensemble of maps, calculates location-wise variance to create an "uncertainty map," and uses this map to guide the UAV trajectory.

        \textcolor{green}{$\checkmark$}: Significantly outperforms baselines; efficiently guides UAV to the most informative sampling locations.

         \textcolor{red}{$\times$}: High computational cost, as it requires generating an ensemble of maps at each time step; complex PnP inner-loop refinement.
    \\
    \hline
    \end{tabular}

\end{table}

\subsubsection{GDM for Sampling-Based RM Interpolation} 
This category of methods treats RM construction as an in-painting or interpolation problem, where the GDM's generative prior is conditioned on a small set of known measurements to reconstruct the full map.
Qiu et al. \cite{qiu2023irdm} first introduce a DDPM for this task, focusing on indoor path loss map interpolation. The model is conditioned on a 3-layer input (geometry, positional encoding, and the sparse map) to generate the complete map. They demonstrated high accuracy using only 10\% of reference points.

Luo et al. \cite{luo2024rm} extend this conditional approach with their RM-Gen framework. RM-Gen is a Conditional DPM that is uniquely conditioned on two low-cost inputs: sparse RSS fragments and transmitter locations. This allows for cost-effective generation for both indoor mmWave and outdoor 5G scenarios, achieving over 95\% accuracy.

Addressing the challenge of ultra-low sampling rates, Liu et al. \cite{liu2025wifi} propose WiFi-Diffusion. This framework is a novel three-block system: 1) a Boost block uses an AttUnet to identify and pre-fill ``critical points"; 2) a Generation block uses a DDIM to generate a diverse candidate set of maps based on this enhanced sparse data; and 3) an Election block uses a physical radio propagation model to select the best and most physically plausible map from the candidate set. This ``generate-and-elect" strategy allows it to achieve high-quality maps with as little as one-fifth the samples required by other methods.

Beyond passively interpolating existing sparse data, GDMs are also being used to actively guide the collection of future samples. Sun et al. \cite{sun2025flow11.13} propose a framework for UAV-based active learning to address limited flight endurance. Their method uses a PnP-refined FM algorithm as the generative prior to reconstruct the map from limited measurements. Its key innovation is leveraging the generative model to quantify uncertainty: by generating an ensemble of $M$ plausible maps from the same sparse data, it computes a location-wise variance map. This ``uncertainty map" is then used to guide the UAV's trajectory, directing it to the most informative (highest uncertainty) regions to sample next.

\subsubsection{GDM for Sampling-Free and Physics-Informed Radio Map Generation} 
This more advanced category of methods removes the need for any on-site RSS measurements. Instead, it models radio map construction as a purely conditional generative problem, synthesizing a map based only on environmental data like building layouts and transmitter locations.

\begin{figure*}
\centering
\includegraphics[width=0.8\textwidth]{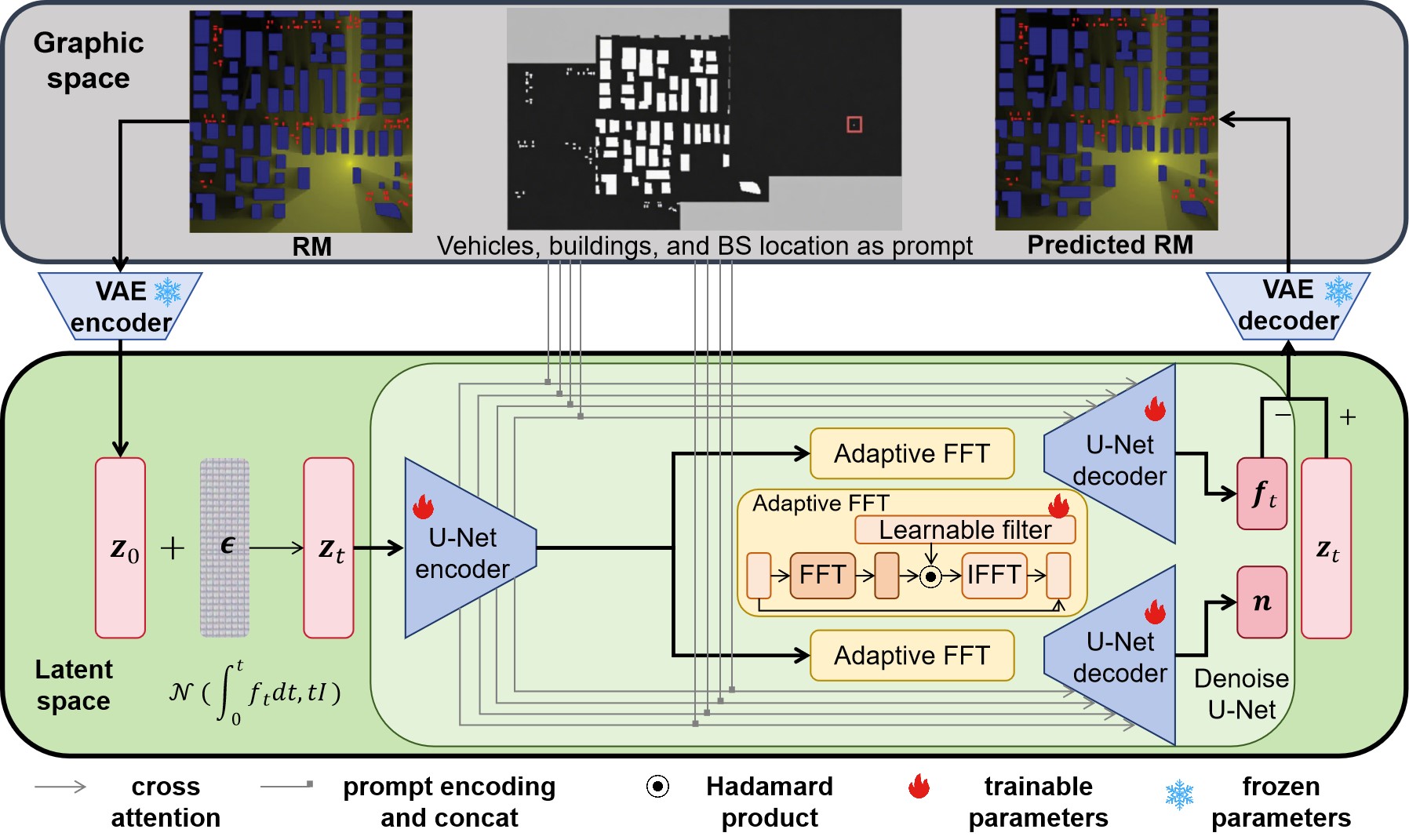}
\caption{
The architecture of the RadioDiff framework\cite{wang2024radiodiff} utilizing a Latent Diffusion Model for efficient dynamic radio map construction. The system begins by employing a pre-trained VAE to compress high-dimensional radio map data into a low-dimensional latent space to significantly reduce the computational resources required for training. Within this latent representation, the model executes a decoupled diffusion process where a U-Net backbone predicts and removes noise conditioned on environmental prompts such as static building layouts and dynamic vehicle positions. A critical innovation embedded in the U-Net is the Adaptive Fast Fourier Transform module which dynamically filters features in the frequency domain to effectively capture high-frequency edge details that are typically missed by standard convolutional layers. By performing the diffusion process in the compressed latent space rather than the pixel space, the LDM achieves a superior balance between training efficiency and generation fidelity to synthesize accurate radio maps for complex dynamic environments.}
\label{picture_RadioDiff}
\end{figure*}

Wang et al. \cite{wang2024radiodiff} are the first to model sampling-free RM construction as a conditional generative problem, as shown is Figure \ref{picture_RadioDiff}. Their proposed method, RadioDiff, is an LDM that takes environmental features (static buildings, dynamic vehicles, and BS location) as a multi-channel prompt. To handle the high-frequency details introduced by dynamic obstacles like vehicles, they designed a novel Attention U-Net backbone that incorporates an Adaptive Fast Fourier Transform (AFT) module. This AFT module operates in the frequency domain to better extract edge features, preventing the blurring seen in standard CNNs.

Jia et al. \cite{jia2025rmdm} enhance the generative approach by introducing physical constraints with GDMs. This framework integrates Physics-Informed Neural Network (PINN) to ensure the generated map adheres to physical laws. RMDM employs a dual U-Net architecture: the first U-Net is trained with a PINN loss function to enforce constraints from the Helmholtz equation; the second U-Net is a standard GDM that refines this physically-consistent output to generate the final, high-fidelity map. This hybrid physics-informed approach significantly enhances accuracy and generalization, especially in sparse data conditions.

While these GDM-based schemes demonstrate high fidelity, their reliance on iterative denoising creates significant inference latency. To solve this latency bottleneck, Jia et al. \cite{jia2025radioflow11.13} propose RadioFlow, a novel framework based on FM. Unlike diffusion's iterative denoising, RadioFlow directly learns the continuous transport trajectories, a vector field, that map noise to the data distribution. This allows for high-fidelity, single-step generation. The authors provide a key theoretical insight: FM's ODE formulation inherently preserves probability mass, which is analogous to energy conservation in EM propagation, whereas SDE  formulation introduces noise that can violate this physical property. This approach achieves SOTA performance with up to 8x fewer parameters and 4x faster inference than the RadioDiff baseline.

\subsection{Summaries and Lessons Learned}
In this section, we reviewed GDM-based schemes for the sensing layer. Our review reveals that the application of GDMs is not merely an incremental improvement but a methodological shift, reframing sensing tasks as solvable Bayesian inverse problems. This approach highlights several key insights and trade-offs, as detailed in Table \ref{tab:gdm_sensing_comparison_optimized}.

\begin{table*}[tbp]
\centering
\caption{Comparison and Insights for GDM's Applications in the Sensing Layer}
\label{tab:gdm_sensing_comparison_optimized}
\renewcommand{\arraystretch}{1.3}
\footnotesize

\begin{tabularx}{\textwidth}{ p{3.5cm} >{\raggedright\arraybackslash}X >{\raggedright\arraybackslash}X }
\toprule
\textnormal{Approach \& Refs} & \textnormal{Core Role \& Best-Suited Scenario} & \textnormal{Pros \& Cons / Trade-offs} \\
\midrule

\multicolumn{3}{l}{\textbf{A. Channel Estimation}} \\
\cmidrule(r){1-3}

Learned Prior for Inverse Problems \cite{ma2024diffusion, fesl2024diffusion, chen2025generative, zhou2025generative, tong2024diffusion} 
& 
\textbf{Core Role:} Acts as a powerful, non-Gaussian prior $P(H)$ to regularize ill-posed inverse problems. 
\par\vspace{4pt} 
\textbf{Scenario:} High-dimensional systems like massive MIMO or systems with complex noise like low-res ADCs and RIS. 
& 
\textbf{Pros:} Enables recovery from limited pilots or complex non-linear noise. 
\par\vspace{4pt} 
\textbf{Cons:} Iterative posterior sampling is computationally latent. Performance can depend on sparse domain assumptions. \\

\noalign{\vskip 4pt}
\cdashline{1-3}
\noalign{\vskip 4pt}

Joint Posterior Distribution Sampling \cite{zilberstein2024joint} 
& 
\textbf{Core Role:} Samples from the joint posterior $P(H, X_D | Y)$ to solve blind inverse problems. 
\par\vspace{4pt} 
\textbf{Scenario:} Blind inverse problems where both channel and data are unknown. 
& 
\textbf{Pros:} Enables simultaneous channel and data recovery. Can reuse data symbols as pilots. 
\par\vspace{4pt} 
\textbf{Cons:} High complexity in handling hybrid continuous and discrete variables. \\
\midrule

\multicolumn{3}{l}{\textbf{B. Channel Generation}} \\
\cmidrule(r){1-3}

High-Fidelity Synthesis \cite{sengupta2023generative, kim2023diffusion} 
& 
\textbf{Core Role:} Learns the full channel distribution $P(H)$ to synthesize diverse, realistic samples. 
\par\vspace{4pt} 
\textbf{Scenario:} Data augmentation for training systems, especially in data-scarce scenarios. 
& 
\textbf{Pros:} Overcomes GAN mode collapse. Accurately captures complex statistical properties like correlation. 
\par\vspace{4pt} 
\textbf{Cons:} Iterative sampling is slow. Generation quality can trade off with sampling speed. \\

\noalign{\vskip 4pt}
\cdashline{1-3}
\noalign{\vskip 4pt}

Conditional and Specialized Generation \cite{lee2024generating, gong2025digital, chi2024rf} 
& 
\textbf{Core Role:} Generates samples from a conditional distribution, such as $P(H|\text{location})$ or $P(\text{RF Signal})$. 
\par\vspace{4pt} 
\textbf{Scenario:} Creating a Digital Twin of Channel. Generating specialized RF time-series datasets. 
& 
\textbf{Pros:} Enables user-specific channel generation. Bypasses pilot estimation. Can model complex time-series signals. 
\par\vspace{4pt} 
\textbf{Cons:} High computational complexity. May require operation in specific domains like beamspace. \\
\midrule

\multicolumn{3}{l}{\textbf{C. Radio Map Construction}} \\
\cmidrule(r){1-3}

Sampling-Based Interpolation \cite{qiu2023irdm, luo2024rm, liu2025wifi, sun2025flow11.13} 
& 
\textbf{Core Role:} Acts as a generative prior to in-paint a full map from sparse measurements. 
\par\vspace{4pt} 
\textbf{Scenario:} Cost-effective map construction where full-site surveys are impractical. UAV-based active sensing. 
& 
\textbf{Pros:} Achieves high accuracy with very few samples. Can actively guide UAVs to informative locations. 
\par\vspace{4pt} 
\textbf{Cons:} Still requires some on-site measurements. Active learning has high computational cost. \\

\noalign{\vskip 4pt}
\cdashline{1-3}
\noalign{\vskip 4pt}

Sampling-Free and Physics-Informed Generation \cite{wang2024radiodiff, jia2025rmdm, jia2025radioflow11.13} 
& 
\textbf{Core Role:} Conditional generator that synthesizes a map based only on environmental features and physical laws. 
\par\vspace{4pt} 
\textbf{Scenario:} Pre-deployment planning. Dynamic environments where maps must be generated, not measured. 
& 
\textbf{Pros:} Requires no on-site RSS measurements. Physics-informed models enhance accuracy. Flow Matching enables single-step generation. 
\par\vspace{4pt} 
\textbf{Cons:} Iterative diffusion models have high latency. FM is faster but still slower than non-generative models. \\

\bottomrule
\end{tabularx}
\end{table*}

\begin{itemize}

\item \textbf{Insight 1: Roles in Inference and Synthesis.}  For channel estimation, GDMs act as a regularizer for ill-posed inverse problems \cite{ma2024diffusion, fesl2024diffusion}. They enable the recovery of high-dimensional channels from corrupted or limited data, such as in low-resolution ADC systems \cite{zhou2025generative} or RIS-aided systems \cite{tong2024diffusion}. For channel generation and radio map construction, GDMs act as a synthesizer. This role is crucial for data augmentation \cite{sengupta2023generative} and, more importantly, for advanced conditional generation, such as generating $P(H|\text{location})$ \cite{lee2024generating}, which forms the basis for a digital twin of channel \cite{gong2025digital}.

\item \textbf{Insight 2: The Path to Practicality through Physics, Activity, and Speed.} While high fidelity is established, the most advanced schemes are evolving toward practical deployment. This is seen in three key trends: (1) \textbf{PINNs} that integrate physical laws, such as the Helmholtz equation, to ensure generated maps are physically consistent \cite{jia2025rmdm}; (2) \textbf{Active Learning} frameworks that use GDMs to quantify uncertainty and guide UAVs to the most informative sampling locations \cite{sun2025flow11.13}; and (3) \textbf{Model Acceleration}, adopting new non-iterative paradigms like FM, to achieve single-step generation and tackle the latency bottleneck \cite{jia2025radioflow11.13}.
\end{itemize}

Despite these advances, significant challenges remain. The primary bottleneck is the inference latency of the iterative GDM process. In the sensing layer, this is a fundamental conflict: GDM's generation time often exceeds the channel's coherence time, making real-time estimation or generation difficult. Furthermore, most GDM backbones, such as U-Nets, are adapted from computer vision and lack an inherent understanding of the physical structure of wireless data. They treat a channel matrix as a generic image, failing to natively capture properties like antenna correlation or mmWave sparsity \cite{lee2024generating}.

Future work should focus on developing ultra-fast samplers, like CM or FM, tailored for sensing tasks. Applying these generative priors to solve new, complex inverse problems, such as compensating for non-linear pilot measurements caused by amplifier distortion, is also a promising direction.

\section{GDM for the Transmission Layer}
\label{section4}

This section systematically analyzes the role of GDMs at the transmission layer. It is crucial to first clarify the taxonomy used in this survey: unlike the sensing layer, which focuses on modeling the wireless environment itself, such as estimating channel parameters $\mathbf{H}$ or generating radio maps, the Transmission Layer is defined by its objective of optimizing the end-to-end delivery and processing of signals and data passing through that environment. Consequently, traditional physical layer functions such as modulation and channel coding, which manipulate the transmitted signal $\mathbf{x}$ rather than detecting the environment, fall within this layer.

Based on our comprehensive literature review, the application of GDMs in this layer exhibits a distinct distribution. The vast majority of current research is concentrated on SemCom \cite{meng2025semantic}. This is a natural development, as GDM's inherent capabilities for iterative denoising and probabilistic generation align perfectly with SemCom's core objective of recovering meaning from distortions rather than exact bitstreams. However, a few pioneering works have recently begun to apply GDMs to traditional bit-level transmission modules. Although these constitute a minority compared to SemCom, they represent a critical innovation in reshaping classical physical layer blocks. Therefore, this section first introduces these emerging physical layer applications, followed by a detailed review of the mainstream SemCom paradigms, including semantic denoiser, auxiliary recovery, semantic-based generation, multimodal transmission, and resource allocation.

\subsection{GDM for Modulation and Coding}
While most generative approaches focus on semantic-level processing, recent studies have successfully integrated GDMs into traditional physical layer modules, specifically for probabilistic constellation shaping and channel error correction, as illustrated in Table \ref{modulation and coding schemes}. These approaches leverage the distribution matching capability of GDMs to optimize signal geometry and decoding trajectories.

\subsubsection{GDM for Probabilistic Constellation Shaping}
Traditional geometric shaping often relies on fixed constellations, while probabilistic shaping requires complex optimization over discrete distributions. Letafati et al.\cite{Letafati202311.25}  propose a novel probabilistic constellation shaping scheme using DDPMs. Instead of solving a discrete optimization problem, they utilize the ``denoise-and-generate" characteristic of GDMs. The transmitter employs a GDM to generate constellation symbols from noise, effectively ``mimicking" the reconstruction process of the receiver. This approach aligns the generated symbol distribution with the receiver's inference capabilities, creating a mutual understanding that enhances resilience. Simulation results demonstrate a threefold improvement in mutual information compared to DNN-based benchmarks, particularly providing robust performance under low-SNR regimes and non-Gaussian noise environments.

\subsubsection{GDM for Channel Error Correction}
In the domain of channel coding, decoding linear codes is classically an NP-hard likelihood problem. Choukroun and Wolf\cite{choukroun2022denoising11.25} introduce Denoising Diffusion Error Correction Codes (DDECCT), framing the decoding process as a reverse diffusion task. They model forward channel corruption such as AWGN as a diffusion process. For decoding, they propose a tailored reverse process conditioned on the received signal’s syndrome, namely parity check errors. A key innovation is the use of a syndrome-based line search to dynamically optimize the reverse diffusion step size. This method enables iterative soft decoding of linear codes at arbitrary block lengths, significantly outperforming state-of-the-art neural decoders and achieving high accuracy even with a single reverse diffusion step.

\begin{table}[tbp]
    \centering
    \caption{Existing GDM-based modulation and coding schemes, where \textcolor{blue}{$\diamondsuit$}, \textcolor{blue}{$\sphericalangle$}, \textcolor{green}{$\checkmark$}, and \textcolor{red}{$\times$} respectively are contributions, the role of GDM, pros, and cons.}
    \label{modulation and coding schemes}
    \renewcommand{\arraystretch}{1.2}
    \normalsize
    \begin{tabular}{|>{\arraybackslash}m{0.025\textwidth}|>{\arraybackslash}m{0.42\textwidth}|}
        \hline
        \textbf{\small{Ref.}} & \textbf{\small{Descriptions}} \\
        \hline

        \footnotesize \cite{Letafati202311.25} & \footnotesize
         \textcolor{blue}{$\diamondsuit$}: Propose a DDPM-based probabilistic constellation shaping scheme to enhance information rate and communication resilience.

         \textcolor{blue}{$\sphericalangle$}: The transmitter employs a DDPM to generate constellation symbols from noise which mimics the receiver reconstruction process to align symbol distributions.

         \textcolor{green}{$\checkmark$}: Achieves a threefold improvement in mutual information compared to DNN-based approaches for 64-QAM and provides robust performance under non-Gaussian noise.

         \textcolor{red}{$\times$}: The iterative symbol generation process at the transmitter introduces higher computational latency compared to static constellation mapping. \\
        \hline

        \footnotesize \cite{choukroun2022denoising11.25} & \footnotesize
         \textcolor{blue}{$\diamondsuit$}: Introduce Denoising Diffusion Error Correction Codes to perform soft decoding of linear codes at arbitrary block lengths.

         \textcolor{blue}{$\sphericalangle$}: The decoder models channel corruption as a diffusion process and reverses it conditioned on the parity check syndrome with an optimal step size line search.

         \textcolor{green}{$\checkmark$}: Outperforms state-of-the-art neural decoders by sizable margins and achieves high accuracy even with a single reverse diffusion step.

         \textcolor{red}{$\times$}: The grid search procedure for determining the optimal reverse step size requires additional computational overhead during the inference phase. \\
        \hline

    \end{tabular}
\end{table}

\subsection{GDM for Semantic Denoiser}

Semantic denoising refers to the process of restoring the original meaning of transmitted content by removing distortions introduced by wireless channels, directly within the semantic feature space. This is different from traditional signal-level denoising, which operates at the waveform or pixel level\cite{gu2019brief7.8}. In SemCom, the core priority lies in ensuring that the recovered message conveys the same intention as the original, even under unpredictable channel conditions.
To achieve this, researchers have explored GDMs that are trained without any external conditioning, as illustrated in Table~\ref{denoisergdm}. GDMs do not require CSI, SNR feedback, or auxiliary labels during training or inference. Instead, they learn to reverse the effects of random channel noise based only on the distribution of clean semantic features\cite{tang2025enabling7.8}.
Specifically, in the training phase, the GDM is optimized solely to remove random distortions in semantic representations, without any additional conditions such as class labels or channel feedback. After training, the GDM-based denoiser can be applied directly during decoding, without retraining or runtime adaptation.

\begin{table}[tbp]
    \centering
    \caption{Existing GDM-based purely trained semantic denoiser schemes, where \textcolor{blue}{$\diamondsuit$}, \textcolor{blue}{$\sphericalangle$}, \textcolor{green}{$\checkmark$}, and \textcolor{red}{$\times$} respectively are contributions, the role of GDM, pros, and cons.}
    \label{denoisergdm}
    \renewcommand{\arraystretch}{1.2}
    \normalsize
    \begin{tabular}{|>{\arraybackslash}m{0.025\textwidth}|>{\arraybackslash}m{0.42\textwidth}|}
        \hline
        \textbf{\small{Ref.}} & \textbf{\small{Descriptions}} \\
        \hline

        \footnotesize \cite{xu2025semantic} & \footnotesize
         \textcolor{blue}{$\diamondsuit$}: Propose SP-Latent-Diff EDNSC, casting adaptive semantic equalising and de-noising as an inverse problem.

         \textcolor{blue}{$\sphericalangle$}: A pre-trained LDM supplies the semantic prior, and it iteratively denoises received latents without explicit SNR knowledge.

         \textcolor{green}{$\checkmark$}: Delivers up to +23.4\% PSNR at SNR = –2 dB.

         \textcolor{red}{$\times$}: Reverse SDE sampling increases on-device computation time and may need finely tuned diffusion steps. \\
        \hline

        \footnotesize \cite{pei2025latent} & \footnotesize
         \textcolor{blue}{$\diamondsuit$}: Develop an LDM-based SemCom system with an end-to-end consistency-distilled denoiser for single-step inference.

         \textcolor{blue}{$\sphericalangle$}: EECD compresses a multi steps LDM into a deterministic mapping that directly outputs clean latents.

         \textcolor{green}{$\checkmark$}: Enables real-time denoising while remaining robust to out-of-distribution sources and semantic adversarial errors.

         \textcolor{red}{$\times$}: Requires auxiliary lightweight adapters and occasional one-shot updates, adding small but non-negligible signalling overhead. \\
        \hline

        \footnotesize \cite{xu2023latent} & \footnotesize
         \textcolor{blue}{$\diamondsuit$}: Present Latent-Diff\,DNSC, which is trained on mixed-SNR semantic vectors.

         \textcolor{blue}{$\sphericalangle$}: During inference, the LDM iteratively refines noisy latents, eliminating the need for channel estimation.

         \textcolor{green}{$\checkmark$}: Gains up to 67\% PSNR and 68\% SSIM over ADJSCC across 0 - 20 dB SNR on LAION2B-EN images.

         \textcolor{red}{$\times$}: Training requires a long forward-diffusion schedule. \\
        \hline

        \footnotesize \cite{wu2024cddm} & \footnotesize
         \textcolor{blue}{$\diamondsuit$}: Introduce CDDM, a DDPM placed after MMSE equalisation to learn the conditional distribution of channel inputs.

         \textcolor{blue}{$\sphericalangle$}: Starts reverse sampling from the equalised signal, drastically shortening the diffusion trajectory.

         \textcolor{green}{$\checkmark$}: Improves JSCC‐PSNR and MSSSIM under both AWGN and Rayleigh fading channels.

         \textcolor{red}{$\times$}: Performance degrades when channel estimates are highly inaccurate; still needs several sampling steps for best quality. \\
        \hline

    \end{tabular}
\end{table}

\subsubsection{Sampling from Posterior Distributions in Semantic Space}

These methods treat semantic denoising as a distinct step, often operating in the latent space for efficiency. For instance, Xu et al. \cite{xu2025semantic} introduce the Latent-Diff DNSC scheme. This approach integrates the GDM as a dedicated de-noiser module positioned after the wireless channel but before the semantic decoder. To adapt the GDM for this task, the authors do not fundamentally modify the GDM model itself, but rather adapt its inference procedure. A standard DDPM is trained offline in the latent space to reverse a full T-step noise schedule. During online inference, a separate SNR estimator is required to measure the channel noise. This SNR value is then used to scale the received noisy semantic vector to an appropriate starting noise level $z_T$. The standard, multi-step DDPM reverse sampling process is then executed from this $z_T$ to recover the clean latent vector. This two-stage integration allows the system to mitigate the mismatch between training and deployment SNRs, but its practical drawback is this added reliance on an accurate SNR estimator.

\begin{figure*}
\centering
\includegraphics[width=0.9\textwidth]{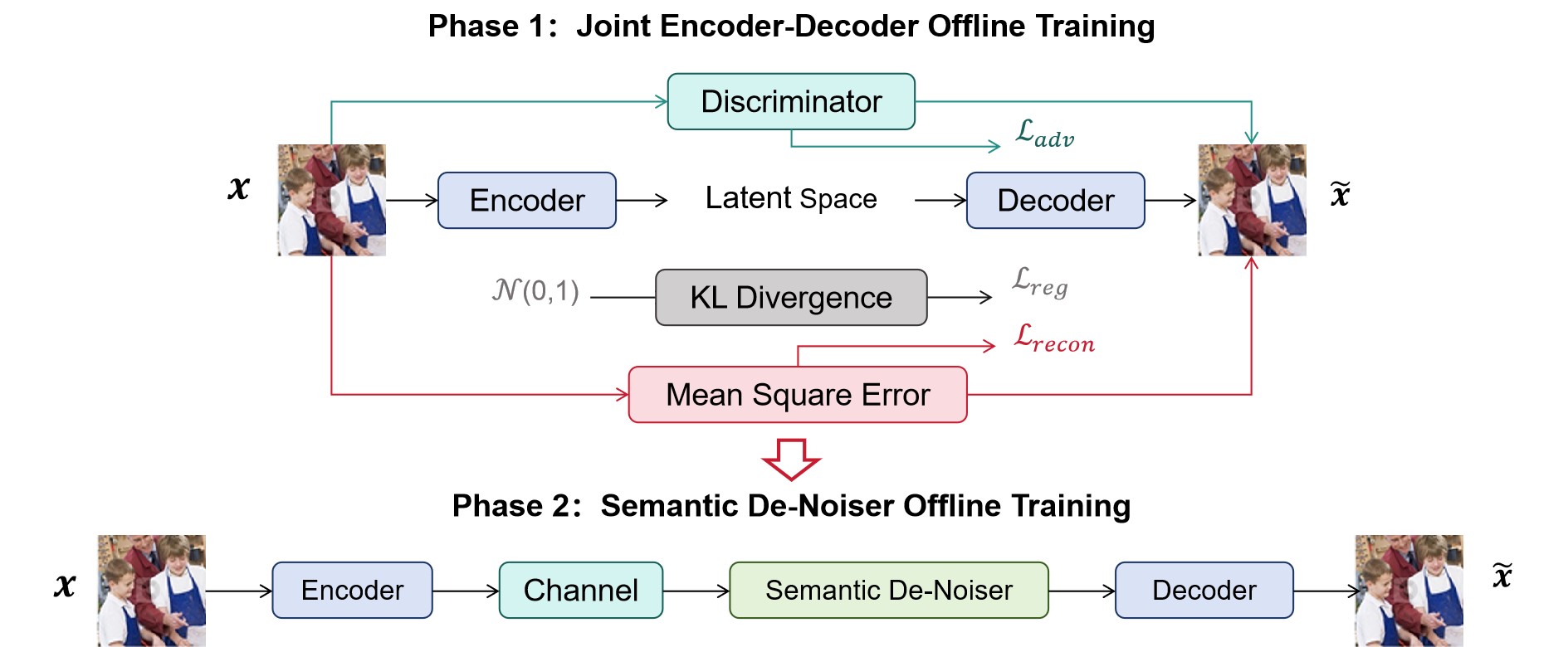}
\caption{Illustration of Latent-Diff DNSC scheme \cite{xu2023latent}. This architecture separates the semantic feature learning from the channel noise modeling to achieve robust SemCom. In Phase 1 shown at the top, a joint encoder-decoder is trained alongside a discriminator to compress images into a latent space while optimizing a composite objective function that includes reconstruction loss for fidelity, adversarial loss for local consistency, and regularization loss to constrain the latent space distribution. In Phase 2 shown at the bottom, a semantic de-noiser operates within this fixed latent space by first gradually adding Gaussian noise to the semantic vectors through a forward diffusion process to simulate physical channel impairments and then employing a U-Net to predict and eliminate these noises via a reverse diffusion process. This design allows the system to effectively model diverse channel noise characteristics during training and enables the adaptive elimination of interference during online transmission without requiring explicit CSI.}
\label{picture_L-Diff DNSC}
\end{figure*}

Building on this, Xu et al. \cite{xu2023latent} later propose the SP-Latent-Diff EDNSC system, as shown in Figure \ref{picture_L-Diff DNSC}, which integrates the GDM in a more theoretically principled manner. This work reframes the entire equalization and denoising process as a single Bayesian inverse problem. Here, the GDM is not an add-on module but serves as a powerful semantic prior, implemented as a pre-trained score-based SDE model. The innovation lies in the sampling process. Instead of a standard reverse chain, this work uses a modified posterior sampling algorithm. At each reverse step, the sampling direction is guided by two forces: first, the pre-trained prior score from the GDM, and second, a likelihood score which is the gradient of the data-consistency term. This inverse problem formulation provides superior channel adaptivity without requiring an SNR estimator. Its main tradeoff is the high computational latency, as this guided posterior sampling is iterative and complex.

\subsubsection{Models Aligned with Channel Characteristics}
Another approach is to design GDM to align with known channel characteristics. Wu et al.\cite{wu2024cddm} propose the Channel Denoising Diffusion Model (CDDM), which integrates the GDM as a dedicated denoising module positioned after the channel equalizer but before the semantic decoder. The core innovation of this work is a fundamental modification of the GDM's forward process. Instead of adopting a standard Gaussian noise schedule, the CDDM's forward diffusion process is mathematically derived to precisely match the statistical distribution of the residual noise found after MMSE equalization on a Rayleigh channel. The U-Net model is then trained to reverse this specific, channel-aware noise. This method is powerful as it embeds physical layer knowledge directly into the diffusion process, and it accelerates inference by starting the reverse sampling from the received signal at an intermediate step $m$ rather than from pure noise. The primary tradeoff is that this still requires multiple iterative sampling steps, which incurs latency, and its performance relies on the availability of accurate CSI at the receiver to define the noise model.

To overcome the critical latency barrier of such multi-step sampling, Pei et al.  \cite{pei2025latent} introduce an LDM-based system that employs a more recent GDM acceleration technique: End-to-End Consistency Distillation (EECD). This work also places the GDM in the latent space after the equalizer, but its goal is to achieve real-time denoising. It begins with a full, multi-step LDM (the teacher model) trained to denoise the semantic features. It then applies EECD to distill this teacher model into a lightweight ``student" consistency model. This student model is trained to execute the entire denoising task in a single, deterministic forward pass. The key modification lies in their EECD loss function, which is optimized end-to-end using perceptual metrics like LPIPS, ensuring the one-shot output is semantically accurate. The clear advantage is the elimination of iterative sampling, enabling genuine low-latency communication. This benefit comes at the cost of a highly complex offline distillation process and a slight, controlled trade-off in reconstruction quality compared to the full multi-step teacher model.

\subsection{GDM for Auxiliary Recovery}

In SemCom, it is not always optimal to rely solely on unconditional generation at the receiver side. A practical and increasingly popular strategy is to transmit a structured but coarse intermediate representation, such as a segmentation map, scene graph, low-resolution image, or raw received signal\cite{li2023diffusion7.8}, and then employ a GDM conditioned on this auxiliary input to produce refined, semantically aligned outputs.
This approach is referred to as conditional semantic restoration with auxiliary inputs. Rather than recovering the content from scratch, GDMs receive partial information that serves as a guiding signal\cite{yang2024semantic}. This guidance effectively anchors the sampling process in a semantically valid region, improving reconstruction quality while allowing for extreme compression and robust operation under noisy or unreliable channels.
Table \ref{Auxiliary Inputs} illustrates recent contributions in this area, and the detailed descriptions are as follows.

\begin{table}[tbp]
    \centering
    \caption{Existing GDM-based conditional semantic restoration schemes, where \textcolor{blue}{$\diamondsuit$}, \textcolor{blue}{$\sphericalangle$}, \textcolor{green}{$\checkmark$}, and \textcolor{red}{$\times$} respectively are contributions, the role of GDM, pros, and cons.}
    \label{Auxiliary Inputs}
    \renewcommand{\arraystretch}{1.2}
    \normalsize
    \begin{tabular}{|>{\arraybackslash}m{0.025\textwidth}|>{\arraybackslash}m{0.42\textwidth}|}
        \hline
        \textbf{\small{Ref.}} & \textbf{\small{Descriptions}} \\
        \hline
    \footnotesize \cite{pezone2024semantic} & \footnotesize
     \textcolor{blue}{$\diamondsuit$}: Propose SPIC, a framework that transmits a semantic map and a low-resolution image for reconstruction.
  
     \textcolor{blue}{$\sphericalangle$}: A conditional GDM is heavily modified to accept dual conditioning. It is conditioned on the low-res image via input concatenation and on the semantic map via SPADE injection into the decoder.

     \textcolor{green}{$\checkmark$}: Deep architectural modification ensures high semantic fidelity and coding efficiency, especially for small objects.

     \textcolor{red}{$\times$}: The primary cost is the increased bandwidth overhead required to transmit both the map and the coarse image. \\
    \hline

    \footnotesize \cite{yilmaz2024high} & \footnotesize
     \textcolor{blue}{$\diamondsuit$}: Propose transmitting only a low-resolution version of the image, using a DDPM at the receiver to fill in missing details.
  
     \textcolor{blue}{$\sphericalangle$}: A pre-trained DDPM acts as a restorer. Its sampling is guided by the DDNM algorithm, which replaces the low-resolution range space at each step, forcing the GDM to only generate high-frequency null space details.

     \textcolor{green}{$\checkmark$}: Improves PSNR and perceptual metrics over DeepJSCC by cleverly guiding a generic GDM.

     \textcolor{red}{$\times$}: The main tradeoff is the high latency of the iterative, multi-step DDNM sampling loop. \\
    \hline

    \footnotesize \cite{chen2024commin} & \footnotesize
     \textcolor{blue}{$\diamondsuit$}: Propose CommIN, a framework treating DeepJSCC as an inverse problem using an INN to simulate degradation and a DDPM to restore details.
  
     \textcolor{blue}{$\sphericalangle$}: An Invertible Neural Network simulates the degradation. A standard pre-trained DDPM is then used only to generate the missing detail component separated by the INN.

     \textcolor{green}{$\checkmark$}: Provides a structured separation of known and unknown information; significantly reduces LPIPS at low bandwidth.

     \textcolor{red}{$\times$}: Extreme system complexity, as it requires training and deploying three separate large models. \\
    \hline
    
    \footnotesize \cite{wang2025diffcom} & \footnotesize
      \textcolor{blue}{$\diamondsuit$}: Propose DiffCom, which uses the raw channel-received signal as a fine-grained condition to guide posterior sampling in a pre-trained GDM.
  
     \textcolor{blue}{$\sphericalangle$}: An SDE-based GDM acts as a prior. Its posterior sampling process is modified with a guidance step derived from a consistency loss that matches the raw received signal.

     \textcolor{green}{$\checkmark$}: Achieves excellent robustness to channel variations, low CSNR, and fading by guiding directly from the received signal.

     \textcolor{red}{$\times$}: High computational latency at the receiver, as each sampling step requires backpropagation through the encoder model.
     \\
    \hline

\end{tabular}
\end{table}

\subsubsection{Image-Level Guidance Coarse Visual Inputs}

These methods transmit a coarse visual representation which the receiver's GDM must refine. Yilmaz et al. \cite{yilmaz2024high} propose an innovative integration where the DeepJSCC encoder is modified to transmit only a low resolution version of the image. The receiver uses a standard, pre-trained DDPM as a powerful restorer. The core innovation here is not a modification of the GDM architecture itself, but rather a sophisticated adaptation of its sampling procedure. The system employs the Denoising Diffusion Null space Model (DDNM) algorithm. This algorithm mathematically decomposes the image into a range space, representing the low resolution content, and a null space, representing the high resolution details. At each step of the reverse diffusion process, the sampler predicts a clean image, but then replaces its range space component with the actual low resolution image received over the channel. This forces the DDPM to only generate new information in the null space, effectively ``filling in" the missing high frequencies while remaining perfectly consistent with the transmitted data. This approach cleverly guides a generic GDM using the received signal, though its main tradeoff is the high latency of the iterative, multi step sampling loop.

\begin{figure*}
\centering
\includegraphics[width=1\textwidth]{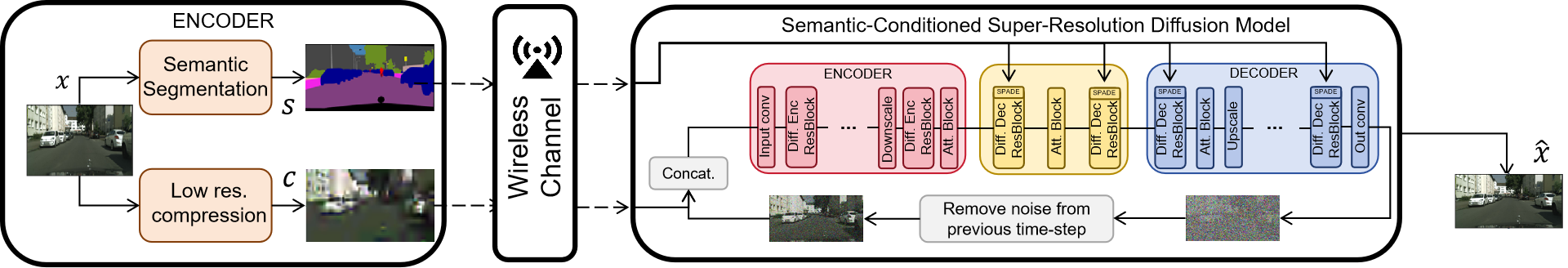}
\caption{Overview of the SPIC framework \cite{pezone2024semantic}. The system employs a modular encoder that transmits a lossless semantic segmentation map along with a compressed low-resolution image to the receiver. The decoding process utilizes a super-resolution CDM which concatenates the upscaled coarse image with the noisy input at each timestep and injects the semantic map features into the U-Net bottleneck and decoder layers using Spatially-Adaptive Normalization modules. This architecture effectively leverages dual conditioning to reconstruct high-resolution images that maintain precise semantic details and object fidelity even under limited bandwidth constraints.}
\label{picture_SPIC}
\end{figure*}

Pezone et al. \cite{pezone2024semantic} adopt a different strategy by transmitting two guiding inputs: a low resolution coarse image and a highly compressed semantic segmentation map, as shown in Figure \ref{picture_SPIC}. This task demands a heavily modified Conditional GDM architecture at the receiver. The GDM's U-Net structure is adapted to accept this dual conditioning. First, the low resolution image is upscaled and concatenated with the input noise tensor at every step, providing a strong structural foundation for the entire generation process. Second, the semantic map is injected deep inside the U-Net's decoder and bottleneck blocks using spatially adaptive normalization. This technique, adapted from GANs, allows the semantic map to control the style and structure of the generated features at a fine grained level. This deep architectural modification ensures high semantic accuracy, particularly for small objects. The primary cost of this superior semantic preservation is the increased bandwidth overhead required to transmit both the map and the coarse image.

\subsubsection{Feature-Level Guidance Signals or Latent Estimates}

A different approach treats the entire DeepJSCC pipeline, including the encoder, channel, and deterministic decoder, as a single complex degradation that must be inverted. Chen et al. \cite{chen2024commin} propose CommIN, which tackles this inversion challenge using a novel combination of two generative models. The system's core modification is the introduction of a separate Invertible Neural Network (INN). This INN is trained offline to simulate the degradation process by learning to split a clean image $x$ into a coarse component $c$, which mimics the blurry DeepJSCC output, and a detail component $d$, which represents the lost information. At the receiver, a standard pre-trained DDPM is then used to perform its generative task, but instead of generating a full image, it is used only to generate the missing detail component $d$. The final high-quality image is reconstructed by the INN's inverse function, which combines the received coarse image $y$ with the GDM generated details $d$. The main advantage is this structured separation of known and unknown information. The significant tradeoff is the extreme system complexity, as it requires training and deploying three separate large models: the DeepJSCC, the INN, and the DDPM.

Wang et al. \cite{wang2025diffcom} introduce DiffCom, which offers a more direct method for inverting the degradation. This approach utilizes a standard pre-trained SDE-based GDM as a generative prior at the receiver, completely replacing the deterministic decoder. The innovation here is a fundamental modification of the GDM's sampling algorithm. At each step of the reverse sampling process, the algorithm performs two operations: first, an unconditional prior step based on the GDM's learned score, and second, a guidance step. This guidance step is a new correction term derived from the gradient of a consistency loss. This loss function, $\|y - \mathcal{W}_{h}(\mathcal{E}(\hat{x}_{0|t}))\|^2$, actively "pulls" the sampling trajectory towards an image $\hat{x}_{0|t}$ that, if passed back through the known encoder $\mathcal{E}$ and channel model $\mathcal{W}_h$, would match the raw received signal $y$. This direct guidance from the received signal provides excellent robustness to channel variations. Its primary cost is the high computational latency at the receiver, as each sampling step now requires a full backpropagation through the encoder model.

\subsection{GDM for Semantic-based Generation}

\begin{table}[tbp]
    \centering
    \caption{Existing GDM-based semantic generation schemes, where \textcolor{blue}{$\diamondsuit$}, \textcolor{blue}{$\sphericalangle$}, \textcolor{green}{$\checkmark$}, and \textcolor{red}{$\times$} respectively are contributions, the role of GDM, pros, and cons.}
    \label{Semantic Reconstruction}
    \renewcommand{\arraystretch}{1.2}
    \normalsize
    \begin{tabular}{|>{\arraybackslash}m{0.025\textwidth}|>{\arraybackslash}m{0.42\textwidth}|}
        \hline
        \textbf{\small{Ref.}} & \textbf{\small{Descriptions}} \\
        \hline
    \footnotesize \cite{li2025goal} & \footnotesize
     \textcolor{blue}{$\diamondsuit$}: Propose a goal oriented SemCom scheme for video transmission. 

     \textcolor{blue}{$\sphericalangle$}: LDM handles denoising and interpolation, improving PSNR and MSE.

     \textcolor{green}{$\checkmark$}: Outperforms JSCC in PSNR and FVD.

     \textcolor{red}{$\times$}: Sensitive to unknown channels even with PSD-GSC. \\
    \hline

    \footnotesize \cite{yan2025semantic} & \footnotesize
     \textcolor{blue}{$\diamondsuit$}: Embed LDM in an FFmpeg-compatible video streaming framework, compressing I-frames into latent vectors and using B/P frames as metadata for efficient video transmission.

     \textcolor{blue}{$\sphericalangle$}: LDM performs denoising and interpolation under varying bandwidth.

     \textcolor{green}{$\checkmark$}: Enhances Quality of Experience(QoE) via adaptive bitrate control.

     \textcolor{red}{$\times$}: Underperforms with strong interference; needs real-time tuning. \\
    \hline

    \footnotesize \cite{duan2024dm} & \footnotesize
     \textcolor{blue}{$\diamondsuit$}: Introduce the DM-MIMO module to mitigate MIMO fading using GDMs, which enhance signal quality through signal distribution learning and joint sampling.

     \textcolor{blue}{$\sphericalangle$}: Integrates with SVD precoding to lower MSE and improve image quality.

     \textcolor{green}{$\checkmark$}: Performs well in low-SNR and complex noise settings.

     \textcolor{red}{$\times$}: Training cost rises with high channel variability.\\
    \hline

    \footnotesize \cite{du2023ai} & \footnotesize
     \textcolor{blue}{$\diamondsuit$}: Propose a full duplex SemCom framework in MR, facilitating the sharing of compact semantic representations for efficient rendering of MR environments.

     \textcolor{blue}{$\sphericalangle$}: CDM ensures accurate spatial visual recovery.

     \textcolor{green}{$\checkmark$}: Saves bandwidth while supporting spatially aligned MR rendering.

     \textcolor{red}{$\times$}: Difficult to scale across large MR user networks. \\
    \hline
    \end{tabular}
\end{table}

Unlike denoising or conditionally guided recovery, which aim to restore distorted transmitted signals, semantic-based generation refers to the use of GDMs to synthesize complete content, such as images or videos based on abstract semantic goals, task specific cues, or minimal symbolic input\cite{grassucci2023generative}. The reconstruction process does not attempt to recover what was exactly transmitted, but rather generates plausible and contextually aligned content that satisfies the intended meaning or function\cite{ren2024generative7.8}.
This approach shifts the objective from low-level fidelity to high-level semantic alignment. It is especially effective in scenarios where the transmitter intentionally omits most of the raw data and sends only symbolic representations, semantic descriptors, or user-intent cues. The GDMs at the receiver then constructs perceptually realistic outputs using its learned prior knowledge.
We organize recent works into three technical directions, based on the type of content and semantic abstraction involved in reconstruction.
Table \ref{Semantic Reconstruction} illustrates existing semantic-based generation schemes, and the detailed descriptions are as follows.

\subsubsection{Video Generation from Semantic Goals or Sparse Motion Descriptions}

These methods do not transmit full frames but instead send abstract semantic information, relying on the receiver's GDM to synthesize the video content. Li et al.\cite{li2025goal}  introduce a goal oriented framework where the transmitter's primary task is to identify and transmit only the most critical keyframes. The system is heavily reliant on a receiver equipped with a powerful, pre-trained Stable Diffusion model, a specific type of LDM. The innovation lies in the receiver's multi-stage GDM pipeline. First, a CDM is used to denoise the received keyframe features, conditioned on the instantaneous channel gain. For unknown channels, a parallel SDE based GDM is proposed to simultaneously estimate the channel gain and denoise the features. Once the keyframes are reconstructed, a separate non GDM module, a motion appearance interpolator, is used to generate the intermediate frames. The benefit is a massive reduction in transmitted data by sending only sparse keyframes. The tradeoff is the high receiver complexity and the system's reliance on the interpolation module's ability to handle complex motion.

Yan et al. \cite{yan2025semantic} propose a different approach by integrating the LDM directly into the established FFmpeg video streaming standard. This system also transmits only key information, but it defines this information in line with traditional video codecs: I-frames (key frames) and motion vectors (for B/P-frames). The primary GDM modification occurs at the receiver, which uses a Conditional LDM to generate non keyframes (B/P-frames). This GDM's sampling process is conditioned on two inputs: the reconstructed latent features of the I-frame, providing content, and the received motion vector metadata, providing temporal guidance. This architecture cleverly replaces the pixel level motion compensation of traditional codecs with a generative latent space motion adjustment. This integration with FFmpeg allows for adaptive bitrate control, while the GDM's generative nature provides robustness against channel noise that would typically corrupt a standard video stream.

\subsubsection{Semantic-Level Image Generation over Wireless Channels}
This approach focuses on generating semantically plausible images over wireless channels. Duan et al. \cite{duan2024dm} extend this concept to complex MIMO channels by introducing DM-MIMO, a specialized GDM module integrated at the receiver. The system is designed to work after standard SVD precoding and equalization. The authors identify a core problem: SVD equalization decomposes the MIMO channel into multiple parallel sub-channels, but each sub-channel possesses a different singular value and thus a different effective noise power. A standard GDM, which assumes a uniform noise level at each step, cannot be directly applied to this non-uniform noise vector.
To solve this, the authors propose a fundamental modification to the GDM's reverse sampling process. The key innovation is to map each sub-channel's unique effective noise power to a corresponding effective sampling step $m_i$. This transforms the problem from denoising a vector with varied noise powers to denoising a vector with varied time steps. A novel joint sampling algorithm is then executed. At any global sampling step $t$, this algorithm dynamically processes each sub-channel: if a sub-channel's noise is already lower than the current step, noise is added back to it. If its noise is higher, the standard GDM denoising U-Net is applied. This modification allows a single GDM to learn the joint distribution of the encoded signal across all sub-channels simultaneously. This method significantly enhances signal quality in MIMO systems, though its main tradeoff is the increased complexity of this adaptive joint sampling logic at the receiver.

\subsubsection{Multiuser Semantic Sharing with Reconstruction from Task Intent}
Du et al. \cite{du2023ai} explore multi-user semantic sharing from a novel perspective, focusing on incentivizing the sharing of computationally heavy results rather than just the transmission method. The paper proposes a framework for Mixed Reality environments where users can share semantic information, such as free space maps or AIGC results, via full duplex D2D links to save redundant computation. GDM is introduced in a highly specialized role: not as a semantic encoder or decoder for the visual data, but as an offline optimization tool to solve the complex incentive mechanism design.
The authors frame the task of finding the optimal payment contract as a high-dimensional generation problem. To solve this, CDM to function as a contract generation policy. This GDM is modified to accept a complex vector $e$ as its input condition. This condition vector $e$ represents the entire state of the wireless environment, including channel conditions, interference, and user costs. The CDM is then trained using a quality network, similar to a Q function in reinforcement learning, to reverse a noise vector into an optimal contract $c$. This generated contract maximizes the utility of the information receiver. The primary advantage is that the GDM proves superior to standard DRL algorithms in generating high-quality, stable solutions for this complex economic optimization. The tradeoff is that the GDM is not part of the real-time communication link itself, but rather a sophisticated meta-level tool for designing the system's economic policy.

\subsection{GDM for Multimodal Transmission}

\begin{table}[tbp]
    \centering
    \caption{Existing GDM-based multimodal SemCom schemes, where \textcolor{blue}{$\diamondsuit$}, \textcolor{blue}{$\sphericalangle$}, \textcolor{green}{$\checkmark$}, and \textcolor{red}{$\times$} respectively are contributions, the role of GDM, pros, and cons.}
    \label{mmgdm}
    \renewcommand{\arraystretch}{1.2}
    \normalsize
    \begin{tabular}{|>{\arraybackslash}m{0.025\textwidth}|>{\arraybackslash}m{0.42\textwidth}|}
        \hline
        \textbf{\small{Ref.}} & \textbf{\small{Descriptions}} \\
        \hline
    \footnotesize \cite{fu2024multimodal} & \footnotesize
     \textcolor{blue}{$\diamondsuit$}: Propose mm-GESCO, a multimodal generative SemCom framework for emergency response systems using visible light and infrared data. 

     \textcolor{blue}{$\sphericalangle$}: Fuses semantic segmentation maps and uses a LDM with contrastive learning for reconstruction at the receiver.

     \textcolor{green}{$\checkmark$}: Achieves a 200x compression ratio with superior downstream task performance, such as object classification and detection.

     \textcolor{red}{$\times$}: Performance may degrade in highly dynamic environments with large scale multimodal data sets. \\
    \hline

    \footnotesize \cite{yin2025generative} & \footnotesize
     \textcolor{blue}{$\diamondsuit$}: Introduce a Generative Video SemCom framework that fuses textual descriptions and visual cues for ultra-low bandwidth video reconstruction.

     \textcolor{blue}{$\sphericalangle$}: LDM-based model to fuse these modalities and ensure high semantic alignment in video reconstruction.

     \textcolor{green}{$\checkmark$}: Achieves CLIP scores exceeding 0.92 under low SNR conditions, enabling effective video transmission under bandwidth constraints.

     \textcolor{red}{$\times$}: Limited by channel capacity and scalability when processing large video datasets. \\
    \hline

    \footnotesize \cite{grassucci2024diffusion} & \footnotesize
     \textcolor{blue}{$\diamondsuit$}: Propose a generative audio framework where audio is represented by lower dimensional semantic forms such as mel-spectrograms and captions.

     \textcolor{blue}{$\sphericalangle$}: Uses DDPM to restore audio while ensuring semantic consistency.

     \textcolor{green}{$\checkmark$}: Robust to transmission noise and errors, maintaining high quality audio restoration even in adverse conditions.

     \textcolor{red}{$\times$}: Challenges with handling multi-modal corruption or large scale audio datasets in real-time applications. \\
    \hline

    \footnotesize \cite{wei2024language} & \footnotesize
     \textcolor{blue}{$\diamondsuit$}: Propose a language-oriented framework for image transmission based on image to text models, utilizing LDM to reconstruct images from textual descriptions.

     \textcolor{blue}{$\sphericalangle$}: Fine-tuned LDM for semantic level restoration based on the received text.

     \textcolor{green}{$\checkmark$}: Reduces data transmission volume significantly while preserving perceptual quality in image reconstruction.

     \textcolor{red}{$\times$}: Dependent on accurate image-to-text models, vulnerable to errors in textual data generation. \\
    \hline

    \end{tabular}
\end{table}

Multimodal SemCom involves the coordinated processing and transmission of data from different types of sources, such as images, text, and audio, to convey rich and comprehensive meanings. In contrast to unimodal systems that rely solely on visual or auditory signals, multimodal setups leverage complementary cues from different modalities to improve communication accuracy and resilience\cite{zhang2024unified7.8,jiang2024large7.8}.

Building on the multimodal conditioning principles outlined in Section II-A, GDMs provide a powerful mechanism to fuse and reconstruct these multimodal signals in a consistent and meaningful way \cite{yang2023diffusion7.8}. Specifically, the generic Cross-Attention mechanism is adapted for wireless tasks: semantic features or channel state information serve as the condition $y$, acting as Keys ($K$) and Values ($V$) to guide the denoising process of the target modality (Query). By aligning different modalities within a shared latent space or guiding generation with these modality-specific cues, GDMs can efficiently reconstruct complex multimodal scenes at the receiver end, even under constrained bandwidth and unpredictable wireless conditions. GDMs typically condition on semantic cues such as segmentation maps, captions, or modality identifiers to ensure semantic coherence across modalities.

Table \ref{mmgdm} encapsulates key GDM-based multimodal SemCom schemes, and the detailed descriptions are as follows.

\subsubsection{Visual Alignment with Modal-Specific Reconstruction}
This line of work focuses on integrating different visual sensing modalities. Fu et al.\cite{fu2024multimodal} introduce mm GESCO, a framework designed for emergency scenarios that fuses visible light and infrared data. The transmitter's task is highly compressive. It does not send the original images, but instead generates a single fused semantic segmentation map representing both modalities and transmits only this compressed map. The receiver's GDM is therefore tasked with a complex generative act: reconstructing both the visible light and the infrared images from this one shared map.
To achieve this, the authors employ a heavily modified LDM architecture that is doubly conditional. The first modification is the use of SPADE injection. The received semantic map is fed deep into the GDM's U-Net decoder to provide fine grained spatial guidance for the reconstruction. The second and more critical modification is the introduction of a modality condition. The LDM sampling process is also conditioned on a simple token that specifies the target output, such as ``RGB'' or ``Infrared''. This allows a single GDM to generate different image types based on this input switch. To make this shared generation possible, the underlying autoencoders for the LDM are trained using a contrastive loss. This loss function explicitly aligns the latent spaces of the visible and infrared encoders, forcing the GDM to learn a unified, cross modal representation. The primary benefit is extreme bandwidth efficiency, while the main cost is the complex two stage training and the need for multiple modality specific decoders at the receiver.

\subsubsection{Cross-Modal Fusion for Video Transmission}

In the domain of cross modal video transmission, Yin et al. \cite{yin2025generative} introduce a generative video SemCom framework (GVSC), that operates at extremely low bandwidths. This system abandons pixel transmission entirely. Instead, the transmitter extracts two distinct semantic modalities from the source video: a text semantic, which is a generated textual description of the video's content, and a structural semantic, which is a visual guide such as the first frame or a sequence of sketches. At the receiver, the system relies on a large, pre-trained GDM to synthesize the final video. The core GDM adaptation lies in how these two different semantic modalities are fused to guide the GDM's sampling process. The authors employ a modification of the classifier free guidance mechanism. At each denoising step, the GDM calculates two separate noise predictions: one conditioned on the received visual semantics $c_v$ and another conditioned on the received textual semantics $c_t$. These two predictions are then combined using a guidance scale $\omega$, creating a fused guidance signal that steers the generation. This clever modification allows a single pre-trained GDM to generate video that is simultaneously consistent with the structural layout of the first frame and the narrative content of the description. The primary advantage is robust, high quality semantic reconstruction at ultra low bit rates. The main tradeoff is the high computational latency of the generative process and the absolute reliance on a massive, non modifiable GDM at the receiver.

\subsubsection{Text-Audio Fusion for Resilient Acoustic Transmission}

In the audio domain, Grassucci et al. \cite{grassucci2024diffusion} propose a generative framework that transmits two modalities: a compressed latent embedding of the audio spectrogram and a latent representation of its textual caption. This system frames the communication task as a dual inverse problem, where the receiver's GDMs simultaneously denoise corrupted data and inpaint missing data segments.
The authors adapt a Conditional LDM where the textual caption embedding serves as the primary condition to guide the generative process. To handle channel impairments, the sampling algorithm is modified based on the range null space decomposition. This algorithm guides the standard reverse diffusion process by enforcing data consistency. At each step, it uses the received corrupted audio embedding to correct the sampling trajectory, effectively restoring the parts of the signal that were received, while using the text guided generative prior to ``hallucinate" or inpaint the parts that were lost. This modification allows a single GDM to robustly perform denoising and packet loss concealment at the same time. The primary benefit is high robustness to severe channel corruption and data loss, while the tradeoff is the computational cost of the iterative reverse sampling process.

\subsubsection{Language-Visual Coupling for Compact Transmission}
This approach radically compresses visual data by converting it entirely into a textual modality. Wei et al. \cite{wei2024language} propose a generative SemCom framework that transmits only text. At the transmitter, an image to text model converts the source image into a descriptive caption. This text is then robustly encoded using a specialized Transformer based codec and sent over the channel. The receiver's task is purely generative, relying on a CDM, to synthesize an image based only on the received text prompt.
They employ a fine tuning technique known as DreamBooth to adapt the pre-trained Stable Diffusion model. This fine tuning process, which is a form of low rank adaptation, retrains a small subset of the GDM weights on a few sample images of the target class, such as portraits. This modification allows the receiver's GDM to generate high fidelity images that are not only semantically consistent with the received text but also stylistically aligned with the specific, task oriented domain. The primary advantage is an extreme reduction in data volume, while the tradeoff is that the system is limited to generative reconstruction and cannot reproduce the exact original image.

\subsection{GDM for Resource Allocation}

\begin{table}[tbp]
    \centering
\caption{Existing generative model-based schemes for SemCom resource allocation, where \textcolor{blue}{$\diamondsuit$}, \textcolor{blue}{$\sphericalangle$}, \textcolor{green}{$\checkmark$}, and \textcolor{red}{$\times$} respectively are contributions, the role of the model, pros, and cons.}
    \label{resource allocation}
    \renewcommand{\arraystretch}{1.2}
    \normalsize
    \begin{tabular}{|>{\arraybackslash}m{0.025\textwidth}|>{\arraybackslash}m{0.42\textwidth}|}
        \hline
        \small{Ref.} & \small{Descriptions} \\
        \hline
    \footnotesize \cite{cheng2024wireless} & \footnotesize
     \textcolor{blue}{$\diamondsuit$}: Propose SemAIGC, a framework that intelligently splits the AIGC generation task between edge and local devices for dynamic workload balancing.
    
     \textcolor{blue}{$\sphericalangle$}: The reverse diffusion process of a DDPM is split. The edge runs the first $T_{\text{total}}-T$ steps, and the local device completes the final $T$ steps. A DRL agent dynamically selects the split point $T$.

     \textcolor{green}{$\checkmark$}: Provides a highly flexible tradeoff between edge computation and transmission latency based on real-time network conditions.

     \textcolor{red}{$\times$}: Adds the complexity of a DRL agent and requires DDPM inference capabilities on both the edge and local sides. \\
    \hline

    \footnotesize \cite{xu2024generative} & \footnotesize
     \textcolor{blue}{$\diamondsuit$}: Propose a semantic-aware power allocation algorithm based on a rigorous analysis of a foundation model's perceptual sensitivity.
    
      \textcolor{blue}{$\sphericalangle$}: A pre-trained LDM acts as the semantic decoder. Rate-distortion-perception theory is used to characterize the LDM's sensitivity to input errors.

     \textcolor{green}{$\checkmark$}: Achieves significant power savings by prioritizing power for the most semantically valuable data streams to meet a perceptual quality target.

     \textcolor{red}{$\times$}: Relies on a pre-trained, non-modifiable foundation model; the analysis framework is specific to this setup. \\
    \hline

    \footnotesize \cite{liu2025generative} & \footnotesize
     \textcolor{blue}{$\diamondsuit$}: Introduce the Age of Semantic Information metric and solve the resource allocation game using a novel DRL algorithm.
    
      \textcolor{blue}{$\sphericalangle$}: Uses two models: 1) A multimodal conditional DDIM serves as the generative semantic decoder. 2) A CDM is used as the DRL policy network to solve the Stackelberg game.

     \textcolor{green}{$\checkmark$}: The CDM-based DRL solver converges faster and finds more robust solutions for the complex resource trading game than traditional DRL algorithms.

     \textcolor{red}{$\times$}: High overall system complexity, as it requires both a DDIM for content generation and a CDM for economic optimization. \\
    \hline

    \end{tabular}
\end{table}

While resource allocation is a broad, cross-stack challenge in general networks, the scope of this subsection is specifically focused on its application within SemCom. In this context, resource allocation refers to the dynamic control of bandwidth, computation, and energy to ensure reliable and timely transmission of semantically meaningful content\cite{yan2022resource7.8}.  When GDMs are deployed at SemCom, resource optimization becomes especially critical. GDMs can consume substantial computational and communication overhead, but also offer flexibility in controlling generation complexity and prioritizing semantic fidelity\cite{liu2025optimizing}.
Recent research has explored the integration of GDMs into semantic-aware resource allocation, using them not just as decoders but as controllable modules that influence how system resources are distributed. GDMs typically fall into four categories: controllable generation, power-efficient transmission, freshness-aware optimization, and incentive-driven resource trading.
Table \ref{resource allocation} illustrates existing GDM-based schemes for resource allocation, and the detailed descriptions are as follows.

\subsubsection{Controllable Generation for Adaptive Workload Balancing}
This category leverages the controllable nature of the diffusion process itself to perform adaptive workload balancing. Cheng et al. \cite{cheng2024wireless} propose the SemAIGC framework, which intelligently splits the AIGC generation task between the edge transmitter and the local receiver. The system is built upon a standard DDPM based generative process. The core innovation is the modification and splitting of the GDM's reverse diffusion process into two parts. The edge transmitter performs the first set of denoising steps, from $T_{\text{total}}$ down to an intermediate step $T$. It then transmits the resulting intermediate noisy latent feature. The local receiver, which has a lightweight fine-tuning GDM module, receives this feature and completes the remaining $T$ steps of denoising to generate the final image.

The key adaptation is that this split point $T$ is not fixed but is a dynamic, controllable parameter. The framework uses a DRL agent, called ROOT, to intelligently decide the optimal value of $T$ based on real-time conditions, including channel quality, edge resource availability, and local device capabilities. Furthermore, the GDM at the receiver is specifically modified. The reverse sampling process is re-derived to account for the additional semantic noise introduced by the wireless channel during the transmission of the intermediate feature. The primary benefit is a highly flexible tradeoff between edge computation and transmission latency. The main cost is the added complexity of the DRL agent and the requirement for GDM inference capabilities on both sides of the link.

\subsubsection{Semantic-Aware Power Allocation in Low-Rate Channels}

This approach leverages pre-trained GDMs as the semantic decoder. Xu et al. \cite{xu2024generative} propose a framework where the transmitter sends multiple semantic streams, such as textual prompts and structural edge maps, to a pre-trained GDM like Stable Diffusion. The authors present a rigorous analysis of the GDM's perceptual sensitivity to input errors. They use rate distortion perception theory to mathematically characterize the relationship between the transmission reliability of the semantic inputs, such as the bit error rate of the text prompt, and the final perceptual quality of the synthesized image.
This analysis allows them to define a ``semantic value" for each distinct data stream. Based on the proven non decreasing property of this perception error function, the paper introduces a semantic aware power allocation algorithm. This algorithm intelligently distributes transmission power between the different semantic streams. It minimizes total power consumption by allocating just enough energy to each stream to collectively satisfy a global perceptual quality requirement. The primary advantage is a significant power saving, reportedly up to ninety percent in channel coded cases. This gain is achieved because the system learns to prioritize power for the most semantically valuable information rather than wasting it on less critical data streams.

\subsubsection{Freshness-Aware Optimization via Semantic Timeliness Metrics}

This method introduces a novel metric called the Age of Semantic Information (AoSI) to quantify the freshness of generated content in mobile AIGC networks. Liu et al.\cite{liu2025generative} integrate AoSI directly into the utility function of a Stackelberg game model, which is used to optimize the resource allocation and pricing between the service provider and users. This paper features two distinct GDM adaptations. First, it uses a multimodal Conditional DDIM as the generative semantic decoder. It is modified using adaptive group normalization, allowing it to fuse received text semantics and image semantics to reconstruct the final AIGC product.
The authors treat the optimal pricing strategy as a generative task. CDM is used as the policy network or actor. It is trained to take the current environment state as its condition and execute a reverse diffusion process, generating the optimal pricing action from an initial Gaussian noise vector. This GDM based solver is shown to converge faster and closer to the optimal equilibrium than traditional DRL algorithms. The primary advantage is its ability to find robust solutions for the complex resource trading game. The main tradeoff is the high overall system complexity, which uses one GDM for content generation and a second GDM for economic optimization.

\subsection{Summaries and Lessons Learned}
\label{sec:trans_summary}

\begin{table*}[tbp]
\centering
\caption{Comparison and Insights for GDM's Applications in the Transmission Layer.}
\label{tab:gdm_transmission_comparison}
\renewcommand{\arraystretch}{1.3}
\footnotesize

\begin{tabularx}{\textwidth}{ p{3.5cm} >{\raggedright\arraybackslash}X >{\raggedright\arraybackslash}X }
\toprule
\textnormal{Approach \& Refs} & \textnormal{Core Role \& Best-Suited Scenario} & \textnormal{Pros \& Cons / Trade-offs} \\
\midrule

\multicolumn{3}{l}{\textbf{A. Modulation and Coding}} \\
\cmidrule(r){1-3}

Probabilistic Shaping \cite{Letafati202311.25} 
& 
\textbf{Core Role:} Generative shaper that aligns symbol distribution with receiver inference. 
\par\vspace{4pt}
\textbf{Scenario:} Systems requiring resilience against non-Gaussian noise or low SNR. 
& 
\textbf{Pros:} Enhances mutual information by mimicking receiver denoising; robust to OOD noise. 
\par\vspace{4pt}
\textbf{Cons:} High transmitter-side computational overhead compared to static mapping. \\

\noalign{\vskip 4pt}
\cdashline{1-3}
\noalign{\vskip 4pt}

Diffusion Decoding \cite{choukroun2022denoising11.25} 
& 
\textbf{Core Role:} Iterative decoder treating corruption as forward diffusion. 
\par\vspace{4pt}
\textbf{Scenario:} Soft decoding of linear block codes using syndrome guidance. 
& 
\textbf{Pros:} Outperforms SOTA neural decoders; effective even with single-step reverse diffusion. 
\par\vspace{4pt}
\textbf{Cons:} Line search for step size optimization increases inference complexity. \\
\midrule

\multicolumn{3}{l}{\textbf{B. Semantic Denoiser}} \\
\cmidrule(r){1-3}

Posterior Sampling \cite{xu2023latent, xu2025semantic} 
& 
\textbf{Core Role:} Acts as a strong semantic prior to guide the sampling process. 
\par\vspace{4pt}
\textbf{Scenario:} Offline tasks where maximum semantic fidelity is critical and latency is tolerable. 
& 
\textbf{Pros:} Theoretically sound with high fidelity and requires no CSI. 
\par\vspace{4pt}
\textbf{Cons:} High computational latency due to the iterative and complex posterior sampling. \\

\noalign{\vskip 4pt}
\cdashline{1-3}
\noalign{\vskip 4pt}

Channel-Aligned Model \cite{wu2024cddm, pei2025latent} 
& 
\textbf{Core Role:} The model is matched to channel statistics or distilled for one-shot inference. 
\par\vspace{4pt}
\textbf{Scenario:} Real-time or low-latency semantic denoising applications. 
& 
\textbf{Pros:} Fast inference is possible, as EECD enables single-step denoising. 
\par\vspace{4pt}
\textbf{Cons:} Requires accurate CSI for the channel-aware model or involves complex offline distillation. \\
\midrule

\multicolumn{3}{l}{\textbf{C. Auxiliary Recovery}} \\
\cmidrule(r){1-3}

Image-Level Guidance \cite{yilmaz2024high, pezone2024semantic} 
& 
\textbf{Core Role:} Conditional generator that refines details from coarse images or semantic maps. 
\par\vspace{4pt}
\textbf{Scenario:} Bandwidth-constrained scenarios where the receiver, such as an edge server, has ample compute power. 
& 
\textbf{Pros:} Massive bandwidth compression is achieved by sending only low-resolution or map data. 
\par\vspace{4pt}
\textbf{Cons:} High computational complexity is shifted to the receiver's GDM. \\

\noalign{\vskip 4pt}
\cdashline{1-3}
\noalign{\vskip 4pt}

Feature-Level Guidance \cite{wang2025diffcom, chen2024commin} 
& 
\textbf{Core Role:} Inverse problem solver that recovers data from raw received signals or features. 
\par\vspace{4pt}
\textbf{Scenario:} Dynamic or unknown channel conditions that require high robustness for physical-layer recovery. 
& 
\textbf{Pros:} Robust to channel variations; DiffCom can even work on unknown channels. 
\par\vspace{4pt}
\textbf{Cons:} Iterative sampling remains slow; the CommIN system is highly complex. \\
\midrule

\multicolumn{3}{l}{\textbf{D. Semantic-based Generation}} \\
\cmidrule(r){1-3}

Video/Image Generation \cite{li2025goal, duan2024dm, wei2024language} 
& 
\textbf{Core Role:} Generative decoder that synthesizes content from abstract goals like keyframes or text. 
\par\vspace{4pt}
\textbf{Scenario:} Extreme low-bandwidth tasks where only the semantic intent matters, not the exact source pixels. 
& 
\textbf{Pros:} Extremely low bandwidth is required, as only abstract semantics are transmitted. 
\par\vspace{4pt}
\textbf{Cons:} Abandons source fidelity entirely; it generates plausible content, not the original. \\
\midrule

\multicolumn{3}{l}{\textbf{E. Multimodal Transmission}} \\
\cmidrule(r){1-3}

Modality Fusion \cite{grassucci2024diffusion, fu2024multimodal, yin2025generative} 
& 
\textbf{Core Role:} Fused generative decoder that unifies the semantic space of different modalities. 
\par\vspace{4pt}
\textbf{Scenario:} Complex tasks needing multi-source fusion, like emergency response or autonomous driving. 
& 
\textbf{Pros:} Enriches semantics and improves downstream task accuracy, such as object detection. 
\par\vspace{4pt}
\textbf{Cons:} High difficulty in modality alignment; requires complex joint training data and methods. \\
\midrule

\multicolumn{3}{l}{\textbf{F. Resource Allocation}} \\
\cmidrule(r){1-3}

Optimization Policy Generation \cite{liu2025generative, xu2024generative, cheng2024wireless} 
& 
\textbf{Core Role:} A controllable module for balancing compute or a policy network for game theory. 
\par\vspace{4pt}
\textbf{Scenario:} Dynamic networks needing joint optimization of semantic quality, AoSI, and power consumption. 
& 
\textbf{Pros:} Enables end-to-end, semantic-aware resource scheduling and optimization. 
\par\vspace{4pt}
\textbf{Cons:} High complexity of the control loop; faces convergence challenges. \\

\bottomrule
\end{tabularx}
\end{table*}

In this section, we reviewed GDM-based schemes for the transmission layer. While predominantly concentrated on SemCom, the scope has expanded to include fundamental physical layer innovations. Our review reveals a fundamental trade-off space for GDM deployment, balancing between bit-level precision (optimizing modulation and coding for exact recovery \cite{Letafati202311.25, choukroun2022denoising11.25}), generative fidelity (recovering clean semantics \cite{xu2025semantic}), physical-layer guidance (adapting to channel conditions \cite{wu2024cddm, wang2025diffcom}), and task-oriented synthesis (generating content from abstract intent \cite{li2025goal, yin2025generative}). As detailed in Table~\ref{tab:gdm_transmission_comparison}, existing approaches explore different optimal points within this trade-off space.

\textbf{Insight 1: Latency vs. Fidelity.} A primary decision point is latency. Methods prioritizing theoretical fidelity, such as SP-Latent-Diff EDNSC \cite{xu2025semantic}, leverage iterative posterior sampling, resulting in high computational latency \cite{xu2025semantic} unsuitable for real-time applications. Conversely, approaches like EECD \cite{pei2025latent} distill the GDM into a single-step model, achieving low latency \cite{pei2025latent} by sacrificing some generative optimality, making them viable for real-time denoising.

\textbf{Insight 2: Bandwidth vs. Computation.} GDMs enable a paradigm shift, trading bandwidth for computation. Conditional restoration schemes demonstrate this clearly. For instance, SPIC \cite{pezone2024semantic} transmits only coarse maps, saving immense bandwidth \cite{pezone2024semantic}, but relies on a powerful GDM at the receiver to perform complex conditional generation \cite{pezone2024semantic}. This is ideal for asymmetric scenarios, such as edge-assisted systems, where the receiver has ample compute.

\textbf{Insight 3: Reconstruction vs. Generation.} The system's objective dictates the GDM's role. If the goal is reconstructing the original source, auxiliary-guided methods like DiffCom \cite{wang2025diffcom} are suitable, using the received signal to guide the GDM towards the original. However, if the goal is conveying intent under extreme bandwidth constraints, semantic-generation methods are superior. They abandon source fidelity entirely, using GDM to synthesize new, plausible content, such as generating an image from text \cite{wei2024language} or video from keyframes \cite{li2025goal}, that satisfies the abstract semantic goal.

\textbf{Insight 4: Beyond Semantics to Physical Precision.} While the current landscape is dominated by semantic-oriented designs, the application of GDMs to fundamental physical layer tasks remains an untapped frontier. Pioneering works in constellation shaping \cite{Letafati202311.25} and error correction \cite{choukroun2022denoising11.25} prove that GDM's generative capabilities are not limited to abstract semantics but can rigorously model bit-level signal distributions. This indicates a broader potential for GDMs to redefine classical communication modules, moving beyond the semantic paradigm to optimize the physical transmission itself.

Despite these advances, significant challenges remain. The iterative sampling required by high-fidelity models \cite{xu2025semantic, wang2025diffcom} remains the primary bottleneck for real-time deployment. Channel mismatch is a critical issue; models trained on synthetic channels, such as in \cite{wu2024cddm}, may fail in dynamic, real-world fading environments \cite{xu2023latent}. Furthermore, in multimodal transmission, ensuring consistent \textbf{semantic alignment} between different modalities, like audio and text \cite{grassucci2024diffusion} or IR and RGB \cite{fu2024multimodal}, is difficult and lacks theoretical guarantees.

Future work can explore lightweight GDM architectures. Additionally, developing channel-adaptive samplers is promising, which would dynamically adjust the number of denoising steps based on real-time CSI or received signal quality. researchers should broaden the scope of GDM applications to more traditional physical layer blocks, and GDM's parameters, such as inference steps, should be treated as optimizable variables in a joint semantic-resource co-design framework \cite{xu2024generative, liu2025generative}, enabling end-to-end optimization of quality, latency, and power.

\section{GDM for Vertical Applications and Services}
\label{section5}
After reviewing the foundational roles of GDMs in the sensing and transmission layers, this section explores how these capabilities are integrated to empower large-scale vertical applications and services.
It is crucial to clarify that the domains discussed herein, such as network digital twins, intelligent transportation, and immersive communication are not treated as a new ``application layer" in the protocol stack sense. Instead, these are complex, task-oriented ecosystems that span multiple functional domains, integrating sensing, computation, and transmission.
Therefore, the focus of this section shifts from the layer of operation to the advanced role GDM plays. We analyze how GDM serves as a key enabling technology, integrating its generative, modeling, and restoration capabilities to meet the specific, end-to-end goals of these services.

\subsection{GDM  for Network Digital Twin Simulation}
\label{sec:gdm_as_dt}

GDM' role is evolving from the passive channel modeling discussed in Section \ref{section3} to the active, generative simulation of an entire network ecosystem. As a ``world model," a GDM becomes the core engine for the network Digital Twin (DT). These models are no longer just fitting existing data distributions; they are learning complex network dynamics, topologies, and user behaviors to generate high-fidelity, interactive virtual network environments.
Recent research in this area has focused on two main streams: 1) leveraging GDMs as a policy optimizer or decision-making engine within the DT, and 2) leveraging them as the core data-generation engine for the DT. Table ~\ref{tab:gdm_dt_summary} summarizes representative schemes for network digital twins.

\subsubsection{GDM for Digital Twin Policy and Decision-Making}
Rather than simply generating passive simulation data, a key stream of research employs specific GDMs as an active, high-dimensional policy and decision-making engine within the DT.
For instance, Huang et al.\cite{huang2024digital} propose a GAI-driven DT architecture where a DDPM is leveraged specifically for closed-loop network decision-making. The core DDPM modification is the inclusion of an inverse dynamics mechanism. This adapts the conditional DDPM to function as a policy network: it first uses state diffusion to generate a future state based on the historical state. Then, the inverse dynamics mechanism takes both historical and generated future states as input to produce a network management action. The primary advantage is that this DDPM-based policy can achieve better convergence performance than standard DRL algorithms. However, the main tradeoff is the significant computational latency and ``black box" nature of using an iterative generative process for a real-time control loop.

\begin{table}[tbp]
    \centering
    \caption{Existing generative model-based schemes for Network Digital Twin applications, where \textcolor{blue}{$\diamondsuit$}, \textcolor{blue}{$\sphericalangle$}, \textcolor{green}{$\checkmark$}, and \textcolor{red}{$\times$} respectively are contributions, the role of the model, pros, and cons.}
    \label{tab:gdm_dt_summary}
    \renewcommand{\arraystretch}{1.2}
    \normalsize
    \begin{tabular}{|>{\arraybackslash}m{0.025\textwidth}|>{\arraybackslash}m{0.42\textwidth}|}
        \hline
        \small{Ref.} & \small{Descriptions} \\  
        \hline
        \footnotesize \cite{huang2024digital} & \footnotesize
         \textcolor{blue}{$\diamondsuit$}: Propose a GAI-driven DT architecture for intelligent closed-loop network management.
         
         \textcolor{blue}{$\sphericalangle$}: A DDPM is leveraged for the network decision-making module, using state diffusion to generate future states and an inverse dynamics mechanism to generate the network management action.
         
         \textcolor{green}{$\checkmark$}: Achieves better convergence performance compared to standard DRL algorithms.
         
         \textcolor{red}{$\times$}: Faces challenges of high computational overhead and the "black box" nature of GAI models.
        \\
        \hline
        \footnotesize \cite{chai2024generative11.8} & \footnotesize
         \textcolor{blue}{$\diamondsuit$}: Propose a GAI-drivenmobile network DT paradigm where GAI is the key enabler for generating DT data.
         
         \textcolor{blue}{$\sphericalangle$}: A conditional SGM functions as an environment simulator for an RL optimizer, accepting the agent's action as a condition to regenerate the next network state.
         
         \textcolor{green}{$\checkmark$}: Enables high-fidelity iterative optimization in virtual space and provides "generative gains" by reducing data upload overhead.
         
         \textcolor{red}{$\times$}: The optimization loop's speed is dependent on the SGM's inference latency, and a potential sim-to-real gap exists.
        \\
        \hline
        \footnotesize \cite{zhou2025user}& \footnotesize
         \textcolor{blue}{$\diamondsuit$}: Propose a data-oriented framework for User-Centric Immersive Communications using User Digital Twins.
         
         \textcolor{blue}{$\sphericalangle$}: GAI is used within UDT Operation Functions to generate atypical, user-specific, or out-of-distribution data for the network model evaluation phase.
         
         \textcolor{green}{$\checkmark$}: Enables fine-grained, robust evaluation of network models against diverse scenarios, improving personalization.
         
         \textcolor{red}{$\times$}: Introduces additional consumption of network resources to support the UDTOFs.
        \\
        \hline
        \footnotesize \cite{zhang2024netdiff} & \footnotesize
         \textcolor{blue}{$\diamondsuit$}: Propose NetDiff, a "world model" to generate high-fidelity network flow traces by explicitly modeling user intent.
         
         \textcolor{blue}{$\sphericalangle$}: Employs a hierarchical, conditional DDPM. The upper layer uses a latent-space DDPM for app intent. The lower layer is a conditional DDPM with a double-layer transformer denoiser.
         
         \textcolor{green}{$\checkmark$}: Avoids pattern collapse, unlike GANs, and accurately captures complex temporal and feature correlations.
         
         \textcolor{red}{$\times$}: High model complexity due to the hierarchical structure and specialized double-layer transformer denoiser.
        \\
        \hline
    \end{tabular}
\end{table}

\subsubsection{GDM for Digital Twin Data Generation}
A second, more direct application stream focuses on using specific models as the engine to actively create the high-fidelity, dynamic network data that constitutes the DT simulation.
Chai et al.\cite{chai2024generative11.8} propose a GAI-driven mobile network DT paradigm, where GAI is the key enabler for generating data, such as user trajectories and network traffic. The core modification is its use as a dynamic simulator for an RL-based optimizer, built on a conditional SGM. The SGM's sampling is explicitly conditioned on the RL agent's action. This allows the SGM to regenerate the next network state in response to the agent's decision, enabling iterative optimization in a high-fidelity virtual space.
Moving from network-state data to user-centric data, the NetDiff model proposed by Zhang et al.\cite{zhang2024netdiff} serves as a sophisticated ``world model" for generating high-fidelity network flow traces by explicitly modeling user intent. The DDPM architecture is heavily modified into a two-layer hierarchical structure. An upper-layer DDPM models app usage intent in a latent space, which then guides a lower-layer conditional DDPM, using a novel double-layer transformer, to generate the corresponding network flow traces. This deep modification allows NetDiff to avoid the pattern collapse common in GANs and capture complex user-behavior correlations.
In a different, more conceptual approach, Zhou et al.\cite{zhou2025user} propose a data-oriented framework using user DTs for personalized immersive communications. Here, GAI is not the live DT itself, but rather an offline evaluation simulator. Its role is to generate atypical network scenarios or user-specific datasets. This data is then used to evaluate the robustness of other network models, such as QoE models, in a virtual space, enabling robust, low-risk testing without live-network trials.

\subsection{GDM for Network Policy and Solution Optimization}

Beyond serving as a data generator for digital twins, GDMs are increasingly being adopted as active decision-making components and policy optimizers. In this capacity, GDMs are leveraged either as direct, end-to-end optimization solvers or as highly effective policy networks within DRL frameworks. This approach capitalizes on GDM's core strength: its ability to learn and sample from a complex distribution of high-quality solutions, rather than merely converging to a single, often locally optimal, point. Table ~\ref{tab:gdm_optimizer_summary_en} presents schemes based on this approach.

\begin{table}[tbp]
    \centering
    \caption{Existing GDM-based schemes for Network Policy and Solution Optimization, where \textcolor{blue}{$\diamondsuit$}, \textcolor{blue}{$\sphericalangle$}, \textcolor{green}{$\checkmark$}, and \textcolor{red}{$\times$} respectively denote contributions, the role of the model, pros, and cons.}
    \label{tab:gdm_optimizer_summary_en}
    \renewcommand{\arraystretch}{1.2}
    \normalsize
    \begin{tabular}{|>{\arraybackslash}m{0.025\textwidth}|>{\arraybackslash}m{0.42\textwidth}|}
        \hline
        \small{Ref.} & \small{Descriptions} \\  
        \hline
        \footnotesize \cite{liang2025diffsg11.8} & \footnotesize
         \textcolor{blue}{$\diamondsuit$}: Propose a DDPM optimizer, DiffSG, to solve complex network problems including MINLP, convex, and non-convex, by learning the distribution of high-quality solutions.
         
         \textcolor{blue}{$\sphericalangle$}: Uses a DDPM with classifier-free guidance. A key modification is setting the guidance strength $\omega$ to an extremely high value to force deterministic convergence toward the optimal solution.
         
         \textcolor{green}{$\checkmark$}: Insensitive to objective function properties such as non-convexity and mixed-integer characteristics. Demonstrates superior robustness on out-of-domain inputs compared to discriminative models.
         
         \textcolor{red}{$\times$}: Relies on a pre-computed supervised dataset of (input, optimal\_solution) pairs for training.
        \\
        \hline
        \footnotesize \cite{zhang2025enhanced11.8} & \footnotesize
         \textcolor{blue}{$\diamondsuit$}: Propose an Actor-Critic GDM (AC-GDM) algorithm to solve the non-convex optimization problem of secure beamforming in an IRS-assisted IoT system.
         
         \textcolor{blue}{$\sphericalangle$}: The AC-GDM replaces the standard DDPG actor network. It acts as a policy generator, conditioned on the channel state, using its reverse diffusion process to denoise noise into an optimal beamforming vector.
         
         \textcolor{green}{$\checkmark$}: Effectively escapes local optima that trap traditional DDPG algorithms. Achieves higher secrecy rates and better convergence stability in dynamic and noisy channel environments.
         
         \textcolor{red}{$\times$}: Increases computational complexity. Action selection requires an iterative denoising process, which adds latency compared to a single forward pass.
        \\
        \hline
        \footnotesize \cite{zhang2024multi11.8} & \footnotesize
         \textcolor{blue}{$\diamondsuit$}: Propose a GDMTD3 algorithm for a complex NP-hard multi-objective optimization problem in a UAV swarm, maximizing secrecy rate while minimizing energy consumption.
         
         \textcolor{blue}{$\sphericalangle$}: A DDPM serves as the actor network in a TD3 DRL framework. It is modified by incorporating sinusoidal position embeddings for the diffusion time step $t$ to better model the temporal denoising sequence.
         
         \textcolor{green}{$\checkmark$}: The model's superior distribution modeling captures the complex trade-offs between the two conflicting objectives more effectively than MLP or Transformer-based actors.
         
         \textcolor{red}{$\times$}: High computational complexity. The iterative denoising process for action selection increases inference latency.
        \\
        \hline
    \end{tabular}
\end{table}

\subsubsection{GDM for Direct Optimization Solving}
Liang et al.\cite{liang2025diffsg11.8,liang2025diffusion11.8} explore the direct application of DDPMs as network optimizers for complex problems like mixed-integer non-linear programming and non-convex optimization. Their optimizer, DiffSG, reframes the problem by learning the distribution of high-quality solutions $p(y|x)$ conditioned on the network input $x$. The key DDPM modification is an adaptation of classifier-free guidance where the strength parameter $\omega$ is set to an exceptionally high value. This forces the denoising process to converge deterministically toward the single, highest-probability solution, making the optimizer robust to non-convex landscapes and OOD inputs.

\subsubsection{GDM  for Policy Networks in DRL}
A more common integration strategy involves using GDM as a powerful policy network within an DRL framework, replacing the standard MLP.
Zhang et al.\cite{zhang2025enhanced11.8} pioneer this with an AC-GDM algorithm to solve the non-convex problem of secure beamforming for an IRS-assisted IoT system. They replace the DDPG's deterministic actor with a conditional GDM. Conditioned on the channel state $h$, this GDM-Actor executes its reverse diffusion process to denoise pure Gaussian noise into an optimal beamforming vector. The Critic's Q-value guides the GDM's training, enabling the policy to escape local optima that trap traditional DRL algorithms.
Building on this hybrid DRL-GDM approach, Zhang et al.\cite{zhang2024multi11.8} tackle a highly complex, NP-hard multi-objective optimization problem for a UAV swarm: maximizing secrecy rate while minimizing flight energy. Their GDMTD3 algorithm integrates a conditional DDPM as the actor into the more advanced TD3 DRL framework. The DDPM-Actor is modified with sinusoidal position embeddings for the time step $t$, enhancing its ability to model the temporal denoising sequence. This superior distribution modeling allows the DDPM to capture the intricate trade-offs between the two conflicting objectives more effectively than standard MLP or Transformer-based actors. For both DRL-GDM models, the primary trade-off is that action selection becomes an iterative denoising process, increasing latency compared to a single forward pass.

\subsection{GDM for Task-Oriented Generative Communications}
Beyond network-wide simulation, GDMs are being directly integrated into the communication and control loops for specific vertical applications. In this context, GDM's generative power is applied to solve complex, domain-specific challenges, often by enabling task-oriented communication or acting as a decision-making engine. This subsection reviews GDM's application in several key domains: UAV networks, vehicular networks, and immersive communication.

\subsubsection{GDM for UAV Network Applications}
UAVs represent a critical application domain for optimization based on GDMs, largely due to their high mobility and the complex, dynamic nature of their operational environments. Tasks such as trajectory planning and resource allocation often translate into high-dimensional, non-convex optimization problems. In this context, GDMs are increasingly utilized not as data generators, but as powerful policy engines, often integrated with DRL frameworks. Table ~\ref{tab:gdm_uav_summary} summarizes key schemes.

For example, Yu et al.\cite{yu2025multi11.8} tackle the multi-UAV trajectory generation problem, aiming to minimize Age-of-Information while maximizing coverage. They replace the traditional MLP actor-policy in a Soft Actor-Critic framework with a DDPM-based predictor. This predictor is conditionally guided by features from a hierarchical graph-transformer, allowing it to capture complex state patterns and converge faster than standard DRL benchmarks.
Addressing a different challenge, Zhang et al.\cite{zhang2025beam11.8} focus on beam tracking for high-speed UAVs to minimize interruptions. They propose the PPBT-AR algorithm, which pairs an LSTM for prediction with a DDPM as a direct optimization solver. The DDPM is structured as a ``solution generation network" ($\epsilon_{\tau}$) trained to denoise a random input into an optimal beam alignment solution, guided by a ``solution evaluation network" ($Q_v$). This approach significantly reduces beam switching frequency.
Taking a more meta-level approach, Sun et al.\cite{sun2024generative11.8} use a GDM to optimize UAV spectrum map estimation and transmission rates. Their framework, SEMG, uniquely employs an LLM with RAG to programmatically design the GDM's network structure and loss function based on high-level prompts. The GDM itself then acts as a policy optimizer, outperforming traditional DRL methods in modeling the complex state-action mappings for spectrum-aware tasks.

\begin{table}[tbp]
    \centering
    \caption{Existing generative model-based schemes for UAV applications, where \textcolor{blue}{$\diamondsuit$}, \textcolor{blue}{$\sphericalangle$}, \textcolor{green}{$\checkmark$}, and \textcolor{red}{$\times$} respectively denote contributions, the role of the model, pros, and cons.}
    \label{tab:gdm_uav_summary}
    \renewcommand{\arraystretch}{1.2}
    \normalsize
    \begin{tabular}{|>{\arraybackslash}m{0.025\textwidth}|>{\arraybackslash}m{0.42\textwidth}|}
        \hline
        \small{Ref.} & \small{Descriptions} \\  
        \hline
        \footnotesize \cite{yu2025multi11.8} & \footnotesize
         \textcolor{blue}{$\diamondsuit$}: Propose a multi-UAV trajectory generation solution to minimize Age-of-Information and maximize data collection coverage.
         
         \textcolor{blue}{$\sphericalangle$}: Replaces the standard actor-policy network in a Soft Actor-Critic framework with a DDPM-based predictor. This DDPM is conditionally guided by features from a hierarchical graph-transformer network that extracts UAV-UAV and UAV-user interactions.
         
         \textcolor{green}{$\checkmark$}: Achieves faster convergence and superior performance in average AoI and user coverage compared to standard SAC and DQN benchmarks. Better captures complex state patterns.
         
         \textcolor{red}{$\times$}: Faces challenges of significant computational complexity, as it requires an iterative diffusion process for action generation and a complex graph transformer for state encoding.
        \\
        \hline
        \footnotesize \cite{zhang2025beam11.8} & \footnotesize
         \textcolor{blue}{$\diamondsuit$}: Propose a position prediction-based beam tracking algorithm with adaptive beam reconstruction to minimize communication interruptions from frequent beam switching.
         
         \textcolor{blue}{$\sphericalangle$}: A DDPM is used as a direct optimization solver to jointly optimize beam width and signal strength. It acts as a "solution generation network" ($\epsilon_{\tau}$) trained to denoise a random input into an optimal beam alignment solution, guided by a "solution evaluation network" ($Q_v$).
         
         \textcolor{green}{$\checkmark$}: Reduces beam switching frequency by up to 76.9\% in high-speed scenarios, successfully balancing stable communication gain against beam adjustment overhead.
         
         \textcolor{red}{$\times$}: High computational complexity, comparable to other DRL-based optimizers, presenting challenges for deployment on resource-limited UAVs.
        \\
        \hline
        \footnotesize \cite{sun2024generative11.8} & \footnotesize
         \textcolor{blue}{$\diamondsuit$}: Propose a GAI-based framework for UAV spectrum map estimation and joint transmission rate optimization.
         
         \textcolor{blue}{$\sphericalangle$}: Functions as the "Generation Part" of the framework, structured as a GDM policy optimizer with a "generation network" and a "value network". Its network structure and loss function are programmatically designed by an LLM+RAG component based on user prompts.
         
         \textcolor{green}{$\checkmark$}: Significantly outperforms LSTM in spectrum estimation accuracy and DDPG in optimizing transmission rates. Better models complex state-action mappings and generates diverse, high-quality samples.
         
         \textcolor{red}{$\times$}: Substantial framework complexity due to the two-stage system (LLM + GDM). GAI inference is resource-intensive, which is a challenge for energy-constrained UAVs.
        \\
        \hline
    \end{tabular}
\end{table}

\subsubsection{GDM for Vehicular Network Applications}

\begin{table}[tbp]
    \centering
    \caption{Existing GDM-based schemes for Vehicular applications, where \textcolor{blue}{$\diamondsuit$}, \textcolor{blue}{$\sphericalangle$}, \textcolor{green}{$\checkmark$}, and \textcolor{red}{$\times$} respectively denote contributions, the role of GDM, pros, and cons.}
    \label{tab:gdm_vehicular_summary}
    \renewcommand{\arraystretch}{1.2}
    \normalsize
    \begin{tabular}{|>{\arraybackslash}m{0.025\textwidth}|>{\arraybackslash}m{0.42\textwidth}|}
        \hline
        \textbf{\small{Ref.}} & \textbf{\small{Descriptions}} \\
        \hline
        \footnotesize \cite{zhang2024generative} & \footnotesize
         \textcolor{blue}{$\diamondsuit$}: Propose a multi-modality semantic-aware V2V framework to improve generative reliability by transmitting both text and a structural image skeleton.
         
         \textcolor{blue}{$\sphericalangle$}: Functions as a denoising U-Net decoder at the receiver. Its image generation process is conditionally guided by both the received text prompt and the structural image skeleton.
         
         \textcolor{green}{$\checkmark$}: Provides more reliable road information for safety decisions because the image skeleton ensures the generated content closely matches the actual scene.
         
         \textcolor{red}{$\times$}: Introduces additional complexity for extracting and transmitting the dual-modality data and places high real-time processing demands on the generative model.
        \\
        \hline
        \footnotesize \cite{lu2024generative11.9} & \footnotesize
         \textcolor{blue}{$\diamondsuit$}: Propose a GAI-enhanced multi-modal SemCom framework for IoV, using BEV Fusion to integrate multi-sensor data.
         
         \textcolor{blue}{$\sphericalangle$}: A DDPM is employed in semantic decoder for two distinct roles: refinement to enhance the clarity of noisy BEV maps and prediction to forecast future BEV images using timestamps as prompts.
         
         \textcolor{green}{$\checkmark$}: Significantly improves IoU and visual clarity, shows strong robustness against channel noise, and can proactively predict future road conditions.
         
         \textcolor{red}{$\times$}: The iterative sampling process of the GDM introduces a generation delay, making it more suitable for non-instantaneous tasks like cloud analysis.
        \\
        \hline
        \footnotesize \cite{wijesinghe2024diff11.8} & \footnotesize
         \textcolor{blue}{$\diamondsuit$}: Propose a Diffusion-based, Goal-Oriented framework based on trading computation for bandwidth, featuring a Local Generative Feedback mechanism.
         
         \textcolor{blue}{$\sphericalangle$}: A full DDPM is embedded at the transmitter for local quality validation. It also uses a quantized noise space, allowing the transmitter to send only a few compressed weights $w$ that represent the optimal noise latent, ensuring an identical reconstruction by the receiver.
         
         \textcolor{green}{$\checkmark$}: Achieves ultra-high spectrum efficiency and guarantees goal-oriented QoS because the transmitter validates the generative output locally before transmission.
         
         \textcolor{red}{$\times$}: Places a massive computational load on the transmitter, which must execute a full forward diffusion, noise projection, and reverse denoising pass for every transmission.
        \\
        \hline
        \footnotesize \cite{grassucci2023generative} & \footnotesize
         \textcolor{blue}{$\diamondsuit$}: Present a generative SemCom framework to synthesize photorealistic scenes from highly compressed and noisy semantic segmentation maps.
         
         \textcolor{blue}{$\sphericalangle$}: A CDM with Spatially-Adaptive Normalization in its decoder generates the image. Its robustness comes from being explicitly trained on semantic maps corrupted by simulated channel noise.
         
         \textcolor{green}{$\checkmark$}: Extremely robust to channel degradation, generating recognizable objects and accurate depth estimations even at 1 dB PSNR. Achieves 92\% bandwidth reduction.
         
         \textcolor{red}{$\times$}: The GDM's iterative sampling process takes several seconds to generate a sample, making the framework more suitable for offline applications or requiring further acceleration.
        \\
        \hline
    \end{tabular}
\end{table}

Vehicular networks are a primary application for task-oriented communication, driven by the significant safety and efficiency gains from GDMs. The environment's high mobility and bandwidth constraints make bit-level sensor data transmission impractical. This has spurred GDM-based approaches focused on transmitting minimal, high-value semantic data. Table ~\ref{tab:gdm_vehicular_summary} details representative works.

A key challenge in vehicular GAI is reliability. Zhang et al.\cite{zhang2024generative} address this by proposing a multi-modal framework that transmits both semantic text and a structural ``image skeleton" to prevent the GDM from generating images that discrepancy from the actual scene. The receiver's GDM is conditioned on both modalities, ensuring the generated road images are structurally reliable for safety decisions.
Similarly, \cite{grassucci2023generative} presents the GESCO framework, which synthesizes photorealistic scenes from highly compressed semantic segmentation maps. The core DDPM modification is the integration of spatially-adaptive normalization to inject the noisy semantic map as a spatial condition at multiple layers. By explicitly training the DDPM on corrupted maps, the system achieves extreme robustness, generating recognizable objects even at 1 dB PSNR.

Moving from single-sensor synthesis to multi-sensor fusion, Lu et al.\cite{lu2024generative11.9} introduce the G-MSC framework for vehicular networks. This system uses Bird's Eye View  Fusion to create a unified spatial representation from cameras and LiDAR/radar. The DDPM at the decoder then plays two roles: ``refinement," to enhance the clarity of the noisy BEV map, and ``prediction," to forecast future BEV maps using timestamps as prompts.

\begin{figure*}
\centering
\includegraphics[width=1\textwidth]{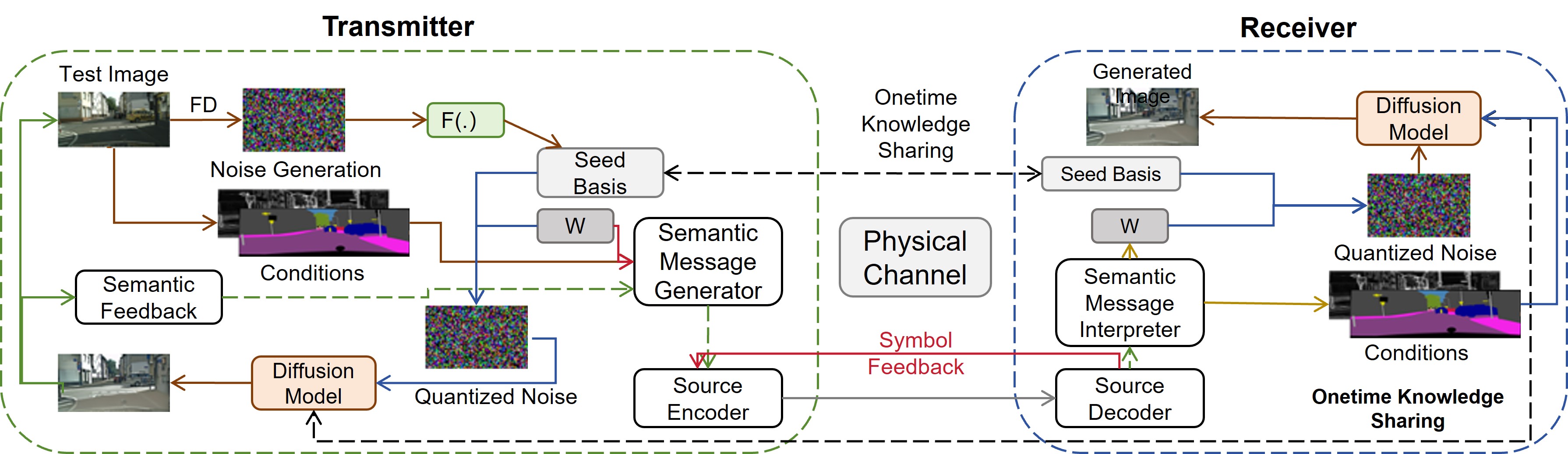}
\caption{Illustration of the Diff-GO system \cite{wijesinghe2024diff11.8}. This framework employs a pre-trained CDM at the transmitter to locally simulate and validate the generation process before transmission. Specifically, the transmitter projects the noise latent of the input image onto a low-dimensional quantized noise space spanned by a shared seed basis and transmits only the corresponding weights along with semantic conditions to the receiver. The receiver then reconstructs the exact noise latent using these lightweight weights and the shared basis to drive its GDM, thereby achieving ultra-high spectrum efficiency while ensuring the generated content meets specific goal-oriented quality requirements.}
\label{picture_Diff_GO}
\end{figure*}

Finally, Wijesinghe et al.\cite{wijesinghe2024diff11.8} propose a novel goal-oriented framework centered on trading computation for bandwidth efficiency, as shown in Figure \ref{picture_Diff_GO}. To solve the problem of uncontrollable generative quality at the receiver, this framework embeds a full DDPM at the transmitter for local generative feedback. The transmitter validates the generated output locally against a QoS metric. It then transmits only the semantic conditions and a small set of compressed weights from a quantized noise space, guaranteeing the receiver reconstructs the exact validated image.

\subsubsection{GDM for Immersive Communication Applications}
\begin{table}[tbp]
    \centering
    \caption{Existing GDM-based schemes for Immersive Communication applications, where \textcolor{blue}{$\diamondsuit$}, \textcolor{blue}{$\sphericalangle$}, \textcolor{green}{$\checkmark$}, and \textcolor{red}{$\times$} respectively denote contributions, the role of GDM, pros, and cons.}
    \label{tab:gdm_immersive_summary}
    \renewcommand{\arraystretch}{1.2}
    \normalsize
    \begin{tabular}{|>{\arraybackslash}m{0.025\textwidth}|>{\arraybackslash}m{0.42\textwidth}|}
        \hline
        \textbf{\small{Ref.}} & \textbf{\small{Descriptions}} \\
        \hline
        \footnotesize \cite{zhang2024diffusion} & \footnotesize
         \textcolor{blue}{$\diamondsuit$}: Propose a DSCVI SemCom system for VR that integrates dual-fisheye image transmission with panoramic stitching.
         
         \textcolor{blue}{$\sphericalangle$}: Acts as a conditional generative decoder at the receiver. It is guided by a Multi-Scale Semantic Condition Extractor that injects feature maps at four different scales into the U-Net to generate a stitched panoramic image.
         
         \textcolor{green}{$\checkmark$}: Achieves higher PSNR, SSIM, and compression efficiency than traditional JPEG plus stitching methods. It is extremely robust to channel noise, maintaining high quality at low SNR.
         
         \textcolor{red}{$\times$}: Incurs computational latency at the receiver due to the iterative denoising process required by the GDM to generate the final stitched image.
        \\
        \hline
        \footnotesize \cite{jiang2024large} & \footnotesize
         \textcolor{blue}{$\diamondsuit$}: Propose a GAM-3DSC system for 3D immersive applications that uses generative models for semantic extraction, compression, and channel estimation.
         
         \textcolor{blue}{$\sphericalangle$}: Used as a CSI refiner within the GDCE component. It takes the coarse CSI generated by a CGAN as input and performs a denoising process to output a high-accuracy, refined CSI.
         
         \textcolor{green}{$\checkmark$}: The GDM-based refinement helps achieve the lowest NMSE in channel estimation compared to all benchmarks, enhancing signal recovery and overall system reliability.
         
         \textcolor{red}{$\times$}: High system complexity due to a multi-stage pipeline of generative models. The DM adds significant computational overhead for the channel estimation task.
        \\
        \hline
    \end{tabular}
\end{table}
Immersive communication, such as VR and 3D applications, presents extreme bandwidth and processing challenges. GDMs are being applied for both end-to-end content generation and to optimize crucial sub-tasks within the immersive pipeline. Table ~\ref{tab:gdm_immersive_summary} summarizes these approaches.

\begin{figure*}
\centering
\includegraphics[width=0.8\textwidth]{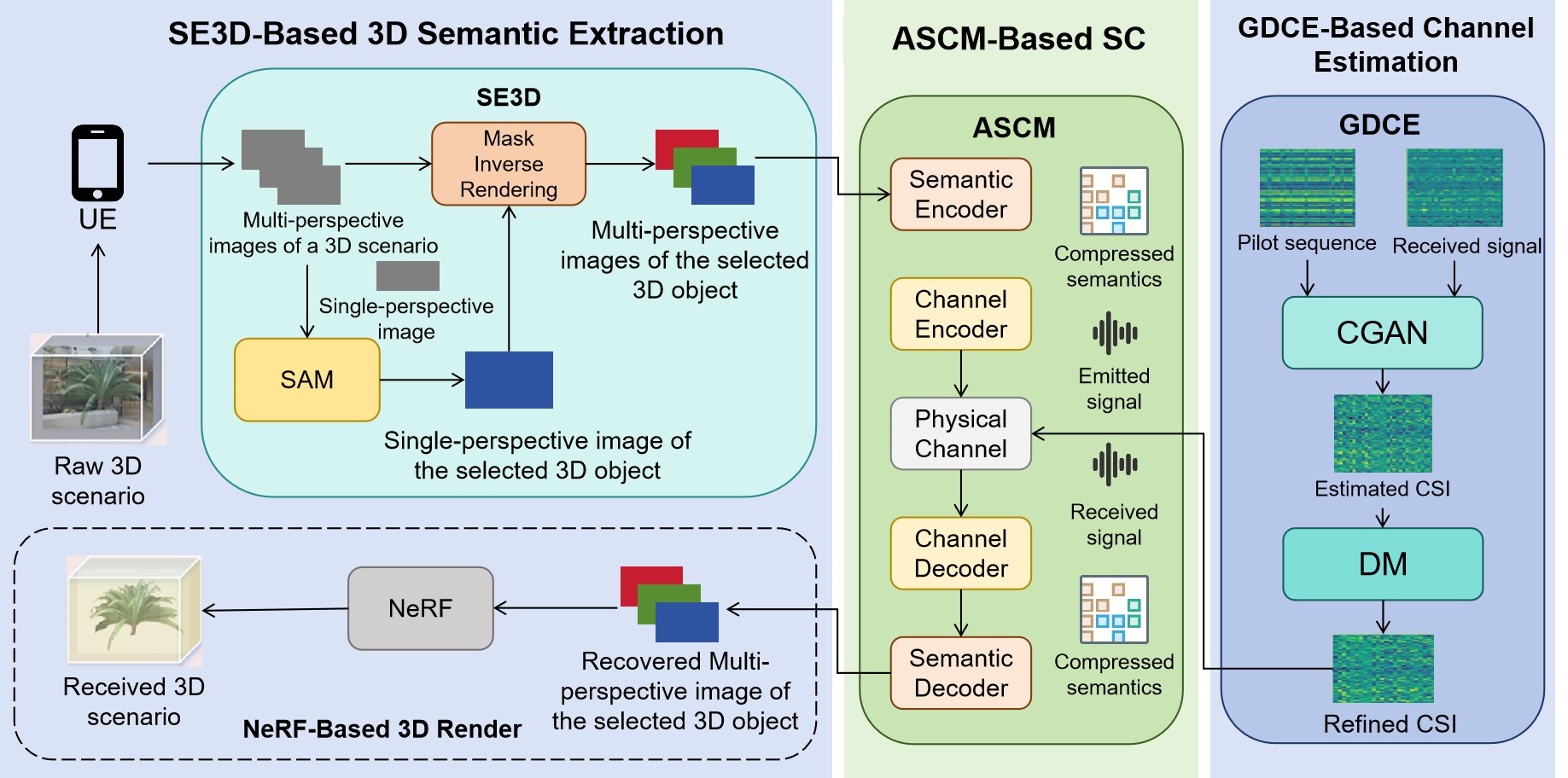}
\caption{The workflow of the GAM-3DSC system\cite{jiang2024large} for 3D semantic communication. This framework integrates multiple GAI models to address key challenges in 3D transmission. It comprises three core modules: (1) SE3D-Based 3D Semantic Extraction (Left): Utilizes the Segment Anything Model for goal-oriented object selection and mask inverse rendering to extract multi-perspective semantic images from the raw 3D scenario; (2) ASCM-Based SemCom (Middle): Employs an Adaptive Semantic Compression Model with dual-head encoders to compress feature-level redundancy and transmit semantics efficiently; and (3) GDCE-Based Channel Estimation (Right): Leverages a Conditional GAN  to estimate CSI from pilots, followed by a DDPM refiner to enhance CSI accuracy for robust signal recovery and final NeRF-based 3D rendering.}
\label{picture_GAM_3DSC}
\end{figure*}

As an example of end-to-end generation, Zhang et al.\cite{zhang2024diffusion} propose the DSCVI system for VR images. This system innovatively integrates panoramic stitching with data transmission. A DDIM acts as a joint decoder and stitcher, guided by a multi-scale semantic condition extractor that injects features at four scales into the U-Net. This allows the GDM to learn the complex geometric deformation required for stitching while simultaneously decoding the image from the channel.
In contrast, Jiang et al.\cite{jiang2024large} utilize GDM to optimize a critical component within an immersive system,as shown in Figure \ref{picture_GAM_3DSC}. Their GAM-3DSC framework for AR/MR applications uses a DDPM in a highly specialized role: as a CSI refiner. A CGAN first provides a coarse CSI estimate, which is then fed into the DM. The DM executes a reverse diffusion process to produce refined CSI. This demonstrates GDM's utility not just for user-facing content, but for enhancing the underlying reliability of the communication link itself.

\begin{table*}[tbp]
\centering
\caption{Comparison and Insights of GDM for Vertical Applications and Services}
\label{tab:gdm_vertical_comparison}
\renewcommand{\arraystretch}{1.3}
\footnotesize

\begin{tabularx}{\textwidth}{ p{3.5cm} >{\raggedright\arraybackslash}X >{\raggedright\arraybackslash}X }
\toprule
\textnormal{Approach \& Refs} & \textnormal{Core Role \& Best-Suited Scenario} & \textnormal{Pros \& Cons / Trade-offs} \\
\midrule

\multicolumn{3}{l}{\textbf{A. Network Digital Twin Simulation}} \\
\cmidrule(r){1-3}

GDM as World Model / Simulator \cite{chai2024generative11.8, zhang2024netdiff, zhou2025user} 
& 
\textbf{Core Role:} Generates high-fidelity simulation data, including network states, user traffic, and atypical scenarios. 
\par\vspace{4pt}
\textbf{Scenario:} Offline network planning, virtual testbeds for RL agents, and robust user-centric model evaluation. 
& 
\textbf{Pros:} Accurately captures complex dynamics and user intent while avoiding GAN pattern collapse. Enables robust offline evaluation. 
\par\vspace{4pt}
\textbf{Cons:} High inference latency during state generation. A potential sim-to-real gap exists. \\

\noalign{\vskip 4pt} 
\cdashline{1-3}
\noalign{\vskip 4pt} 

GDM as DT Decision-Maker \cite{huang2024digital} 
& 
\textbf{Core Role:} Functions as a policy network within the DT that uses inverse dynamics to generate management actions. 
\par\vspace{4pt}
\textbf{Scenario:} Intelligent closed-loop network management where decision quality outweighs real-time latency. 
& 
\textbf{Pros:} Achieves better convergence in closed-loop management compared to standard DRL algorithms. 
\par\vspace{4pt}
\textbf{Cons:} High computational overhead and a "black box" nature. The iterative process adds significant latency. \\
\midrule

\multicolumn{3}{l}{\textbf{B. Network Policy \& Solution Optimization}} \\
\cmidrule(r){1-3}

GDM as Direct Solver \cite{liang2025diffsg11.8, liang2025diffusion11.8} 
& 
\textbf{Core Role:} Acts as an end-to-end optimizer that learns the distribution of high-quality solutions for complex problems. 
\par\vspace{4pt}
\textbf{Scenario:} Complex non-convex optimization problems, such as MINLP, where solution data can be precomputed. 
& 
\textbf{Pros:} Remains insensitive to non-convexity and robust on out-of-domain inputs, outperforming discriminative models. 
\par\vspace{4pt}
\textbf{Cons:} Relies on a precomputed supervised dataset of input-solution pairs and suffers from iterative latency. \\

\noalign{\vskip 4pt}
\cdashline{1-3}
\noalign{\vskip 4pt}

GDM as DRL Policy Network \cite{zhang2025enhanced11.8, zhang2024multi11.8, yu2025multi11.8, sun2024generative11.8} 
& 
\textbf{Core Role:} Replaces the DRL actor network by denoising a random vector into an optimal action, such as a beamforming vector. 
\par\vspace{4pt}
\textbf{Scenario:} High-dimensional dynamic control problems where solution quality is more critical than single-pass latency. 
& 
\textbf{Pros:} Effectively escapes local optima and models complex multi-objective trade-offs more effectively than MLPs. 
\par\vspace{4pt}
\textbf{Cons:} Action selection becomes an iterative denoising process, which adds significant latency. \\
\midrule

\multicolumn{3}{l}{\textbf{C. Task-Oriented Generative Comm.}} \\
\cmidrule(r){1-3}

Task-Oriented Generator \cite{zhang2024generative, lu2024generative11.9, wijesinghe2024diff11.8, grassucci2023generative, zhang2024diffusion} 
& 
\textbf{Core Role:} A generative decoder that synthesizes high-fidelity content from minimal, semantic-only inputs. 
\par\vspace{4pt}
\textbf{Scenario:} Bandwidth-starved applications, such as vehicular networks or immersive communication, where semantic intent is sufficient. 
& 
\textbf{Pros:} Achieves massive bandwidth reduction by trading bandwidth for computation and remains robust to channel noise. 
\par\vspace{4pt}
\textbf{Cons:} Shifts an enormous computational load to the receiver. Generation latency can be high and the model may fail to reconstruct exact details. \\

\bottomrule
\end{tabularx}
\end{table*}

\subsection{Summaries and Lessons Learned}

In this section, we reviewed GDM-based schemes for vertical applications and services. Our review reveals GDM's role evolving from a foundational component in the sensing and transmission layers to a key enabling technology for large-scale, 6G-oriented ecosystems. This shift highlights a new set of trade-offs, primarily balancing GDM's function as a high-fidelity environment simulator, an active policy optimizer, and a task-specific content generator. As detailed in Table \ref{tab:gdm_vertical_comparison}, existing approaches demonstrate how GDMs are being adapted for these advanced, system-level roles.

\begin{itemize}
\item 
\textbf{Insight 1: GDM as a Passive Simulator vs. an Active Agent.} A primary distinction in this section, not seen previously, is GDM's dual role as both an environment and an agent. In network digital twins, GDMs function as passive ``world models" \cite{chai2024generative11.8, zhang2024netdiff, zhou2025user}. They leverage their high-fidelity generative power to simulate complex, dynamic network states and user behaviors, serving as a virtual testbed for other algorithms. Conversely, in network policy optimization, the GDM becomes an active decision-maker. It functions as the agent itself, replacing traditional DRL actors to generate optimal policies or decisions, such as beamforming vectors or UAV trajectories, by sampling from a learned distribution of high-quality solutions \cite{huang2024digital, zhang2025enhanced11.8, yu2025multi11.8}.
\item 
\textbf{Insight 2: The Trade-off between Optimization and Generation.} When GDMs are used as optimizers, they introduce a powerful but costly paradigm. By modeling the entire distribution of good solutions, GDM-based actors can effectively escape local optima \cite{zhang2025enhanced11.8} and model complex, multi-objective trade-offs in NP-hard problems \cite{zhang2024multi11.8}, yielding superior solutions. This high optimization quality, however, comes at a significant cost: decision generation is no longer a single, fast forward pass but an iterative denoising process. This introduces substantial computational latency, creating a fundamental bottleneck for real-time control loops in dynamic environments like UAV networks \cite{zhang2024multi11.8, zhang2025beam11.8}.
\item 
\textbf{Insight 3: From ``Communication-as-Reconstruction'' to ``Communication-as-Generation''.} This insight clearly distinguishes this section from the transmission layer. The transmission layer focuses on reconstruction of using GDM to recover a noisy source signal with high fidelity. In contrast, vertical applications employ GDM for task-oriented generation. The system no longer aims to perfectly reconstruct an original image; instead, it transmits minimal semantic goals, such as a segmentation map for a vehicular system \cite{grassucci2023generative, zhang2024generative} or a layout for a VR scene \cite{zhang2024diffusion}. The receiver's GDM then generates entirely new, photorealistic content that is useful for the task, even if it differs from the original. This provides extreme robustness and bandwidth efficiency \cite{wijesinghe2024diff11.8}, but the core metric shifts from source fidelity to task utility.
\end{itemize}

The pervasive challenge across all three roles is the high iterative latency of GDM's inference process. Whether generating the next network state in a DT \cite{chai2024generative11.8}, generating an optimal policy action \cite{liang2025diffusion11.8}, or generating a photorealistic scene \cite{lu2024generative11.9}, this multi-step sampling remains the primary bottleneck. For DTs and policy optimizers, this latency makes real-time network control currently unfeasible. Furthermore, a critical sim-to-real gap emerges; policies trained within a GDM-generated virtual environment may fail when deployed against the physics and unmodeled dynamics of the real world \cite{chai2024generative11.8}.

Future work should therefore center on accelerating GDM-based policy generation, likely by developing consistency models explicitly for decision-making tasks, not just image synthesis. For DTs, grounded simulation frameworks that can assimilate real-world data streams to correct the sim-to-real gap are essential. Finally, hybrid architectures may offer a practical path forward, perhaps using a fast, traditional algorithm for initial policy and a GDM for iterative refinement, balancing the urgent speed requirements of 6G services with the high solution quality GDMs can provide.

\section{GDM for the Security Plane}
\label{section6}
This section provides an overview of existing GDM-based schemes for protecting wireless network security, as well as existing schemes for protecting wireless networks that support GDM.

\subsection{Applications of GDM in Wireless Network Security}
With unique complex distribution learning, robust denoising, and controllable generation capabilities, GDM has become an important technical support for strengthening wireless network security protection. It can adapt to the dynamic and changing environmental characteristics of wireless networks, providing flexible and efficient protection ideas against various security threats. From ensuring the trustworthy operation of key network links to identifying abnormal risks and resisting malicious interference, it comprehensively helps to enhance the security resilience of wireless networks and provides new solutions to address diverse network security challenges.

\subsubsection{Defenses of Eavesdropping Attacks}

\begin{table}[tbp]
    \centering
\caption{Defenses of Eavesdropping Attacks, where \textcolor{blue}{$\diamondsuit$}, \textcolor{blue}{$\sphericalangle$}, \textcolor{green}{$\checkmark$}, and \textcolor{red}{$\times$} respectively are contributions, the role of the model, pros, and cons.}
    \label{Defenses of Eavesdropping Attacks}
    \renewcommand{\arraystretch}{1.2}
    \normalsize
    \begin{tabular}{|>{\arraybackslash}m{0.025\textwidth}|>{\arraybackslash}m{0.42\textwidth}|}
        \hline
        \small{Ref.} & \small{Descriptions} \\
        \hline
    \footnotesize \cite{he2025diffusion} & \footnotesize
     \textcolor{blue}{$\diamondsuit$}: Propose a GDM empowered anti eavesdropping secure SemCom framework, which balances privacy protection and communication quality by combining artificial noise with the forward and backward processes of the GDM.
    
     \textcolor{blue}{$\sphericalangle$}: The pluggable encryption module generates artificial noise and adds it to the output of the semantic transmitter, while the pluggable decryption module before the semantic receiver utilizes DDPM to generate detailed semantic information by removing artificial noise and channel noise.

     \textcolor{green}{$\checkmark$}: The pluggable encryption module prevents semantic eavesdropping, while the pluggable decryption module achieves high-quality SemCom.

     \textcolor{red}{$\times$}: It does not integrate more advanced GAI technologies such as Transformer, and there is still room for further improvement in noise estimation and denoising efficiency of DDPM.\\
    \hline

    \footnotesize \cite{du2024generative} & \footnotesize
     \textcolor{blue}{$\diamondsuit$}: Propose GAI assisted secure SemCom framework without joint training and resource allocation scheme based on GDM.
    
      \textcolor{blue}{$\sphericalangle$}: By utilizing the inverse deterministic process of DDIM, the original image is encoded as a noisy image as a visual cue, which complements the textual cue to improve reconstruction accuracy.

     \textcolor{green}{$\checkmark$}: Solving the high computational overhead in SemCom while ensuring the security and accuracy of data transmission.

     \textcolor{red}{$\times$}: The adaptability of the scene is relatively limited, and current research mainly focuses on the semantic transfer of facial images, without verifying and expanding the adaptability of other SemCom scenarios such as text and speech. \\
    \hline

    \footnotesize \cite{gao2025semstediff} & \footnotesize
     \textcolor{blue}{$\diamondsuit$}: Design a steganographic SemCom scheme based on CDM , and verify its anti-eavesdropping effectiveness through case studies.
    
      \textcolor{blue}{$\sphericalangle$}: CDM directly generates encrypted images, eliminating the reliance on traditional steganography on carrier images, while controlling secret information recovery through private keys to confuse eavesdroppers.

     \textcolor{green}{$\checkmark$}: Outstanding anti-intelligent eavesdropping capability and strong compatibility and integration.

     \textcolor{red}{$\times$}: The size and quality of the carrier image directly limit the capacity of secret information. Improper selection and updating of the carrier can easily arouse suspicion from eavesdroppers and be detected by advanced steganalysis techniques. \\
    \hline

    \end{tabular}
\end{table}

In this and the next subsubsection, we focus on the application of GDMs in securing SemCom. In SemCom, the transmitted semantic features often retain distinct statistical patterns closely aligned with the original data, enabling eavesdroppers to infer sensitive information without full decoding\cite{guo2024survey,meng2025secure,zhang2025srec,meng2026secure,meng2025survey}. To mitigate this risk, GDM-based defenses have evolved beyond simple noise injection by leveraging the unique capabilities of the model in distribution mapping, latent encoding, and conditional generation. These schemes are summarized in Table \ref{Defenses of Eavesdropping Attacks}.

He et al. \cite{he2025diffusion} propose a diffusion enabled pluggable encryption framework. The core innovation lies in redefining the forward diffusion process where the model maps AN such as adversarial perturbations and physical channel noise directly into the forward diffusion steps. Consequently, the legitimate receiver employs a tailored reverse denoising process trained to reconstruct detailed semantic information from this specific noisy mixture while eavesdroppers perceive only unintelligible noise.

Securing the transmission of multi-modal prompts including text and visual cues requires protecting both the content and the transmission behavior. Du et al. \cite{du2024generative} introduce a GAI-aided framework that utilizes GDMs in two distinct roles. First, for content security, they employ the deterministic inversion of DDIM to encode images into latent noise maps serving as abstract visual prompts that conceal structural details. Second, for transmission security, they utilize a conditional GDM as a policy generator. By conditioning on the CSI, the GDM iteratively denoises a random distribution to generate optimal resource allocation schemes such as jamming and transmit power thereby ensuring the communication remains covert against warden detection.

\begin{figure*}[htbp]
\centering
\includegraphics[width=1\textwidth]{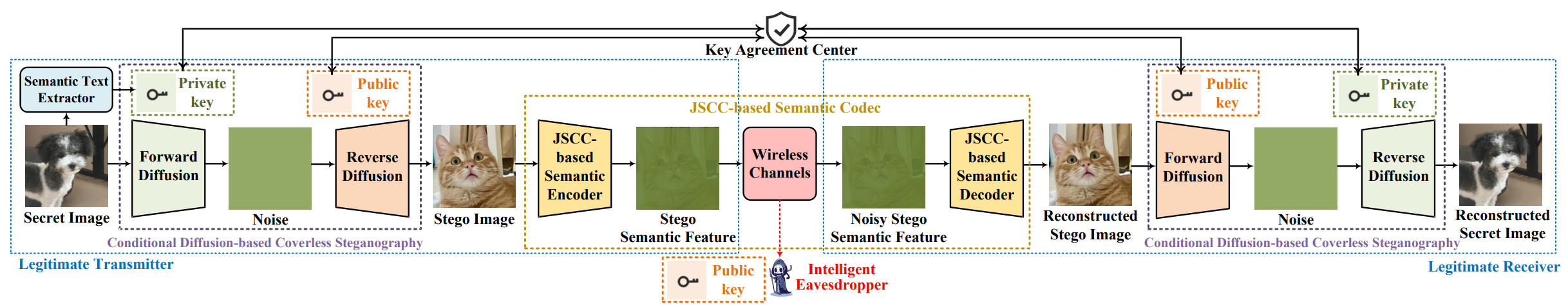}
\caption{Illustration of the CDM-based coverless steganography SemCom \cite{gao2025semstediff}, where a secret image is first encoded into a visually natural stego image through CDM guided by private and public keys. The stego image is then transmitted over channels using the JSCC. At the receiver, the secret image is recovered from the reconstructed stego image using CDM, which employs public and private keys as the conditions of the forward and reverse processes.}
\label{GDM for Eavesdropping Protection and Secure Transmission}
\end{figure*}

Unlike the method of directly encrypting and transmitting content,
steganography conceals the very existence of the message\cite{wang2025image}. Motivated by this, Gao et al.\cite{gao2025semstediff} propose a coverless steganography scheme based on CDMs, as shown in Figure \ref{GDM for Eavesdropping Protection and Secure Transmission}. This approach abandons traditional carrier modification and instead leverages the high sensitivity of CDMs to conditional inputs. The transmitter generates a realistic-looking stego image based on a public semantic prompt serving as a public key to confuse eavesdroppers. The legitimate receiver possessing the correct private prompt as a private key guides the reverse diffusion process to reconstruct the hidden secret image. This method achieves invisible encryption by ensuring that without the specific semantic condition the generative process cannot converge to the target secret.

\subsubsection{Defenses of Adversarial Attacks}

\begin{table}[tbp]
    \centering
\caption{Defenses of Adversarial Attacks, where \textcolor{blue}{$\diamondsuit$}, \textcolor{blue}{$\sphericalangle$}, \textcolor{green}{$\checkmark$}, and \textcolor{red}{$\times$} respectively are contributions, the role of the model, pros, and cons.}
    \label{Defenses of Adversarial Attacks}
    \renewcommand{\arraystretch}{1.2}
    \normalsize
    \begin{tabular}{|>{\arraybackslash}m{0.025\textwidth}|>{\arraybackslash}m{0.42\textwidth}|}
        \hline
        \small{Ref.} & \small{Descriptions} \\
        \hline
    \footnotesize \cite{ren2023asymmetric} & \footnotesize
     \textcolor{blue}{$\diamondsuit$}: Propose a secure SemCom system DiffuSeC to cope with adversarial disturbances.
    
     \textcolor{blue}{$\sphericalangle$}: The sender uses a dynamically adjustable diffusion step size to gradually add Gaussian noise to fuse and dilute the adversarial perturbations caused by data source attacks. The receiving end uses a reverse denoising process based on DDPM to simultaneously remove noise and adversarial interference, restoring the original image semantics.

     \textcolor{green}{$\checkmark$}: Strong anti attack generalization and good channel adaptability.

     \textcolor{red}{$\times$}: High system complexity and limited adaptation to scenarios and modalities. \\
    \hline

    \footnotesize \cite{ren2024diffusion} & \footnotesize
     \textcolor{blue}{$\diamondsuit$}: Propose a secure SemCom system based on diffusion purification to address security issues caused by semantic perturbations.
    
      \textcolor{blue}{$\sphericalangle$}: The sender gradually adds Gaussian noise with a fixed step size, fuses and dilutes the invisible adversarial perturbations caused by data source attacks. The receiving end uses a symmetric fixed reverse process based on DDPM, while removing noise and adversarial perturbations, to restore the original image semantics.

     \textcolor{green}{$\checkmark$}: Strong anti-attack generalization and good compatibility.

     \textcolor{red}{$\times$}: Single scene adaptation and insufficient flexibility in step size. \\
    \hline

    \footnotesize \cite{qiu2025plugging} & \footnotesize
     \textcolor{blue}{$\diamondsuit$}: Propose a PBNet defense system to address the two core issues of retraining and difficulty in adapting to time-varying channels in DLSC system defense.
    
      \textcolor{blue}{$\sphericalangle$}: The pluggable protector uses DDPM as the backbone and achieves adversarial disturbance purification through forward noise addition and reverse noise reduction, restoring the original semantics of the attacked signal.

     \textcolor{green}{$\checkmark$}: Defend against known and unknown physical layer adversarial attacks without retraining, and have good adaptability to time-varying channels.

     \textcolor{red}{$\times$}: Insufficient adaptation of the system to semantically ambiguous scenarios, as well as limitations in protecting against latency when faced with high concurrency and high throughput practical communication scenarios. \\
    \hline

    \footnotesize \cite{zhao2025secdiff} & \footnotesize
     \textcolor{blue}{$\diamondsuit$}: Propose a plug and play, diffusion assisted decoding framework called SecDiff.
    
      \textcolor{blue}{$\sphericalangle$}: Based on DDIM, the number of denoising steps is reduced through skip sampling, which not only reduces latency but also utilizes generative capabilities to recover damaged semantic features and adapt to the distortion characteristics of wireless channels.

     \textcolor{green}{$\checkmark$}: Significantly enhance the security and robustness of JSCC in adversarial wireless environments.

     \textcolor{red}{$\times$}: Modal and scene adaptation are single and hyperparameter dependent on manual tuning. \\
    \hline

    \end{tabular}
\end{table}

Adversarial attacks disrupt the integrity and confidentiality of communication by maliciously constructing interference signals or tampering with transmitted data which causes the receiver to misinterpret information or fail to communicate normally \cite{do2025security,10159517}.
To mitigate these threats, GDM-based defense utilizes forward diffusion and backward denoising models to purify disturbances. Related schemes are summarized in Table \ref{Defenses of Adversarial Attacks}.

The first strategy employs an asymmetric diffusion process to neutralize semantic perturbations from both data sources and channels. Ren et al. \cite{ren2023asymmetric, ren2024diffusion} propose the DiffuSeC system which deviates from the standard symmetric diffusion paradigm. The sender injects Gaussian noise to submerge adversarial perturbations while the receiver executes a tailored asymmetric reverse denoising process to recover the original semantics. To address the instability of wireless channels, the system integrates a DRL-based controller that adaptively adjusts the diffusion steps based on real-time CSI. This dynamic step selection mechanism enables the model to balance the trade-off between thorough perturbation removal and semantic fidelity preservation under fluctuating channel conditions.

\begin{figure*}
    \centering
    \includegraphics[width=0.85\textwidth]{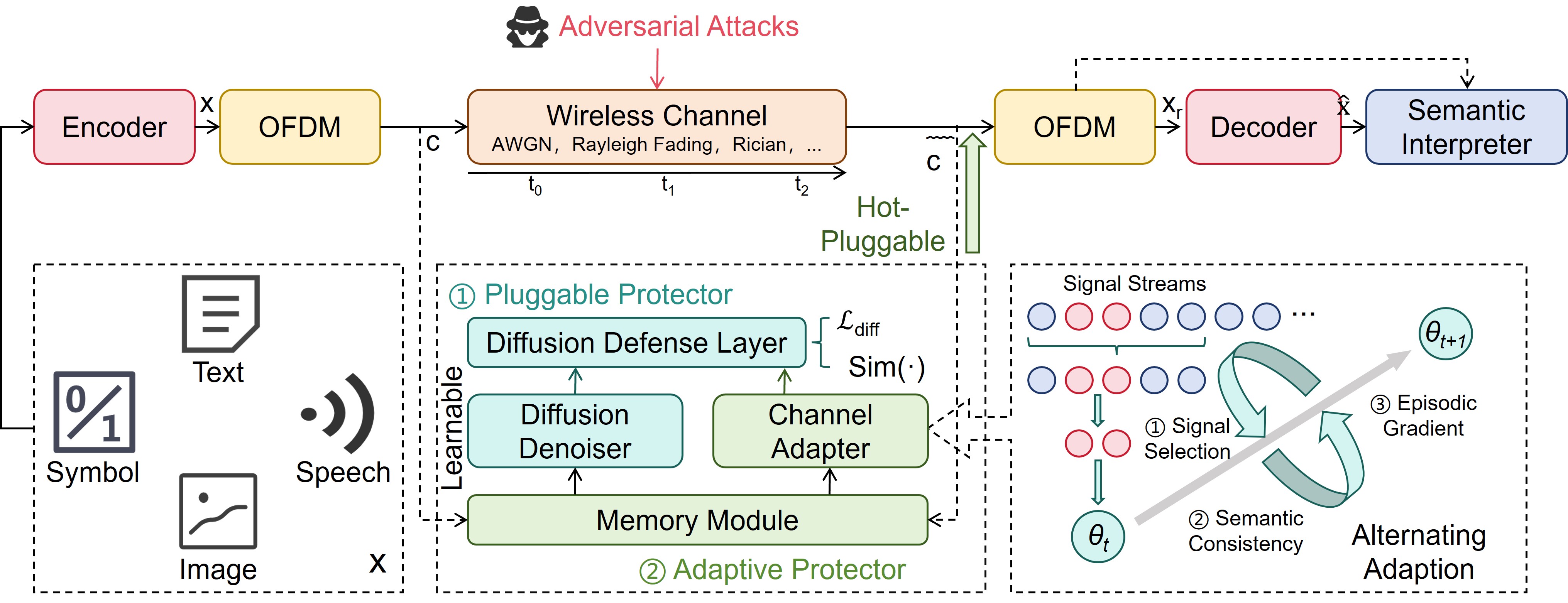}
    \caption{
    The architecture of the PBNet framework for securing wireless SemCom \cite{qiu2025plugging}. This system deploys a hot-pluggable defense mechanism at the receiver composed of two coordinated units to ensure robustness against physical-layer attacks and channel variations. The Pluggable Protector shown in the center utilizes a learnable Diffusion Defense Layer and Denoiser to purify received signals corrupted by adversarial attacks and channel noise into clean semantic features without interrupting online transmission. The Adaptive Protector depicted at the bottom and right facilitates continual learning over time-varying channels by employing a Memory Module and a Channel Adapter. This adapter executes an Alternating Adaption strategy that iteratively updates the model parameters through Signal Selection, Semantic Consistency checks mimicking working memory, and Episodic Gradient replay mimicking stable memory to dynamically balance plasticity and stability in fluctuating wireless environments.}
    \label{GDM for Defense System with Plug-and-Play Function}
\end{figure*}

Addressing the operational challenge where traditional defenses require retraining the entire semantic model, the second strategy focuses on pluggable defense mechanisms that maintain online service continuity. Qiu et al. \cite{qiu2025plugging} introduce the PBNet framework which integrates a pluggable protector with an adaptive protector, as shown in Figure \ref{GDM for Defense System with Plug-and-Play Function}. In this architecture, the GDM functions as an independent denoising module trained with a hybrid objective combining diffusion loss and semantic similarity loss. This design allows the protector to be hot-plugged into existing DLSC systems to filter physical layer adversarial attacks without altering the core codec parameters. Furthermore, an alternating adaptation strategy enables the model to update its defense capabilities against time-varying channels and unknown attack patterns.

The third strategy reformulates the defense against specific physical layer threats as a guided generative inverse problem. Targeting vulnerabilities in OFDM systems such as subcarrier jamming and pilot spoofing, Zhao et al. \cite{zhao2025secdiff} propose the SecDiff framework. This method modifies the standard sampling process by introducing pseudoinverse guidance which leverages the physical channel structure to steer the generation trajectory. For jamming attacks, the framework utilizes a power-based masking strategy to treat corrupted subcarriers as missing regions and solves the recovery as a masked inpainting task. For pilot spoofing, an EM-driven algorithm alternates between signal reconstruction and channel operator refinement within the diffusion process to ensure robust decoding even when CSI is compromised.

\subsubsection{Securing ISAC Networks} 

\begin{table}[tbp]
    \centering
\caption{Securing ISAC Networks, where \textcolor{blue}{$\diamondsuit$}, \textcolor{blue}{$\sphericalangle$}, \textcolor{green}{$\checkmark$}, and \textcolor{red}{$\times$} respectively are contributions, the role of the model, pros, and cons.}
    \label{Securing ISAC Networks}
    \renewcommand{\arraystretch}{1.2}
    \normalsize
    \begin{tabular}{|>{\arraybackslash}m{0.025\textwidth}|>{\arraybackslash}m{0.42\textwidth}|}
        \hline
        \small{Ref.} & \small{Descriptions} \\
        \hline
    \footnotesize \cite{wang2024generative} & \footnotesize
     \textcolor{blue}{$\diamondsuit$}: Propose a security awareness system DFSS based on GDM.
    
     \textcolor{blue}{$\sphericalangle$}: Generate graph structures through discrete CDM to accurately activate optimal links and nodes. Continuous CDM uses activated links, nodes, and user positions as conditions to gradually denoise and generate composite protection signals from Gaussian noise.

     \textcolor{green}{$\checkmark$}: DFSS can reduce the activity recognition accuracy of unauthorized devices by about 70\%, effectively protecting users from illegal monitoring.

     \textcolor{red}{$\times$}: The cost of obtaining high-quality training data is high, the GDM is time-consuming and resource intensive, and it affects the estimation of some signal parameters. \\
    \hline

    \footnotesize \cite{zhang2025enhanced} & \footnotesize
     \textcolor{blue}{$\diamondsuit$}: Propose a novel actor-critic algorithm combined with a GDM named AC-GDM.
    
      \textcolor{blue}{$\sphericalangle$}: The AC-GDM algorithm utilizes denoising process to recover the optimal BF solution from the inherent Gaussian noise in multi-user wireless channel environment.

     \textcolor{green}{$\checkmark$}: Significantly reduce the impact of channel spatial correlation on the security of IoT communication systems.

     \textcolor{red}{$\times$}: Performance is limited in extreme scenarios, computational complexity is high and relies on hardware resources, attack scenario coverage is single and intensive, and user scenario optimization is insufficient.\\
    \hline

    \end{tabular}
\end{table}

ISAC systems face dual security challenges regarding the eavesdropping on communication data and the illegitimate sensing of user privacy via CSI. To address these threats, GDM-based approaches have been developed to synthesize protective signals and optimize complex beamforming strategies, which are summarized in Table \ref{Securing ISAC Networks}.

Targeting the threat of unauthorized sensing, Wang et al. \cite{wang2024generative} propose the DFSS system which leverages a layered generative approach to shield user activities. The core innovation lies in the deployment of two distinct GDMs. A discrete CDM is first employed to generate graph structures that guide the optimal activation of sensing links and nodes. Subsequently, a continuous CDM synthesizes diverse safeguarding signals which are modulated onto the pilot symbols at the transmitter. These signals effectively mask the CSI fluctuations caused by user movements. Since only authorized devices possessing the synchronized generation seed can cancel out these safeguarding signals, the system effectively prevents illegitimate sensing without requiring additional hardware jammers.

As one of the core technologies of ISAC, beamforming can not only achieve high-precision perception and reliable communication, but also be used to ensure the security and privacy of ISAC. Addressing the security of the transmission component within intelligent reflective environments, Zhang et al. \cite{zhang2025enhanced} propose the AC-GDM framework for IRS-assisted secure beamforming. The spatial correlation between legitimate and eavesdropping channels often compromises traditional physical layer security measures. To overcome the non-convexity of maximizing the secrecy rate under imperfect CSI, this method integrates GDM into an Actor-Critic reinforcement learning architecture. The GDM functions as a policy generation network which iteratively denoises random Gaussian noise conditioned on the channel state to reconstruct the optimal precoding matrix and IRS phase shift matrix. This generative optimization approach allows the system to escape local optima and significantly improve the minimum achievable secrecy rate in dynamic wireless environments.

\subsubsection{Identity Management} 

\begin{table}[tbp]
    \centering
\caption{Identity Management, where \textcolor{blue}{$\diamondsuit$}, \textcolor{blue}{$\sphericalangle$}, \textcolor{green}{$\checkmark$}, and \textcolor{red}{$\times$} respectively are contributions, the role of the model, pros, and cons.}
    \label{Identity Management}
    \renewcommand{\arraystretch}{1.2}
    \normalsize
    \begin{tabular}{|>{\arraybackslash}m{0.025\textwidth}|>{\arraybackslash}m{0.42\textwidth}|}
        \hline
        \small{Ref.} & \small{Descriptions} \\
        \hline
    \footnotesize \cite{meng2025generative} & \footnotesize
     \textcolor{blue}{$\diamondsuit$}: Design a GDM based PLA case study to verify the superiority of GAI in PLA.
    
     \textcolor{blue}{$\sphericalangle$}: Implementing fingerprint estimation, enhancement, and denoising reconstruction through GDM, and eliminating fingerprint triggering features through denoising process to defend against backdoor attacks.

     \textcolor{green}{$\checkmark$}: Resolve the core pain points of PLA and have strong adaptability to multiple scenarios and tasks.

     \textcolor{red}{$\times$}: High computational requirements and significant latency issues, inherent security risks in the model, difficulty balancing security performance and complexity, and insufficient adaptability to emerging scenarios.\\
    \hline

    \footnotesize \cite{yin2025noise} & \footnotesize
     \textcolor{blue}{$\diamondsuit$}: Effectively restoring RFF in low SNR scenarios using GDMs.
    
     \textcolor{blue}{$\sphericalangle$}: The network uses forward noise to simulate signal states under different SNRs. The reverse denoising process predicts noise through an improved HDT model. The receiving end takes the noise signal as the intermediate state input for diffusion, while removing the noise and preserving the unique RFF features caused by hardware defects in the device.

     \textcolor{green}{$\checkmark$}: Efficient and faithful denoising, and adaptable to complex noisy environments.

     \textcolor{red}{$\times$}: The scene and equipment coverage are single, and the experimental conditions are idealized. \\
    \hline

    \footnotesize \cite{wang2023diffusion} & \footnotesize
     \textcolor{blue}{$\diamondsuit$}: Propose an integrated open set recognition architecture that includes diffusion module, denoising module, out-of-distribution detection module, and classifier.
    
     \textcolor{blue}{$\sphericalangle$}: The diffusion module destroys the input signal into a Gaussian prior distribution, while the denoising module restores the corresponding Gaussian distribution to the original data.

     \textcolor{green}{$\checkmark$}: Strong adaptability to open environments, and higher stability and reliability.

     \textcolor{red}{$\times$}: The dataset and scene coverage are single, and the anti-interference scene coverage is insufficient. \\
    \hline

    \end{tabular}
\end{table}

Identity management ensures network security by verifying device legitimacy through physical layer authentication. While effective, traditional PLA faces challenges from environmental dynamics\cite{meng2024survey}, high noise levels, and unknown threats. GDM-based approaches address these issues by leveraging capabilities in conditional prediction, signal denoising, and distribution reconstruction \cite{cheng2026apeg}. Related approaches are illustrated in Table \ref{Identity Management}.

Addressing the challenge of fingerprint obsolescence in dynamic environments, Meng et al. \cite{meng2025generative} propose an adaptive PLA framework integrating channel extrapolation with CDMs. Instead of relying on static historical data, the model learns the joint distribution of channel fingerprints between a legitimate user and a collaborator. During the authentication phase, the CDM functions as a conditional generator that predicts the legitimate user's current CSI fingerprints based on the collaborator's real-time fingerprints. This generative extrapolation mechanism allows the system to maintain high authentication reliability even as signal-to-noise ratios and channel conditions fluctuate rapidly.

Targeting the extraction of subtle hardware impairments in low-SNR scenarios, Yin et al. \cite{yin2025noise} employ a CDM-based noise predictor to restore radio frequency fingerprints. The core innovation is an SNR mapping algorithm that aligns the noise level of the received signal with a specific diffusion timestep. By initiating the deterministic reverse process from this mapped timestep, the model effectively strips away environmental noise to recover distinct device-specific features. This generative denoising approach significantly enhances robustness against noise and achieves superior accuracy compared to traditional discriminative methods at 0 dB SNR.

For scenarios involving unauthorized or unknown emitters, Wang et al. \cite{wang2023diffusion} utilize the generative reconstruction capability of DDPMs for open-set identification. The architecture consists of a diffusion module that degrades input signals into a Gaussian prior and a denoising module tasked with restoring them. The system leverages the reconstruction error as an out-of-distribution detection metric where known emitters yield low errors and unknown emitters result in high errors due to the distribution mismatch. This method effectively identifies unknown transmitters in dynamic wireless environments and outperforms traditional threshold-based classifiers.

\subsection{Securing GDM-based Wireless Networks}
With the deep integration of GDM in wireless networks, potential security risks emerging from their training and deployment processes have attracted significant attention. The core objective of this direction is to build a security guarantee system tailored to the characteristics of GDM which avoids introducing new vulnerabilities while ensuring stable operation. As illustrated in Table \ref{Protecting Networks based on GDM}, existing research primarily focuses on securing the federated deployment environment and preserving the privacy of training data.

\begin{table}[tbp]
    \centering
\caption{Protecting Networks based on GDM, where \textcolor{blue}{$\diamondsuit$}, \textcolor{blue}{$\sphericalangle$}, \textcolor{green}{$\checkmark$}, and \textcolor{red}{$\times$} respectively are contributions, methods to protect GDM, pros, and cons.}
    \label{Protecting Networks based on GDM}
    \renewcommand{\arraystretch}{1.2}
    \normalsize
    \begin{tabular}{|>{\arraybackslash}m{0.025\textwidth}|>{\arraybackslash}m{0.42\textwidth}|}
        \hline
        \small{Ref.} & \small{Descriptions} \\
        \hline
    \footnotesize \cite{he2024securing} & \footnotesize
     \textcolor{blue}{$\diamondsuit$}: Proposed a comprehensive GDM architecture SS-Diff to prevent trigger based security threats.
    
     \textcolor{blue}{$\sphericalangle$}: Compressing local model parameters through dynamic quantization mechanism to reduce the risk of leakage and communication energy consumption during model transmission. Simultaneously embedding a trigger detection module to filter malicious noise input.

     \textcolor{green}{$\checkmark$}: Universal and efficient security protection, strong deployment compatibility, and optimal balance between energy consumption and performance.

     \textcolor{red}{$\times$}: Limited device and scene adaptation, and insufficient real-time performance. \\
    \hline

    \footnotesize \cite{wangprivacy} & \footnotesize
     \textcolor{blue}{$\diamondsuit$}: Propose a hybrid training method applied to the GDM of Wi-Fi data as a defense against MIA.
    
     \textcolor{blue}{$\sphericalangle$}: Pre-train a CDM on the raw data to preserve key signal features, selectively fine tune the attention module and embedding module by applying differential privacy, and then perform parallel collaborative optimization process to counteract the impact of noise.

     \textcolor{green}{$\checkmark$}: Reduce the success rate of member inference attacks while maintaining high fidelity data generation performance.

     \textcolor{red}{$\times$}: The complexity of parameter tuning is high, and the denoising model's ability is limited. \\
    \hline

    \end{tabular}
\end{table}

\subsubsection{Securing Federated Deployment}
The deployment of GDMs in multi-access AIoT environments often relies on federated learning which introduces vulnerabilities to backdoor attacks from malicious devices and imposes severe communication energy costs due to large parameter sizes. To simultaneously address security and efficiency, He et al. \cite{he2024securing} propose the SS-Diff architecture. In the training phase, the framework incorporates dynamic quantization into the federated aggregation process to minimize communication overhead. For security, the authors exploit the multi-step nature of the diffusion process by designing a detection-based defense strategy during collaborative sampling. By analyzing the Euclidean distance between inputs and their intermediate denoising trajectories, the system identifies and filters out malicious trigger inputs associated with backdoor attacks. This approach effectively secures the federated diffusion pipeline while significantly enhancing energy efficiency suitable for resource-constrained IoT systems.

\subsubsection{Preserving Data Privacy}
Generative models trained on sensitive wireless data such as Wi-Fi signals face the risk of privacy leakage through Membership Inference Attacks (MIA) where attackers determine if specific samples were used in training. Standard differential privacy methods often degrade the high-fidelity requirements of complex-valued wireless data. To resolve the trade-off between privacy and utility, Wang et al. \cite{wangprivacy} propose a hybrid training framework combining CDMs with differential privacy. The method begins with standard pre-training to capture essential signal characteristics followed by a fine-tuning phase where Differentially Private Stochastic Gradient Descent (DP-SGD) is selectively applied only to the attention and embedding modules. Furthermore, a joint optimization module is introduced to counteract the quality degradation caused by the injected noise. This selective application of privacy constraints effectively reduces the success rate of MIA while maintaining the physical integrity of the generated wireless data.

\subsection{Summaries and Lessons Learned}
\begin{table*}[tbp]
\centering
\caption{Comparison and Insights of GDM's Applications in the Security Plane}
\label{tab:gdm_security_comparison}
\renewcommand{\arraystretch}{1.3}
\footnotesize

\begin{tabularx}{\textwidth}{ p{3.2cm} >{\raggedright\arraybackslash}X >{\raggedright\arraybackslash}X }
\toprule
\textnormal{Approach \& Refs} & \textnormal{Core Role \& Best-Suited Scenario} & \textnormal{Pros \& Cons / Trade-offs} \\
\midrule

\multicolumn{3}{l}{\textbf{A. Defenses of Eavesdropping Attacks}} \\
\cmidrule(r){1-3}

Forward Process Mapping \cite{he2025diffusion} 
& 
\textbf{Core Role:} Redefines the forward diffusion process to map artificial noise and channel noise into the generation steps for active encryption. 
\par\vspace{4pt}
\textbf{Scenario:} Physical layer key generation scenarios. 
& 
\textbf{Pros:} Achieves high security by confusing eavesdroppers with unintelligible noise. 
\par\vspace{4pt}
\textbf{Cons:} Requires precise estimation of noise parameters to ensure successful reverse denoising. \\

\noalign{\vskip 4pt}
\cdashline{1-3}
\noalign{\vskip 4pt}

Generative Steganography \& Covert Encoding \cite{du2024generative, gao2025semstediff} 
& 
\textbf{Core Role:} Uses conditional generation or deterministic inversion to hide secret semantics within cover features. 
\par\vspace{4pt}
\textbf{Scenario:} Covert communication. 
& 
\textbf{Pros:} Achieves invisible encryption and covertness by generating realistic cover content. 
\par\vspace{4pt}
\textbf{Cons:} May limit the embedding capacity of secret information; susceptible to advanced steganalysis. \\
\midrule

\multicolumn{3}{l}{\textbf{B. Defenses of Adversarial Attacks}} \\
\cmidrule(r){1-3}

Asymmetric \& Pluggable Purification \cite{ren2023asymmetric, ren2024diffusion, qiu2025plugging} 
& 
\textbf{Core Role:} Acts as an independent purification module that recovers clean semantics. 
\par\vspace{4pt}
\textbf{Scenario:} Online communication systems requiring hot-pluggable defense against physical layer attacks. 
& 
\textbf{Pros:} Offers strong generalization against various attack patterns; enables adaptive defense strategies via DRL. 
\par\vspace{4pt}
\textbf{Cons:} Introduces additional inference latency due to the iterative purification process. \\

\noalign{\vskip 4pt}
\cdashline{1-3}
\noalign{\vskip 4pt}

Guided Inverse Solving \cite{zhao2025secdiff} 
& 
\textbf{Core Role:} Reformulates defense as a guided inverse problem. 
\par\vspace{4pt}
\textbf{Scenario:} OFDM systems facing specific physical layer threats such as subcarrier jamming and pilot spoofing. 
& 
\textbf{Pros:} High robustness in high-interference environments by leveraging physical channel structures for guidance. 
\par\vspace{4pt}
\textbf{Cons:} High complexity in tuning the guidance strength and solving the inverse problem. \\
\midrule

\multicolumn{3}{l}{\textbf{C. Securing ISAC Networks}} \\
\cmidrule(r){1-3}

Protective Signal Generation \cite{wang2024generative} 
& 
\textbf{Core Role:} Generates safeguarding signals via CDMs to mask CSI fluctuations caused by user activities. 
\par\vspace{4pt}
\textbf{Scenario:} Privacy-sensitive sensing environments. 
& 
\textbf{Pros:} Effectively prevents illegitimate sensing without requiring additional hardware jammers. 
\par\vspace{4pt}
\textbf{Cons:} Generating high-fidelity protective signals is resource-intensive. \\

\noalign{\vskip 4pt}
\cdashline{1-3}
\noalign{\vskip 4pt}

Secure Beamforming Optimization \cite{zhang2025enhanced} 
& 
\textbf{Core Role:} Functions as a policy generation network to optimize beamforming matrices under secrecy constraints. 
\par\vspace{4pt}
\textbf{Scenario:} IRS-assisted networks. 
& 
\textbf{Pros:} Capable of escaping local optima to maximize secrecy rates in non-convex optimization landscapes. 
\par\vspace{4pt}
\textbf{Cons:} High computational complexity for training. \\
\midrule

\multicolumn{3}{l}{\textbf{D. Identity Management}} \\
\cmidrule(r){1-3}

Dynamic \& Robust Fingerprint Modeling \cite{meng2025generative, yin2025noise} 
& 
\textbf{Core Role:} Uses conditional generation to restore fingerprints from low-SNR signals. 
\par\vspace{4pt}
\textbf{Scenario:} High-mobility environments with rapid channel variation or low-SNR scenarios where RFF is submerged. 
& 
\textbf{Pros:} Maintains high authentication accuracy in dynamic environments by adapting to channel changes and noise. 
\par\vspace{4pt}
\textbf{Cons:} Requires real-time collaboration or precise SNR mapping which adds system overhead. \\

\noalign{\vskip 4pt}
\cdashline{1-3}
\noalign{\vskip 4pt}

Open-Set Identification \cite{wang2023diffusion} 
& 
\textbf{Core Role:} Leverages the reconstruction error of GDM as an out-of-distribution metric to identify unknown emitters. 
\par\vspace{4pt}
\textbf{Scenario:} Dynamic open environments where unauthorized or unknown devices frequently attempt access. 
& 
\textbf{Pros:} Superior capability in distinguishing unknown devices compared to threshold-based classifiers. 
\par\vspace{4pt}
\textbf{Cons:} Performance depends heavily on the diversity of the training distribution. \\
\midrule

\multicolumn{3}{l}{\textbf{E. Protecting Networks based on GDM}} \\
\cmidrule(r){1-3}

Securing Federated Deployment \cite{he2024securing} 
& 
\textbf{Core Role:} Filters backdoor triggers and utilizes dynamic quantization for efficient updates. 
\par\vspace{4pt}
\textbf{Scenario:} Federated learning environments with potential malicious nodes and resource constraints. 
& 
\textbf{Pros:} Simultaneously ensures model integrity against backdoors and improves communication efficiency. 
\par\vspace{4pt}
\textbf{Cons:} Defense mechanisms may impact the diversity of generated samples. \\

\noalign{\vskip 4pt}
\cdashline{1-3}
\noalign{\vskip 4pt}

Privacy-Preserving Training \cite{wangprivacy} 
& 
\textbf{Core Role:} Applies selective differential privacy to specific modules to prevent membership inference. 
\par\vspace{4pt}
\textbf{Scenario:} Training generative models on sensitive wireless datasets where data privacy is paramount. 
& 
\textbf{Pros:} Significantly reduces the success rate of membership inference attacks. 
\par\vspace{4pt}
\textbf{Cons:} Trade-off between the strength of privacy protection and the fidelity of generated data. \\

\bottomrule
\end{tabularx}
\end{table*}

In this section, we systematically sort out two core applications of GDM in wireless network security: one is GDM serving as a security enhancement tool to empower protection across all network layers, and the other is constructing protection mechanisms for the security risks brought by GDM’s own deployment. Centered on the dual roles of defending and being defended, it presents distinct technical trade-offs and application characteristics. As shown in Table \ref{tab:gdm_security_comparison}, various methods collectively demonstrate the dual enabling value of GDM in the field of wireless network security.

\textbf{Insight 1: Empowerment as Defender and Protected Object.} GDM exhibits distinct dual attributes in the wireless network security system: it actively participates in security defense while needing to resist risks itself. As a defender, its core capabilities stem from robust denoising that filters maliciously interfered data \cite{ren2023asymmetric, ren2024diffusion, qiu2025plugging}, controllable generation that constructs secure verification environments \cite{wang2024generative, gao2025semstediff}, and complex distribution learning that accurately models normal network behavior patterns \cite{wang2023diffusion}.
As a protected object, GDM is vulnerable to attacks from malicious devices during the training phase \cite{he2024securing}. Such attacks can be successfully implemented with only a small number of devices, seriously affecting the model’s reliability. Meanwhile, the wireless network data generated by GDM may be memorized by the model, leading to privacy leakage risks \cite{wangprivacy}. In response, dynamic quantization \cite{he2024securing} and differential privacy \cite{wangprivacy} techniques have been introduced to form targeted defense measures.

\textbf{Insight 2: Trade-off between security capability and cost.} The security enhancement brought by GDM is not cost-free, with core trade-offs focusing on the matching degree between resource consumption and security benefits. Firstly, complex distribution learning and robustness optimization require substantial computing power support, and model training and real-time inference will occupy more server resources \cite{wang2024generative}. Then, in order to improve protection accuracy, GDM needs to conduct multi-level analysis and verification of network data, which may increase data transmission latency and bandwidth occupancy, affecting the real-time response capability of wireless networks \cite{ren2023asymmetric, zhao2025secdiff}. Finally, GDM needs to be compatible with existing network architectures and security devices, which increases its deployment costs \cite{qiu2025plugging}. Therefore, dynamic adjustments should be made according to scenario requirements in practical applications. For example, in high-security-level scenarios, priority can be given to ensuring security capabilities with moderate tolerance for cost increases; in ordinary civil scenarios, the model architecture needs to be simplified to balance security and usability.

\textbf{Insight 3: The particularity of scene adaptation.} The effectiveness of GDM’s security applications is highly dependent on scenario characteristics and data modalities, and existing schemes still have obvious adaptation limitations \cite{du2024generative, zhao2025secdiff}. Security requirements vary significantly across different wireless network scenarios, so GDM’s model parameters and protection strategies need to be deeply bound to scenario requirements; otherwise, security redundancy or insufficient protection will occur. In addition, most studies focus on single scenarios or single modalities, lacking adaptation designs for cross-scenario and cross-modal scenarios. In complex dynamic environments, network topology, data types, and attack methods are all in dynamic changes. GDM needs to possess adaptive \cite{ren2023asymmetric}, lightweight, and cross-modal characteristics to improve versatility and stability in complex scenarios.

In the future, the development of GDM in the field of wireless network security will not only strive to enhance its protection effectiveness as a defender but also optimize the matching degree between security benefits and costs in different scenarios. It will break through the application limitations of single scenarios and single modalities, improve versatility and stability in complex dynamic network environments, and ultimately become a multi-scenario-adaptive, low-overhead, and highly reliable core supporting technology for wireless network security.

\section{Challenges and Future Research Directions}
\label{section7}

\subsection{Conflict Between Inference Latency and Real-Time Constraints}

\subsubsection{Challenges}
GDM's ability to generate high-fidelity samples stems from its iterative denoising process, which often requires tens to hundreds of sequential steps. In image synthesis, a user may tolerate this delay. However, in wireless networks, the environment state, such as CSI, is governed by physics and evolves rapidly, often within a coherence time of just a few milliseconds.
This creates a critical failure point: for example, a GDM used for channel estimation or generation, is still executing step $t$ of its reverse process, the actual physical channel may have already evolved from $H_1$ to $H_2$. By the time the GDM generates a ``high-fidelity" sample of $H_1$ hundreds of milliseconds later, the sample is already ``Stale on Arrival." Any decision based on this outdated information, such as beamforming or resource allocation, will be suboptimal or even detrimental to the network's performance.

Alongside latency, a more prohibitive barrier for edge deployment is the sheer computational cost and energy consumption. Unlike traditional discriminative models like ResNet-50 ($\approx$ 4 GFLOPs), GDMs entail computational complexity that is orders of magnitude higher \cite{tang2025diff11.22}. For instance, a single inference step of the LDM U-Net requires approximately 730 GOPs, and a standard 50-step generation involves tens of Tera-operations \cite{yang2024sda11.22}. Quantitative experiments reveal that generating a single sample on an embedded ARM Cortex-A53 CPU takes roughly 3.5 hours and consumes excessive energy, only 0.0014 iter/s/W \cite{yang2024sda11.22}. This computational burden far exceeds the power budget of battery-operated IoT nodes, rendering ``out-of-the-box'' deployment of full-precision GDMs infeasible for resource-constrained wireless devices.

\subsubsection{Potential Solutions and Future Directions}
\begin{itemize}
    \item \textbf{Ultra-Fast Samplers for Wireless Utility:} Future research can move beyond optimizing for perceptual quality and focus on wireless task utility. This involves validating few-step or single-step samplers, such as CMs mentioned in Section II-A-3\cite{luo2023latentconsistencymodelssynthesizing11.14,song2023consistencymodels}, and evaluating if their generated samples produced under extreme latency constraints are sufficiently accurate for downstream tasks.
    \item \textbf{Integrating GDM's Latency as Network Parameters:} The GDM's inference time, $T_{infer}$, can be treated as a critical network parameter, not an afterthought. Research is needed to model the relationship between $T_{infer}$ and the channel's coherence time $T_{coherent}$. Future GDM architectures should be capable of dynamically adjusting iteration steps based on channel volatility, creating a real-time trade-off between generation accuracy and timeliness.

    \item \textbf{GDM-Specific Hardware Acceleration with Low-Bit Quantization:} To bridge the feasibility gap, future research can explore dedicated hardware accelerators, like FPGA or ASIC, that leverage aggressive quantization and specialized dataflows, compressing weights to 4-bit and activations to 8-bit integers. Recent works like SDA \cite{yang2024sda11.22} and Diff-Acc \cite{tang2025diff11.22} demonstrate that executing GDMs on edge FPGAs using low-bit quantization can reduce model parameter size by approximately $8\times$ and improve energy efficiency by $12.5\times \sim 27.9\times$ compared to edge GPUs and CPUs. By adopting novel architectures such as Hybrid Systolic Arrays \cite{yang2024sda11.22} or Group-wise Parallelism \cite{tang2025diff11.22}, the inference speed can be accelerated by nearly two orders of magnitude, $97.3\times$ speedup over ARM CPU. Co-designing these hardware accelerators with channel-aware, few-step algorithms will be the key to driving per-iteration latency into the sub-millisecond range for real-time wireless control.
\end{itemize}

\subsection{Architectural Mismatch: CV Backbones vs. Wireless Physical Structure}

\subsubsection{Challenges}
GDM's practical success relies heavily on its U-Net backbone, an architecture originally designed for image segmentation. However, directly applying this architecture, which is optimized for 2D ``pixel grids," to the wireless domain creates a profound representation gap: a mismatch between the U-Net's convolutional nature and the physical structure of wireless data, such as CSI.
A MIMO CSI matrix or a beamforming vector is not merely a collection of pixels. It is a high-dimensional physical entity whose elements possess strong, physics-based correlations across the antenna, spatial, and frequency domains.
Standard U-Nets, which use stacked convolutional layers to capture local spatial features, are effective for image textures but may be inefficient and non-interpretable when tasked with understanding the correlation between antennas in a CSI matrix or the propagation patterns in a radio map. This inefficiency forces GDMs to  require more data to learn these underlying physical relationships, thereby exacerbating the high demand for computational resources.

\subsubsection{Potential Solutions and Future Directions}
\begin{itemize}
    \item \textbf{Development of New GDM Backbones:} Future research can move beyond the standard U-Net. This involves exploring the integration of Transformers\cite{peebles2023scalablediffusionmodelstransformers}, Graph Neural Networks\cite{li2025gnnenabledrobusthybridbeamforming}, or Tensor Networks directly into the GDM's denoising step. For instance, a GNN could explicitly model the topology of an antenna array, while a Transformer's self-attention mechanism might be more adept at capturing the long-range dependencies within a CSI matrix.
    \item \textbf{Physics-Aware Representation Learning:} Investigate methods to inject the prior physical structure of wireless data, such as the low-rank properties of CSI or channel covariance, directly into the GDM's representation space. This might involve designing new normalization layers or attention modules that "understand" antenna indices and physical locations, rather than treating them as simple grid coordinates.
\end{itemize}

\subsection{Inconsistency Between Gaussian Noise Assumption and Channel Physics}

\subsubsection{Challenges}

The core theory of GDMs is built on a forward process that incrementally corrupts data by injecting Gaussian noise. The learned reverse process is then trained to invert this specific Gaussian process. However, the ``damage" inflicted on a signal by a wireless channel is far from simple Gaussian noise. It is a complex, structured, and often non-Gaussian physical process, encompassing multi-path fading, non-Gaussian interference, phase noise, and quantization noise from Analog-to-Digital Converters (ADCs).
Consequently, a GDM trained to reverse Gaussian noise is fundamentally mismatched when asked to reverse physical fading. This mismatch may prevent the GDM from accurately modeling the true channel distribution in the sensing layer or effectively recovering channel-corrupted semantic features in the transmission layer.

\subsubsection{Potential Solutions and Future Directions}
\begin{itemize}
    \item \textbf{Enhancement of Physics-Informed GDMs:} Future work can further explore physics-informed or channel-aware GDMs\cite{böck2025physicsinformedgenerativemodelingwireless11.14}. This involves embedding the physical laws of wireless channels, such as fading models or Maxwell's equations, as a prior into the GDM's training objective, a concept that has seen initial exploration in radio map construction\cite{jia2025rmdm}.
    \item \textbf{Redefining the Forward Process:} Research is needed to replace the standard Gaussian noise injection \cite{nachmani2021nongaussiandenoisingdiffusion11.14} with new forward processes that more accurately simulate physical channel impairments, such as a process that models Rayleigh fading or ADC quantization.
    \item \textbf{Training of the Inverse Physical Process:} Correspondingly, the GDM's reverse process should be re-designed and trained to learn the inverse of the complex physical channel. Initial efforts in this direction, such as the CDDM designed to match specific channel noise statistics, already exist\cite{wu2024cddm}.
\end{itemize}

\subsection{Endogenous Security Risks and Process Vulnerabilities}

\subsubsection{Challenges}
The iterative denoising mechanism of GDMs could introduce a novel attack that do not present in conventional DL models: attacks targeting the dynamic sampling process itself\cite{dai2024advdiffgeneratingunrestrictedadversarial}.
In a traditional DL model, a backdoor attack is often static: one trigger leads to one incorrect output. However, in a GDM, an adversary can design a process amplification attack. They may only need to inject a subtle, adversarially-crafted trigger noise that is imperceptible at any single step. The GDM's U-Net, in its multi-step denoising process, may not only fail to remove this perturbation but instead progressively amplify and steer it toward a malicious, attacker-defined state\cite{11036312}.
For example, in an intelligent transportation application, a minor perturbation to a GDM might be amplified over denoising steps to generate an incorrect traffic sign. In secure beamforming, a manipulated GDM could be steered to output a vector that leaks information to an eavesdropper. Such attacks, which leverage the dynamics of the generation process, are far more covert and difficult to defend against than static backdoors.

\subsubsection{Potential Solutions and Future Directions}
\begin{itemize}
    \item \textbf{Trajectory Integrity Verification against Process Amplification:} Current defenses largely focus on static inputs which are ineffective against process amplification attacks where errors accumulate dynamically. Future research should develop trajectory integrity verification mechanisms that scrutinize the intermediate states of the denoising chain. By establishing mathematical bounds for acceptable inter-step deviations such as analyzing the Euclidean distance of intermediate trajectories in federated settings, defenders can identify and prune malicious branches where adversarial noise is being progressively amplified rather than waiting to detect the final corrupted sample \cite{he2024securing}.

    \item \textbf{Selective Privacy Preservation for High-Fidelity Wireless Data:} Standard DP often degrades the physical integrity of complex wireless data like CSI phase information required for downstream tasks. A critical future direction is the development of selective privacy-preserving architectures. Instead of applying global noise, research should focus on hybrid training frameworks that selectively apply DP-SGD only to sensitive modules including attention layers or embeddings while leaving physical feature extractors intact \cite{wangprivacy}. This approach aims to minimize MIA success rates without compromising the physical validity of the generated channel.

    \item \textbf{Secure and Efficient Federated Deployment:} As GDMs move to the wireless edge via federated learning, they face dual challenges of backdoor attacks and communication overhead. Future solutions can move beyond simple weight aggregation. Research should explore dynamic quantization-integrated aggregation protocols that simultaneously compress model parameters for energy efficiency and filter out malicious trigger updates from compromised edge devices during the collaborative training process.
\end{itemize}

\subsection{Evaluation Metric Misalignment: From Perceptual Fidelity to Task Utility}

\subsubsection{Challenges}

In image synthesis, metrics like PSNR, SSIM, or FID are used to quantify the ``visual fidelity" of a generated sample. Influenced by this, many wireless network studies have also adopted these metrics. However, a CSI matrix that is ``high-fidelity" under PSNR may be nonsensical in its physical meaning. For example, the matrix might have the correct magnitude but be entirely wrong in its phase information, a component that has low weight in a PSNR calculation but is decisive for beamforming.
Therefore, a GDM-generated channel sample might look perfect according to statistical metrics, but when used for a downstream wireless task, such as calculating a beamforming vector, the resulting system BER may be difficult to meet network requirements.

\subsubsection{Potential Solutions and Future Directions}
\begin{itemize}
    \item \textbf{Transitioning from Perceptual Fidelity to Wireless Utility:} While metrics like PSNR and SSIM provide a baseline for signal reconstruction quality, relying solely on them is insufficient for wireless networks where physical properties, like phase alignment, are paramount. Future research should treat these visual metrics as preliminary indicators rather than definitive proof of performance, shifting the primary evaluation focus toward domain-specific validity.
    
    \item \textbf{Establishing Task-Utility Benchmarks:} Future research can establish a new suite of evaluation benchmarks oriented around downstream wireless task utility \cite{baur2024evaluationmetricsmethodsgenerative}. For instance, a GDM for channel generation should not be judged by the MSE of its CSI matrix, but by the final BER performance of a beamforming network trained on its generated data\cite{guler2025multitaskfoundationmodelwireless,kim2023diffusion}. Similarly, a GDM for radio map construction can be evaluated by the localization accuracy supported by its generated map.
\end{itemize}

\section{Conclusions}
\label{section8}
In this survey, we first introduced the concept, advantages, and mathematical principles of GDMs. Subsequently, we presented a structured taxonomy that systematically categorizes GDM-based schemes into the sensing layer, transmission layer, vertical applications, and security plane. Furthermore, we comprehensively surveyed existing GDM-based schemes for wireless networks, covering channel estimation, channel generation, and radio map construction in the sensing layer; semantic denoiser, auxiliary recovery, semantic-based generation, multimodal transmission, and resource allocation in the transmission layer; network digital twin simulation, network policy optimization, and task-oriented generative communications in vertical applications; and specific defense mechanisms alongside GDM-enabled network protection in the security plane. Ultimately, we highlighted existing challenges and provided directional guidance for future research in GDM-enhanced wireless networks.


\section{Acknowledgment}
We would like to sincerely thank Zhidi Zhang, Xiqi Cheng, and Haixiao Gao from BUPT for their contributions to this article.

\bibliography{ref.bib}
\bibliographystyle{IEEEtran}

\vfill

\end{document}